\documentclass[preprint,5p,twocolumn,number]{elsarticle}

\usepackage{subcaption}
\usepackage{calc}
\usepackage{amssymb}
\usepackage{amstext}
\usepackage{amsmath}
\usepackage{amsthm}
\usepackage{multicol}

\usepackage{paralist}
\usepackage{csquotes} 

\usepackage{pifont}

\usepackage[table]{xcolor}
\usepackage{tabularx}
\usepackage{multicol}
\usepackage{multirow}
\usepackage{enumitem}
\usepackage[flushleft]{threeparttable}

\usepackage{tabularx,booktabs}
\newcolumntype{b}{X}
\newcolumntype{s}{>{\hsize=.4\hsize}X}
\newcolumntype{C}{>{\centering\arraybackslash}X} 

\usepackage{lineno,hyperref}
\usepackage[capitalise,nameinlink,noabbrev]{cleveref}

\usepackage{color}
\usepackage{xspace}

\newcommand{\eg}{e.g.,\xspace}
\newcommand{\ie}{i.e.,\xspace}

\newcommand{\managerial}{\textit{business}\xspace}
\newcommand{\devops}{\textit{technical}\xspace}

\newcommand{\license}[0]{\textit{Licensing}\xspace}
\newcommand{\licName}[0]{\textit{License}\xspace}
\newcommand{\licType}[0]{\textit{Type}\xspace}
\newcommand{\installation}[0]{\textit{Installation}\xspace}
\newcommand{\instType}[0]{\textit{Type}\xspace}
\newcommand{\instTargetHosts}[0]{\textit{Target Hosts}\xspace}
\newcommand{\sourceCode}[0]{\textit{Source Code}\xspace}
\newcommand{\srcAvailability}[0]{\textit{Availability}\xspace}
\newcommand{\srcHostingPlatform}[0]{\textit{Open Source Repository}\xspace}
\newcommand{\srcProgrammingLanguage}[0]{\textit{Programming Language}\xspace}
\newcommand{\release}[0]{\textit{Release}\xspace}
\newcommand{\relStatus}[0]{\textit{Status}\xspace}
\newcommand{\interface}[0]{\textit{Interface}\xspace}
\newcommand{\intType}[0]{\textit{Type}\xspace}

\newcommand{\intApplicationManagement}[0]{\textit{Application Management}\xspace}
\newcommand{\intPlatformAdministration}[0]{\textit{Platform Administration}\xspace}
\newcommand{\community}[0]{\textit{Community}\xspace}
\newcommand{\commGitHub}[0]{\textit{GitHub}\xspace}
\newcommand{\commStackoverflow}[0]{\textit{Stackoverflow}\xspace}
\newcommand{\documentation}[0]{\textit{Documentation}\xspace}
\newcommand{\docFunctions}[0]{\textit{Functions}\xspace}
\newcommand{\docPlatform}{\textit{Platform}\xspace}
\newcommand{\restrictions}[0]{\textit{Quotas}\xspace}
\newcommand{\resDeployment}[0]{\textit{Deployment}\xspace}
\newcommand{\resRuntime}[0]{\textit{Runtime}\xspace}
\newcommand{\dev}[0]{\textit{Development}\xspace}
\newcommand{\devruntime}[0]{\textit{Function Runtimes}\xspace}
\newcommand{\devruntimeextension}[0]{\textit{Runtime Customization}\xspace}
\newcommand{\devIDE}[0]{\textit{IDEs and Text Editors}\xspace}
\newcommand{\devSDK}[0]{\textit{Client Libraries}\xspace}
\newcommand{\vers}[0]{\textit{Version Management}\xspace}
\newcommand{\versapps}[0]{\textit{Application versions}\xspace}
\newcommand{\versfuncs}[0]{\textit{Function versions}\xspace}
\newcommand{\testdeb}[0]{\textit{Testing and Debugging}\xspace}
\newcommand{\testing}[0]{\textit{Testing}\xspace}
\newcommand{\debugging}[0]{\textit{Debugging}\xspace}

\newcommand{\eventsources}[0]{\textit{Event Sources}\xspace}
\newcommand{\eventendpoint}[0]{\textit{Endpoint}\xspace}
\newcommand{\eventendpointsync}[0]{\textit{Synchronous Call}\xspace}
\newcommand{\eventendpointAsync}[0]{\textit{Asynchronous Call}\xspace}
\newcommand{\eventendpointCust}[0]{\textit{Endpoint Customization}\xspace}
\newcommand{\eventendpointTLS}[0]{\textit{TLS Support}\xspace}
\newcommand{\eventscheduler}[0]{\textit{Scheduler}\xspace}
\newcommand{\eventdatastore}[0]{\textit{Data Store}\xspace}
\newcommand{\eventdatastorefile}[0]{\textit{File Level}\xspace}
\newcommand{\eventdatastoreDB}[0]{\textit{Database Mode}\xspace}
\newcommand{\eventmq}[0]{\textit{Message Queue}\xspace}
\newcommand{\eventstreaming}[0]{\textit{Stream Processing Platform}\xspace}
\newcommand{\eventspecservice}[0]{\textit{Special-purpose Service}\xspace}
\newcommand{\eventintegration}[0]{\textit{Event Source Integration}\xspace}
\newcommand{\orchestration}[0]{\textit{Function Orchestration}\xspace}
\newcommand{\orchestrationWF}[0]{\textit{Workflow Definition}\xspace}
\newcommand{\orchestrationConstructs}[0]{\textit{Control Flow Constructs}\xspace}
\newcommand{\orchestrationQuota}[0]{\textit{Quotas}\xspace}
\newcommand{\orchestrationQuotaExecTime}[0]{\textit{Execution Time}\xspace}
\newcommand{\orchestrationQuotaIO}[0]{\textit{Task Input and Output Size}\xspace}
\newcommand{\delivery}[0]{\textit{Application Delivery}\xspace}
\newcommand{\deliveryDepAut}[0]{\textit{Deployment Automation}\xspace}
\newcommand{\deliveryCICD}[0]{\textit{CI/CD Pipelining}\xspace}
\newcommand{\observability}[0]{\textit{Observability}\xspace}
\newcommand{\observeLog}[0]{\textit{Logging}\xspace}
\newcommand{\observeMon}[0]{\textit{Monitoring}\xspace}
\newcommand{\observeInt}[0]{\textit{Tooling Integration}\xspace}
\newcommand{\codereuse}[0]{\textit{Code Reuse}\xspace}
\newcommand{\codereuseMarkets}[0]{\textit{Function Marketplace}\xspace}
\newcommand{\codereuseRepos}[0]{\textit{Code Sample Repository}\xspace}
\newcommand{\accessmgmt}[0]{\textit{Access Management}\xspace}
\newcommand{\accessAuth}[0]{\textit{Authentication}\xspace}
\newcommand{\accessControl}[0]{\textit{Access Control}\xspace}

\newcommand{\ok}{\checkmark}
\newcommand{\ko}{$\times$}
\newcommand{\NA}{n/a\xspace}
\newcommand{\NS}{n/s\xspace}

\newcommand{\faastener}{\textsc{FaaStener}\xspace}
\journal{The Journal of Systems and Software}

\begin{document}

\begin{frontmatter}

\title{FaaSten Your Decisions: Classification Framework and \\ Technology Review of Function-as-a-Service Platforms}

\author[stuttgart]{Vladimir Yussupov}
\ead{yussupov@iaas.uni-stuttgart.de}

\author[pisa]{Jacopo Soldani}
\ead{soldani@di.unipi.it}

\author[stuttgart]{Uwe Breitenb{\"u}cher}
\ead{breitenbuecher@iaas.uni-stuttgart.de}

\author[pisa]{Antonio Brogi}
\ead{brogi@di.unipi.it}

\author[stuttgart]{Frank Leymann}
\ead{leymann@iaas.uni-stuttgart.de}

\address[stuttgart]{Institute of Architecture of Application Systems, University of Stuttgart, Germany}
\address[pisa]{Department of Computer Science, University of Pisa, Italy}

\begin{abstract}
Function-as-a-Service (FaaS) is a cloud service model enabling developers to offload event-driven executable snippets of code.
The execution and management of such functions becomes a FaaS provider's responsibility, hereby included their on-demand provisioning and automatic scaling.
Key enablers for this cloud service model are FaaS platforms, \eg AWS Lambda, Microsoft Azure Functions or OpenFaaS.
At the same time, the choice of the most appropriate FaaS platform for deploying and running a serverless application is not trivial, as various organizational and technical aspects have to be taken into account.
In this work, we present (i) a FaaS platform classification framework derived using a mixed method study and (ii) a systematic technology review of the ten most prominent FaaS platforms, based on the proposed classification framework.
Moreover, we present (iii) a FaaS platform selection support system, called \faastener, which helps researchers and practitioners to choose the FaaS platform most suited for their requirements.
\end{abstract}


\end{frontmatter}



\section{Introduction}
\label{sec:intro}
\noindent
In the context of cloud computing, the term \textit{serverless} is typically used to describe a paradigm focusing on cloud architectures that comprise provider-managed components~\cite{serverless:baldini2017serverless}.
From the developer's perspective, the decreased control over the infrastructure gives the impression that servers are no longer needed, whereas in fact servers become a provider's burden.
One exemplary serverless use case is when different non-managed component types, \eg{}~Database-as-a-Service and Software-as-a-Service, are combined to implement a Backend-as-a-Service~\cite{cncf:serverless-wp}.

Function-as-a-Service~(FaaS) is a cloud service model enabling the hosting of business logic in a serverless fashion, which makes this model an essential instrument for developing serverless applications.
By deploying event-driven, stateless and often short-lived functions to FaaS platforms, developers outsource the maintenance efforts to the corresponding platform provider.
As a consequence, functions are automatically scaled without any imposed limits on the amount of new instances.
Moreover, in case functions are no longer needed, they are scaled to zero instances, hence eliminating the need to pay for idle application components unlike in other service models that continuously run components, e.g., Platform-as-a-Service~\cite{Yussupov2019_SystematicMappingStudyFaaS}.

The FaaS service model started gaining a lot of attention after the release of AWS Lambda~\cite{aws:lambda} in 2015, which started the overall serverless trend. 
Afterwards, all major cloud providers introduced their FaaS offerings including notable examples such as Microsoft Azure Functions~\cite{ms:azure-functions}, Google Cloud Functions~\cite{google:cloud-functions}, and IBM Cloud Functions~\cite{ibm:cloud-functions}, with the latter based on the open source FaaS platforms Apache OpenWhisk~\cite{apache:openwhisk}.
The landscape of open source FaaS platforms is also blooming: Various alternatives such as OpenFaaS~\cite{openfaas}, Apache OpenWhisk~\cite{apache:openwhisk}, Kubeless~\cite{kubeless}, or Knative~\cite{knative} are being actively developed and maintained by the community.

However, in most cases the FaaS programming model is the only common denominator for all available offerings since the underlying technical platform characteristics and supported feature sets vary significantly.
For example, one of the main strength of proprietary FaaS platforms lies in the out-of-the-box integration with provider-specific services, which is typically not the case for open source platform offerings.
For instance, AWS Lambda can natively be combined with Amazon SQS to use message queues as sources triggering function execution.
In contrast, open source platforms such as Kubeless or OpenFaaS provide a more portable way to develop serverless applications.
They indeed reduce platform lock-in, favouring more portable serverless applications that are not locked into specific cloud offerings, \eg by relying on based on Kubernetes~\cite{kubernetes}, a portable application orchestrator, rather than on Amazon's or Microsoft's clouds.

As a result, choosing the most suitable platform becomes a decision problem involving multiple different dimensions, which affect both the high-level business perspective of project managers and the low-level technical perspective of application developers and operators.
For instance, high-level requirements include the license used by a FaaS platform, which heavily influence the choice of the platform when the latter is intended to be incorporated as an internal task-scheduling component for a company's product.
Various low-level technical details have to be analyzed too, which might become a serious obstacle for deciding which platform to choose. 
Examples of such technicalities are available event source integrations, platform-supported tooling, or restrictions on function execution times or on supported language runtimes.

Our objective here is precisely to provide a way to uniformly classify FaaS platforms with the goal of simplifying the decision-making process for specialists with different levels of responsibilities, i.e., managers and technical specialists.
To come up with an efficient classification mechanism, we conducted a mixed method study which combines the results from a systematic literature review and the documentation analysis of existing platforms to derive a \textit{FaaS Platform Classification Framework}, which we present as the first contribution of this paper.

Our classification framework clearly distinguishes between two views: (i) the  \managerial view of project managers and (ii) the \devops view of developers and operators.
These two different views can help organisations in choosing the FaaS platform best suited for their requirements, with the \managerial view helping project managers to first select the subset of FaaS platforms complying with the project and business requirements with a high-level analysis, and without delving into all technicalities of the platforms themselves.
Complementarily, the \devops view can then be exploited by the technical specialists working on the development and operation of a FaaS-based project to analyze the technical features of FaaS platforms, while at the same time focusing \textit{only} on those already complying with the project and business requirements. 

Based on our classification framework, we conducted a \textit{FaaS Platform Technology Review}, which constitues the second contribution of this paper.
The technology review classifies and compares the ten most popular FaaS Platforms, \ie the three most used proprietary platforms and the seven highest-ranked open source platforms.
Notably, our review provides a first systematic knowledge base that managers and DevOps can exploit to choose the platform best suited for their requirements.

To further support the selection process, we also provide a \textit{FaaS Platform Selection Support System}, called \faastener, constituting the third contribution of this paper.
\faastener is an open source web-based application exploiting the results of our technology review and enabling to search for FaaS platforms through multi-attribute queries.
This enables, for example, to look for platforms with a certain license and supporting given monitoring and logging solutions.

\smallskip
To summarise, the main contributions of this paper are threefold:
\begin{itemize}
\item[(i)] A \textit{FaaS Platform Classification Framework}, which enables characterising FaaS platforms under two different perspectives, \ie the high-level business view of FaaS project managers, and the low-level technical view of application developers and operators.
\item[(ii)] A \textit{FaaS Platform Technology Review} of the ten most popular FaaS platforms, \ie the three most used commercial platforms and the seven highest-ranked open source platforms.
The review exploits our classification framework to provide a characterisation of the investigated platforms under both the business and the technical perspectives.
\item[(iii)] A \textit{FaaS Platform Selection Support System} in the form of a web-based application called \textsc{FaaStener}, which enables researchers and practitioners to look for the FaaS platforms most suited for the requirements of their serverless projects and to browse through the systematic knowledge base formed by our technology review.
\end{itemize}

\noindent
The rest of the paper is organised as follows. 
\cref{sec:bg} provides background on FaaS and related technologies.
\cref{sec:research-design} illustrates the reseach approach for conducting our study.
\cref{sec:framework} introduces the \textit{FaaS Platform Classification Framework}.
\cref{sec:tech-results} presents the \textit{FaaS Platform Technology Review}, while \cref{sec:portal} illustrates the \textit{FaaS Platform Selection Support System} materialising the results of the review.
\cref{sec:threats-to-validity} and \cref{sec:related} discuss potential threats to the validity of our study and related work, respectively.
\cref{sec:conclusion} finally draws concluding remarks.

\section{Background}
\label{sec:bg}
\noindent
The popularity of serverless computing started to rise after Amazon introduced its FaaS platform offering called AWS Lambda~\cite{aws:lambda}.
While the term \textit{serverless} was also used previously in other contexts, it gained most popularity in the context of cloud-native application development~\cite{serverless:fox2017status}.
Essentially, serverless applications are composed of components that are managed by third-parties, which significantly reduces control over the infrastructure, thus, minimizing management efforts~\cite{cncf:serverless-wp}.

Function-as-a-Service is an integral part of the serverless world as it enables hosting business logic in the form of functions that are typically stateless and driven by events.
This means that the deployed function code can be triggered by events originating from multiple heterogeneous event sources such as databases, message queues, or streaming platforms.
Moreover, functions can be exposed as HTTP endpoints via API Gateways, or invoked on a scheduled basis, e.g., using cron jobs.
The actual list of supported event sources and possible ways to integrate events from third party services depends on the employed FaaS platform.
Since the provider is responsible for managing functions, autoscaling comes out-of-the-box, also including scaling to zero instances when functions are not needed.
This introduces a new, more flexible cost model, where users do not need to pay for idle components as for Platform-as-a-Service deployments.
However, in some cases the costs might become less appealing than with classic cloud service models~\cite{serverless:Eivy2017}.
Moreover, aside from its benefits, FaaS also has some well-known limitations~\cite{serverless:hellerstein2018serverless} such as the cold start issue, limited execution time which might also vary on different FaaS platforms, or tighter-coupling with provider's specifics due to outsourced management efforts~\cite{Yussupov2019_FaaSPortability}.
However, the combination of the above mentioned properties and limitations is what essentially affects the ways how FaaS platforms need to be chosen.

The FaaS technology landscape comprises a variety of heterogeneous platforms, which are offered as-a-service, or can be installed on-premises.
For example, Microsoft introduced Azure Functions~\cite{ms:azure-functions} while IBM started offering its Cloud Functions~\cite{ibm:cloud-functions} based on the open source FaaS platform Apache Openwhisk~\cite{apache:openwhisk}.
The landscape of open source FaaS platforms started to evolve rapidly too, also due to the popularity of container orchestration brought by Kubernetes~\cite{kubernetes}.
Multiple open source products such as OpenFaaS~\cite{openfaas}, Kubeless~\cite{kubeless}, or Fission~\cite{fission} that provide serverless, FaaS-based application development experience on top of Kubernetes.
Typically, since functions are event-driven, FaaS platforms can be integrated with multiple possible event sources such as databases, messaging and streaming platforms, or provider-specific services such as AWS Alexa~\cite{alexa}.
As a result of this heterogeneity, there is an ongoing work on maintaining the list of available technologies with high-level details such as web-site and documentation pointers curated by the Cloud Native Computing Foundation~(CNCF)~\cite{cncf:serverless-landscape}.
Moreover, CNCF is also curating the work on standardization of event specification format called CloudEvents~\cite{cncf:cloudevents}.

\begin{figure*}
    \centering
    \includegraphics[width=.97\textwidth]{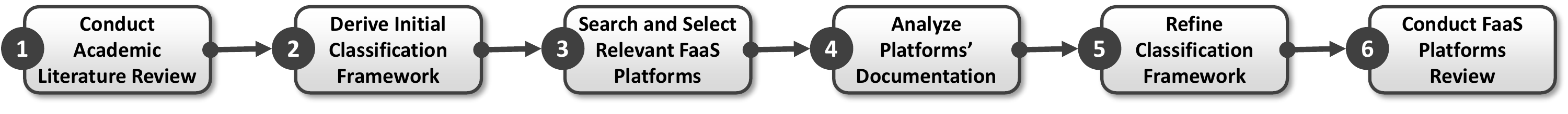}
    \caption{The sequence of steps to derive the FaaS Platforms Classification Framework}
    \label{fig:research-steps}
\end{figure*}

\section{Research Method}
\label{sec:research-design}
\noindent
In this section, we first recap terminology for core concepts related to the world of serverless and FaaS, which we use throughout this paper. 
We then proceed by describing the research method's steps.

\subsection{Terminology}
\label{subsec:terminology}
\noindent
In this section, we define the serverless- and FaaS-related terminology used in the subsequent sections.

\paragraph{Function-as-a-Service Platform / FaaS Platform}
We use the term \textit{FaaS Platform} to describe an environment that enables deploying, managing, and observing function instances, similar to the notion of a FaaS platform used in the CNCF Serverless Landscape~\cite{cncf:serverless-landscape}.
This definition also covers FaaS platforms that require a specific underlying platform, e.g., when a platform must run on top of a container orchestration platform such as Kubernetes, since these kinds of restrictions do not influence the overall purpose of the given product, i.e., to enable the usage of user-provided functions in a serverless fashion.

\paragraph{Event}
We base our the definition of an \textit{event} on the CloudEvents specification by CNCF.
More precisely, by using the term \textit{event} we mean a data record capturing an occurrence of a particular fact in a software system during its operating time~\cite{cncf:cloudevents}.

\paragraph{Event Source}
As for events, we rely on the definition from CloudEvents specification~\cite{cncf:cloudevents}, i.e., an \textit{event source} is the context where events happen, e.g., a remote database.

\paragraph{Function}
We base our definition of a \textit{function} on the definition from the serverless whitepaper by CNCF~\cite{cncf:serverless-wp}.
We use the term function to describe an executable snippet of code which can be hosted on the FaaS platform and triggered by events originating from supported event sources, or invoked directly, e.g., via programmatic access provided by available client libraries.

\paragraph{Serverless Application}
The term \textit{serverless application} is used to describe a combination of one or more functions that interact with a set of resources managed by  third parties such as cloud providers or other external as-a-service offerings.
Essentially, a serverless application represents a logical grouping of resources and functions in particular, which facilitates, e.g., versioning for development and deployment of applications.


\paragraph{Function Orchestrator}
To describe a specialized software that enables composing multiple functions by means of workflows, we use the term \textit{function orchestrator}.
This definition is related only to specialized products explicitly tailored for FaaS, e.g., AWS Step Functions~\cite{step-functions} or Azure Durable Functions~\cite{durable-functions}, and does not include general-purpose workflow engines such as Apache ODE.

\paragraph{Function Marketplace}
We use the term \textit{Function Marketplace} to describe dedicated marketplace platforms for offering functions and FaaS-based applications that are distributed by third parties under proprietary or open source licenses for commercial and non-commercial use, e.g., applications that represent a common serverless use case and can be purchased for educational or development purposes.
A function marketplace is managed by a marketplace provider, who also defines packaging and description formats of marketplace's products, for instance.
A notable example of function marketplace is the AWS Serverless Application Repository~(AWS SAR)~\cite{aws:sar}.

\paragraph{Code Samples Repository}
A \textit{Code Samples Repository} is a publicly-available and officially-maintained collection of function and application examples developed for a given FaaS platform, which can be used as a training material or reused for application development.
Unlike function marketplaces, code samples are always distributed under open source licenses and typically provided as-is.
Technically, it might be a standalone repository or a part of the main platform's repository maintained by the community or the platform producer, e.g., function examples stored in a \emph{examples} folder with the platform repository.

\subsection{Research Method's Steps}
\noindent
\Cref{fig:research-steps} shows the multi-step process we followed based on a combination of academic literature review and documentation analysis.
As an initial step we analyzed existing academic publications that focus on criteria-based review of FaaS platforms and we derived an initial FaaS platform classification framework.
Afterwards, we selected ten existing FaaS platfoms, we analyzed them, and we refined the initial classification framework based on the newly-discovered data, and used it for the review of these platforms.
In the following, we elaborate on the research method's design.

\subsubsection{Step 1: Academic Literature Review}
\label{sec:lit-review}
\noindent
To derive a classification framework that covers both academic and industrial views on FaaS platforms, as a first step, we analyzed the existing literature that focuses on reviewing Function-as-a-Service platforms by searching the initial set of publications using well-known electronic research databases, namely ACM Digital Library, arXiv.org, Google Scholar, IEEE Xplore, Springer Link, Science Direct and Wiley Online Library.


\paragraph{Selection criteria}
\label{subsec:selection-criteria}
To identify publications relevant to FaaS platforms analysis and comparison, we defined a set of selection criteria.
The initial dataset was screened by the authors using adaptive reading depth~\cite{systematic:petersen2015guidelines} to identify publications' relevance.
We defined the inclusion~(\checkmark) and exclusion~($\times$) criteria as follows:
\begin{itemize}
    \item[\checkmark] Publications that evaluate and compare existing FaaS platforms and function orchestration technologies.
    \item[\checkmark] Publications that are written in English.
    \item[$\times$] Publications not accessible as full-text or not in the form of a full research paper, e.g. extended abstract, presentation, tutorial, PhD research proposal, demo paper, as they do not provide enough details.
\end{itemize}
As a result, we identified five~\cite{related:8605774, related:8605772, related:8591002, related:8457830, related:8241104} relevant publications after applying the selection criteria to the identified initial set of publications.

\def\tabularxcolumn#1{m{#1}}
\begin{table*}[t]
		\caption{List of selected FaaS platforms (sorted alphabetically).}
		\label{tab:platforms-documentation}
   \footnotesize
		\centering
		\begin{tabular}{ll}
        \hline
        \textbf{FaaS Platform} &
        	\textbf{Documentation Sources} \\
        \hline
        \rowcolor[HTML]{EFEFEF}
    			\textit{Apache Openwhisk} 
        		& \url{https://openwhisk.apache.org}, \url{https://github.com/apache/openwhisk} \\ 
        \textit{AWS Lambda} 
        		& \url{https://docs.aws.amazon.com/lambda} \\ 
        \rowcolor[HTML]{EFEFEF}
    			\textit{Fission} 
        		& \url{https://docs.fission.io}, \url{https://github.com/fission/fission} \\ 
        \textit{Fn} 
        		& \url{https://fnproject.io/}, \url{https://github.com/fnproject} \\ 
        \rowcolor[HTML]{EFEFEF}
    			\textit{Google Cloud Functions} 
        		& \url{https://cloud.google.com/functions/docs} \\ 
        \textit{Knative} 
        		&  \url{https://knative.dev/docs}, \url{https://github.com/knative} \\ 
        \rowcolor[HTML]{EFEFEF}
    			\textit{Kubeless} 
        		&  \url{https://kubeless.io/docs}, \url{https://github.com/kubeless} \\ 
        \textit{MS Azure Functions} 
        		&  \url{https://docs.microsoft.com/en-us/azure/azure-functions}, \url{https://github.com/Azure/Azure-Functions} \\ 
        \rowcolor[HTML]{EFEFEF}
    			\textit{Nuclio} 
        		&  \url{https://nuclio.io/docs}, \url{https://github.com/nuclio/nuclio} \\ 
        \textit{OpenFaaS} 
        		&  \url{https://docs.openfaas.com}, \url{https://github.com/openfaas/faas} \\      
        \hline
    \end{tabular}    
\end{table*}

\paragraph{Snowballing and Combination}
As additional means to identify relevant literature, we applied the snowballing technique~\cite{systematic:wohlin2014guidelines}.
More specifically, we used Google Scholar for forward snowballing, i.e., analyzed all research papers that cite each of the selected publication, and applied closed recursive backward snowballing, i.e., analyzed all research papers cited by each of the selected publication.
Afterwards, we applied the selection criteria defined previously in~\Cref{subsec:selection-criteria}, which lead to the identification of six additional publications~\cite{related:snowballing:rajan2018serverless,related:snowballing:palade2019evaluation,related:snowballing:kumar2019serverless,related:snowballing:kalnauz2019productivity,related:snowballing:gand2020serverless,related:snowballing:bortolini2019investigating}.
Due to the relatively small amount of relevant publications~(11 in total), the combination step was not necessary.

\subsubsection{Step 2: Derive Initial Classification Framework}
\noindent 
To derive an initial classification framework, we used the keywording technique~\cite{systematic:petersen2008systematic}.
The overal process is as follows:
\begin{inparaenum}[(1)]
    \item at least two researchers separately analyzed each publication from the selected set to identify common concepts and define keywords for them.
    \item The resulting keywords were discussed and clustered to form the initial classification framework.
\end{inparaenum}
At this stage we had a set of generic, high-level categories, e.g., \textit{Licensing}, \textit{Installation}, \textit{Documentation}, \textit{Development}, \textit{Interfaces}, and \textit{Observability}, each coming with a set of concrete criteria.
For example, \textit{Observability} included \textit{Logging} and \textit{Monitoring} categories, which listed supported tooling and available integration mechanisms for adding new tools.

\subsubsection{Step 3: Search and Select Relevant FaaS Platforms}
\label{subsec:platform-search-selection}
\noindent
As a next step, we selected ten relevant general-purpose FaaS platforms for subsequent refinement of our initial classification framework.
We defined the following requirements that had to be fulfilled by a FaaS platform to be included for the documentation analysis:
\begin{itemize}
    \item[\checkmark] A platform must be \textit{general-purpose}, meaning that it should not tailored for a specific use case (\eg training AI models).
    \item[\checkmark] A platform has to be \textit{actively-maintained}, i.e., there exist one or more parties consistently contributing towards new platform releases and the code repository is not stale or archived.
\end{itemize}
We structured the overall process of searching and selecting relevant FaaS platforms using approaches from existing work, \eg search engine hits analysis~\cite{related:iaas:Wurster2019_EDMM}.

\paragraph{Phase 1: Platforms Search}
To select relevant FaaS platforms, we used white and gray literature sources, namely the list of platforms reviewed in the publications chosen during the literature review step described in~\Cref{sec:lit-review}, and the serverless landscape~\cite{cncf:serverless-landscape} maintained by CNCF that describes the platforms and tooling related to serverless computing.
We aggregated the lists of platforms obtained from both sources while checking if the aforementioned inclusion criteria are fulfilled to ensure that we do not miss relevant platforms.
For example, we excluded Snafu~\cite{related:spillner2017snafu}, a research prototype evaluated in one of the publications as it is not a general-purpose and actively-maintained FaaS platform~(no commits in the last 2 years).
Afterwards, we sorted the remaining platforms by popularity using separate criteria for hosted and installable platforms~(described next), and apply the \textit{effort bounded} stopping criteria~\cite{systematic:garousi2019guidelines} by selecting ten platforms in total.

\paragraph{Phase 2: Platforms Selection}
A first crucial selection aspect is how to decide on proper ratio between commercial and open source platforms.
It is important to mention that at the moment of writing CNCF serverless landscape lists 33 FaaS platforms, 19 of which are hosted (mostly closed-source) and 14 are installable and open source.
With the aim of putting more emphasis on open source installable FaaS platforms, which source code is publicly available and which can be installed on premises, we decided to cover 50\% of them, \ie seven open source installable FaaS platforms.
Taking into account our bounded effort stopping criteria~\cite{systematic:garousi2019guidelines}, according to which we limited our analysis to ten FaaS platforms, we hence decided to analyse three commercial platforms and seven open source platforms.

To pick the most representative commercial and open source FaaS platforms, we then decided to select the most popular of them.
We used different popularity quantifiers for commercial and open source platforms since
\begin{inparaenum}[(i)]
    \item commercial platforms typically have popularity quantifiers, like quantity of search engine hits or Stackoverflow questions, which are higher with respect to those of open source platforms by orders of magnitude, 
    \item such popularity quantifiers can also be misleading for open source platforms, as their name may not be branded and correspond to other projects, and
    \item the developers' interest in open source platforms can be estimated by analyzing publicly available source code-related metrics, while commercial platforms are typically proprietary closed-source products maintained by one party, \eg AWS Lambda.
\end{inparaenum}

We searched for the three most popular commercial FaaS platforms by sorting commercial platforms by Google Search hits, and by using the full platform name as a search query (\eg \enquote{aws lambda}, \enquote{azure functions}, \enquote{google cloud functions}, and \enquote{ibm cloud functions}).
As a result, we selected AWS Lambda, Microsoft Azure Functions, and Google Cloud Functions
It is worth mentioning that for all three selected platforms the amount of Google Search hits is above 500.000 hits, whereas there is a significant drop in number of hits for the remaining commercial platforms, all of which have less than 150.000 hits.

In contrast, we approximated the overall interest in open source platforms by measuring the amount of GitHub stars given by GitHub users since all open source FaaS platforms are hosted on GitHub~\cite{cncf:serverless-landscape}.
The amount of stars associated with the GitHub repositories of FaaS platforms were extracted using GitHub's API, which resulted in the following list: OpenFaaS, Apache Openwhisk, Nuclio, Fission, Fn, Kubeless, and Knative.
This decision was driven by several factors, namely
\begin{inparaenum}[(i)]
    \item to highlight the trends in open source FaaS platform development community, and
    \item to provide more data for identifying the gaps between open source and commercial platforms, since the latter generally focus less on supporting integration with third parties.
\end{inparaenum}

\Cref{tab:platforms-documentation} shows the final list of FaaS platforms selected for the documentation analysis together with the respective documentation sources we considered.

\subsubsection{Step 4: Analyze Platforms' Documentation}
\label{subsec:platform-doc-analysis}
\noindent
We analyzed only the documentation provided via official sources such as a dedicated website maintained by the official producer of the platform and its GitHub repository, in case the platform is open source.
The documentation was analyzed separately by the authors using the initial classification framework as a baseline.
After the analysis, the results were discussed to identify potential refinements for the classification framework.

\subsubsection{Step 5: Refine Classification Framework}
\label{sec:refine-classification}
\noindent 
After the classification framework was modified using the refinements resulting from the analysis in~\Cref{subsec:platform-search-selection}, the final version of the classification framework was further analyzed and cross-checked among authors to reduce potential bias (\cref{sec:threats-to-validity}).
The resulting set of concepts was used to derive our \textit{FaaS Platform Classification Framework} which we discuss in-detail in~\Cref{sec:framework}.

\subsubsection{Step 6: Conduct FaaS Platforms Review}
\label{sec:review-platforms}
\noindent 
We exploited our classification framework derived in ~\Cref{sec:refine-classification} to conduct a technology review of the ten FaaS platforms listed in~\Cref{tab:platforms-documentation}.
As a first step, each platform was reviewed separately by two authors.
Afterwards, the results were verified by the other authors and merged.
The conflicting results were discussed and resolved until a unanimous consensus was achieved.
The results of the review are presented in~\Cref{sec:tech-results}.


\section{FaaS Platform Classification Framework}
\label{sec:framework}
\noindent 
When choosing the FaaS platform best suited for a software project, various different aspects have to be taken into account, and two different views can be identified.
On the one hand, \textit{project managers} focus on project- and business-oriented aspects when choosing the possible FaaS platforms for shipping the projects they manage, \eg whether to select a proprietary or open source platform, and its actual licensing.
%
On the other hand, aspects like whether a FaaS platform natively supports certain function triggers or whether it can be integrated with given monitoring and logging solutions are typically not analysed by project managers.
Such aspects are rather considered by the \textit{technical experts} actually developing and operating the project itself.

Following the above idea, we hereafter present our framework for classifying FaaS platforms (derived in Step 4 of \cref{sec:research-design}), by clearly distinguishing the \managerial view of project managers from the \devops view of developers and operators\footnote{As already explained in \cref{sec:research-design}, the classification framework was derived by extracting self-declared information available in FaaS platforms' online documentation. As a result, information that is not self-declared (\eg vendor lock-in) is not included in our framework.}.
Classification categories pertaining to the \managerial view are presented in \cref{ssec:managerial-view}, while those pertaining to the \devops view are presented in \cref{ssec:devops-view}.


\begin{figure*}[tb]
\centering
\footnotesize
\noindent
\def\arraystretch{1.2}
\begin{tabular}{|r
>{\columncolor[HTML]{EFEFEF}}l
>{\columncolor[HTML]{C0C0C0}}r 
>{\columncolor[HTML]{9B9B9B}}l
>{\columncolor[HTML]{9B9B9B}}l|}
\hline
\textbf{\license}      & \licName                      & \multicolumn{2}{l}{\cellcolor[HTML]{9B9B9B} Apache 2.0, GNU GPL 1.0, GNU GPL 2.0, GNU GPL 3.0, MIT, ...}     	& \\
                       & \licType                      & \multicolumn{2}{l}{\cellcolor[HTML]{9B9B9B} public domain, permissive, copyleft, freeware, proprietary, ...} 	& \\
\hline
\textbf{\installation} & \instType                     & \multicolumn{2}{l}{\cellcolor[HTML]{9B9B9B} as-a-service, installable}                                       	& \\
                       & \instTargetHosts              & \multicolumn{2}{l}{\cellcolor[HTML]{9B9B9B} Docker, Kubernetes, Linux, MacOS, OSX, Windows, ...}                         & \\
\hline
\textbf{\sourceCode}   & \srcAvailability              & \multicolumn{2}{l}{\cellcolor[HTML]{9B9B9B} open source, closed source}                                      	& \\
                       & \srcHostingPlatform           & \multicolumn{2}{l}{\cellcolor[HTML]{9B9B9B} BitBucket, GitHub, SourceForge, ...}                                & \\
                       & \srcProgrammingLanguage       & \multicolumn{2}{l}{\cellcolor[HTML]{9B9B9B} C, C\#, F\#, Go, Java, Javascript, Python, Ruby, Scala, ...}                            	& \\
\hline
\textbf{\release}      & \relStatus                    & \multicolumn{2}{l}{\cellcolor[HTML]{9B9B9B} pre-alpha, alpha, beta, release candidate, stable release, rtm, ga, production }  	& \\
\hline
\textbf{\interface}    & \intType                      & \multicolumn{2}{l}{\cellcolor[HTML]{9B9B9B} CLI, API, GUI }                                               & \\
                       & \intApplicationManagement     & \multicolumn{2}{l}{\cellcolor[HTML]{9B9B9B} creation, retrieval, update, deletion }                        & \\
                       & \intPlatformAdministration    & \multicolumn{2}{l}{\cellcolor[HTML]{9B9B9B} deployment, configuration, enactment, termination, undeployment }              & \\
\hline
\textbf{\community}    & \commGitHub                   & Stars                                                   & \textit{number}                                   	            & \\
                       &                               & Forks                                                   & \textit{number}                                   	            & \\
                       &                               & Issues                                                  & \textit{number}                                   	            & \\
                       &                               & Commits                                                 & \textit{number}                                   	            & \\
                       &                               & Contributors                                            & \textit{number}                                   	            & \\
                       & \commStackoverflow            & Questions                                               & \textit{number}                                   	            & \\
\hline
\textbf{\documentation}& \docFunctions                 & \multicolumn{2}{l}{\cellcolor[HTML]{9B9B9B} development, deployment }                                      & \\
                       & \docPlatform                  & \multicolumn{2}{l}{\cellcolor[HTML]{9B9B9B} usage, development, deployment, architecture }     & \\
\hline
\textbf{\restrictions} & \resDeployment                & Code Size                                               & limited, unbounded                      	    & \\
                       &                               & Package Size                                            &  limited, unbounded                        	    & \\
                       & \resRuntime                   & CPU                                                     &  limited, unbounded                        	    & \\
                       &                               & Memory                                                  &  limited, unbounded                        	    & \\
                       &                               & Storage                                                 &  limited, unbounded                        	    & \\
                       &                               & Execution Time                                          &  limited, unbounded                        	    & \\
\hline
\end{tabular}
\caption{The \managerial view of our classification framework. White cells contain categories, lighter grey cells contain classification dimensions, while dark grey cells contain values that can possibly be associated with classification dimensions.}
\label{fig:managerial-view}
\end{figure*}

\subsection{A Business View for Classifying FaaS Platforms}
\label{ssec:managerial-view}
\noindent 
The \managerial view of our \textit{FaaS Platform Classification Framework} comprises categories and dimensions of interest for project managers aiming to identify the FaaS platforms complying with the high-level project requirements.
These include, for instance, the license under which a FaaS platform is released and whether the platform can be installed on premise or not, as both dimensions can impact on the integration of the platform with the rest of the software developed in a project~\cite{Yussupov2019_SystematicMappingStudyFaaS,Laurent04_OpenSource}.

All dimensions pertaining to the \managerial view of our classification framework are listed and categorised in \cref{fig:managerial-view}, and explained hereafter.

\paragraph{\license}
Each FaaS platform is released under some existing licensing, whether open or proprietary.
The category \license enables classifying platforms under such dimension by allowing to indicate the actual \licName and the corresponding license \licType.
\cref{fig:managerial-view} reports some possible values for classifying the name and type of the license under which a platform is released, which can anyhow be any existing open source license name and type (such as those recapped by Laurent in his book~\cite{Laurent04_OpenSource}, for instance), or any proprietary licensing option under which a vendor releases its software.

\paragraph{\installation}
FaaS platforms currently come in two different forms, as they can be hosted by vendors, installed on-premise, or both.
The purpose of the \installation category is precisely to enable classifying FaaS platforms under this dimension by allowing to distinguish their installation \instType (\ie \textit{as-a-service}, \textit{installable}).
The \installation category also enables to indicate the set of \instTargetHosts where a FaaS platform can be installed, which is given by a subset of the set formed by existing OSs and cluster orchestrators (some examples of which are in \cref{fig:managerial-view}).
Of course, if a platform only comes \textit{as-a-service}, then the set of \instTargetHosts will be empty.

\paragraph{\sourceCode}
The \sourceCode category enables classifying the \srcAvailability of the sources of a FaaS platform, whether it is \textit{open source} or \textit{closed source}.
In the former case, the \textit{FaaS Platform Classification Framework} also enables classifying the \srcHostingPlatform and main \srcProgrammingLanguage used for the \sourceCode.
\cref{fig:managerial-view} lists some possible values for classifying the \srcHostingPlatform and main \srcProgrammingLanguage, which can anyhow be any existing open source repository and programming language.

\paragraph{\release}
The \release category enables to classify the release \relStatus of FaaS platforms.
Possible values for \release \relStatus are listed in \cref{fig:managerial-view}, following the possible release statuses discussed in the book on software delivery by Humble and Farley~\cite{Humble10_ContinuousDelivery}.

\paragraph{\interface}
Existing FaaS platforms offer different ways for interacting with them, and the purpose of the \interface category is precisely to enable classifying platform under this dimension.
The \interface category allows to list the supported interface \intType{}s, \ie whether they offer a command line interface (\textit{CLI}), an application programming interface (\textit{API}), and/or a graphical user interface (\textit{gui}).
It also enables to classify whether a FaaS platform supports \intApplicationManagement, \ie which subset of CRUD operations it offers to manage applications.
Finally, the \interface category includes the \intPlatformAdministration dimension, which enables indicating whether a FaaS platform offers operations for its own \textit{deployment}, \textit{configuration}, \textit{enactment}, \textit{termination} and \textit{undeployment}.

\paragraph{\community}
The \community category enables to classify FaaS platforms based on the size, activity, and popularity of their development community.
Given that the seven most popular open source FaaS platforms (which drove the development of our classification framework) are all hosted on GitHub, the obtained \community category explicitly enables to classify platforms based on quantitative information taken from their \commGitHub repository, \ie the amount of \textit{Stars}, \textit{Forks}, \textit{Issues}, \textit{Commits} and \textit{Contributors}.
All such numeric information gives an indication of the size, activity, and popularity of the corresponding repository~\cite{Guidotti19_HelpingDockerImages}.
Another dimension explicitly included in the \community category is the amount of high-scored platform-related \textit{Questions} on \commStackoverflow, which also indicates the usage and popularity of the platform.


\paragraph{\documentation}
The available official documentation for currently existing FaaS platforms is various in comprehensibility and nature.
This is the main reason why the \managerial view of our classification framework includes the \documentation category.
The latter enables to indicate whether a FaaS platform comes with an official documentation for function \textit{development} and \textit{deployment} (with the  \docFunctions dimension), \ie whether it documents how to develop functions that can be executed by the platform, and the actual processes for suitably deploying them on the platform.
The \documentation category also enables to indicate whether a FaaS platform officially documents its \textit{usage}, \textit{development}, \textit{deployment} and \textit{architecture} (with the \docPlatform dimension), \ie whether it documents the processes to use the platform for running function-based applications, to develop extension to the platform, to deploy the platform, and whether it provides information on the platform's architecture.

\paragraph{\restrictions}
Finally, some FaaS platforms comes with upperbounds on the size of applications that can be deployed, as well as on the computing resources and execution time that application can get at runtime.
The purpose of the \restrictions category is precisely to qualitatively indicate whether such upperbounds apply to a given FaaS platform.
The \restrictions category indeed includes the \resDeployment dimension, which enables indicating whether the source \textit{Code Size} and the deployment \textit{Package Size} are limited or unbounded.
It also includes the \resRuntime dimension, which enables specifying whether \textit{CPU}, \textit{Memory}, \textit{Storage} and \textit{Execution Time} are \textit{limited} or \textit{unbounded}.

\subsection{A Technical View for Classifying FaaS Platforms}
\label{ssec:devops-view}
\noindent
The \devops view in our classification framework comprises a set of categories and dimensions that have to be considered by software development and operations specialists willing to identify if a given FaaS platform suffices the low level, technical requirements.
For example, a runtime for the programming language used in the company or project might not be supported by the platform which would complicate the development.
\Cref{fig:devops-view} presents a set of categories helping to classify a given FaaS platform from the \devops perspective.
In the following, we discuss all categories and their respective dimensions in-detail.


\begin{figure*}[tb]
    \centering
    \footnotesize
    \newlength\q
    \setlength\q{\dimexpr .22\textwidth -2\tabcolsep}
    \noindent
    \def\arraystretch{1.2}
    \begin{tabular}{|r
            >{\columncolor[HTML]{EFEFEF}}p{\q} 
            >{\columncolor[HTML]{C0C0C0}}l 
            >{\columncolor[HTML]{9B9B9B}}l
            >{\columncolor[HTML]{9B9B9B}}l|}
        \hline
        \textbf{\dev}
        & \devruntime & \multicolumn{2}{l}{\cellcolor[HTML]{9B9B9B}Go, Java, JavaScript, Python, Docker Image, \dots} 	&  \\
        & \devruntimeextension & \multicolumn{2}{l}{\cellcolor[HTML]{9B9B9B}supported, not supported} 	&  \\
        & \devIDE & \multicolumn{2}{l}{\cellcolor[HTML]{9B9B9B}IntelliJ IDEA, Eclipse, Visual Studio Code, \dots} 	& \\
        & \devSDK & \multicolumn{2}{l}{\cellcolor[HTML]{9B9B9B}Go, Java, JavaScript, Python, Docker Image, \dots} 	& \\
        
        \hline
        \textbf{\vers} & \versapps & \multicolumn{2}{l}{\cellcolor[HTML]{9B9B9B}dedicated mechanisms, implicit versioning, no support} & \\
                       & \versfuncs & \multicolumn{2}{l}{\cellcolor[HTML]{9B9B9B}dedicated mechanisms, implicit versioning} & \\
        
        \hline
        \textbf{\eventsources}
        & \eventendpoint  & \eventendpointsync & HTTP, gRPC, \dots  & \\
        &                 & \eventendpointAsync & HTTP, gRPC, \dots  & \\
        &                 & \eventendpointCust & supported, not supported & \\
        &                 & \eventendpointTLS & supported, not supported & \\        
        & \eventdatastore & \eventdatastorefile & AWS S3, Min.io \dots & \\
        &                 & \eventdatastoreDB & Azure CosmosDB, MySQL \dots & \\
        & \eventscheduler & \multicolumn{2}{l}{\cellcolor[HTML]{9B9B9B}supported, not supported} & \\
        & \eventmq        & \multicolumn{2}{l}{\cellcolor[HTML]{9B9B9B}AWS SQS, RabbitMQ \dots } & \\
        & \eventstreaming & \multicolumn{2}{l}{\cellcolor[HTML]{9B9B9B}AWS Kinesis, Apache Kafka \dots} & \\
        & \eventspecservice & \multicolumn{2}{l}{\cellcolor[HTML]{9B9B9B}AWS Alexa, GitHub, IBM Watson, \dots} & \\
        & \eventintegration & \multicolumn{2}{l}{\cellcolor[HTML]{9B9B9B}plugin development, messaging-based integration, \dots } & \\
        
        \hline
        \textbf{\orchestration}
        & \orchestrationWF & \multicolumn{2}{l}{\cellcolor[HTML]{9B9B9B}standard language, custom DSL, orchestrating function, \dots} & \\        
        & \orchestrationConstructs & \multicolumn{2}{l}{\cellcolor[HTML]{9B9B9B}documented, not documented} & \\
        & \orchestrationQuota & \orchestrationQuotaExecTime & present, not present & \\
        &                     & \orchestrationQuotaIO & present, not present & \\
        
        \hline
        \textbf{\testdeb}
        & \testing   & \textit{Functional}     &  platform-native tooling, 3rd party tooling &  \\
        &            & \textit{Non-functional} &  platform-native tooling, 3rd party tooling &  \\
        & \debugging & \textit{Local}          &  platform-native tooling, 3rd party tooling & \\
        &            & \textit{Remote}         &  platform-native tooling, 3rd party tooling & \\
        
        \hline
        \textbf{\observability}        
        & \observeLog & \multicolumn{2}{l}{\cellcolor[HTML]{9B9B9B}platform-native tooling, 3rd party tooling }	&  \\
        & \observeMon & \multicolumn{2}{l}{\cellcolor[HTML]{9B9B9B}platform-native tooling, 3rd party tooling}	&  \\
        & \observeInt & \multicolumn{2}{l}{\cellcolor[HTML]{9B9B9B}push-based, pull-based, plugin development, ...} 	& \\
        
        \hline
        \textbf{\delivery}        
        & \deliveryDepAut & \multicolumn{2}{l}{\cellcolor[HTML]{9B9B9B}platform-native tooling, 3rd party tooling} &  \\
        & \deliveryCICD & \multicolumn{2}{l}{\cellcolor[HTML]{9B9B9B}supported, not supported}	&  \\
        
        \hline
        \textbf{\codereuse}        
        & \codereuseMarkets & \multicolumn{2}{l}{\cellcolor[HTML]{9B9B9B}official marketplace, 3rd party marketplaces}	&  \\
        & \codereuseRepos & \multicolumn{2}{l}{\cellcolor[HTML]{9B9B9B}present, not present}	&  \\
        
        \hline
        \textbf{\accessmgmt}        
        & \accessAuth & \multicolumn{2}{l}{\cellcolor[HTML]{9B9B9B}built-in, external, \dots}	&  \\
        & \accessControl & \multicolumn{2}{l}{\cellcolor[HTML]{9B9B9B}functions, resources, \dots}	&  \\
        
        \hline
    \end{tabular}
    \caption{The \devops view of our classification framework. White cells contain categories, lighter grey cells contain classification dimensions, while dark grey cells contain values that can possibly be associated with classification dimensions.}
    \label{fig:devops-view}
\end{figure*}

\paragraph{\dev}
Essentially, a FaaS platform might provide various mechanisms that facilitate the overall function and application development experience.
Firstly, to start developing functions in a specific programming language, e.g., Python or Java, the platform must support the required \devruntime.
Note that even in cases when the platform does not support a particular language, it still might be possible to develop function in such language if the platform supports \devruntimeextension, e.g., by supplying custom container images as a function runtime.
However, since a custom container image has to be created first following the specific set of requirements (\eg implementing a required interface), instead of simply providing the source code, this option also introduces additional management efforts.

As FaaS platforms typically impose varying requirements on, for example, the way functions have to be implemented~\cite{Yussupov2019_FaaSPortability}, the source code development can also be simplified via plugins for popular \devIDE such as IntelliJ IDEA~\cite{intellij} or Visual Studio Code~\cite{vscode} that provide support, e.g., for platform-specific syntax highlighting or automated code packaging.
Moreover, a platform can provide \devSDK for a set of programming languages that wrap the platform's APIs as a part of platform's Software Development Kit~(SDK) to facilitate the development process, e.g., usage of platform-specific libraries for working with typed events' data.

\paragraph{\vers}
An important aspect in FaaS platforms which can simplify several distinct phases of the application lifecycle is \vers of serverless applications and functions.
Apart from facilitating the development process, versioning is also helpful for application deployment, e.g., to support canary releases~\cite{fowler:canary} or blue-green deployments~\cite{fowler:blue-green}.
It is hence advantageous if a FaaS platform provides mechanisms for managing versions on the level of single \versfuncs, or entire \versapps, i.e., a combination of functions and resources tracked together.
It is worth mentioning that while it is always possible, for instance, to encode version identifiers as part of function or application names or namespaces, however such \textit{implicit} versioning is less advantageous than the presence of \textit{dedicated mechanisms} for version management.
Additionally, a platform might not support version management on the serverless application level in case no notion of an application is present.

\paragraph{\eventsources}
Due to the event-driven nature of FaaS the event sources support plays an important role in deciding whether the platform is suitable for given development requirements.
For instance, one of the common FaaS use cases is when a function needs to be exposed as an \eventendpoint via an API Gateway~\cite{Yussupov2019_SystematicMappingStudyFaaS}.
Subsequently, several endpoint-related aspects can be considered here, e.g., whether a \eventendpointsync or an \eventendpointAsync is supported and which protocols can be used for performing such calls, e.g., HTTP or gRPC.
Moreover, it might be needed to use a custom endpoint's name instead of the default resource name assigned when exposing the function as an endpoint, i.e., \eventendpointCust must be supported. 
In addition, in case a secure communication via HTTPS is required, a platform must provide \eventendpointTLS.

The next dimension of event sources encapsulates different \eventdatastore types.
Since serverless applications comprise non-managed components, we consider only the higher-level storage types~\cite{dbtypes:mansouri2018data} such as \eventdatastorefile which includes object stores like AWS S3, and \eventdatastoreDB that covers SQL and NoSQL databases like Azure CosmosDB.

Another relevant event source is the \eventscheduler, which allows invoking functions on a scheduled basis.
Typically, internally these sources are implemented using cron jobs, hence developers might need to understand the subtle differences in the required format of cron expressions.

Such event source dimensions as \eventmq and \eventstreaming are the next important parts for FaaS platforms.
Being one of the main integration mechanisms, messaging plays a crucial role in serverless component integration.
Examples of messaging and streaming solutions include AWS SQS~\cite{aws:sqs}, Apache Kafka~\cite{kafka}, and RabbitMQ~\cite{rabbitmq}.

Generally, multiple event sources such as AWS Alexa~\cite{alexa}, IBM Watson~\cite{ibm:watson}, or GitHub~\cite{github} are not related to particularly-large dimensions and instead represent particular use cases, which is covered by the \eventspecservice dimension.
The list of supported specialized service offerings varies drastically from platform to platform and might be a major factor influencing the choice of a suitable platform~\cite{Yussupov2019_FaaSPortability}.

Finally, a platform might provide \eventintegration options, which makes it possible to trigger functions using custom event sources.
Examples of integration options might include plugin development or webhook-based integration.
Essentially, the integration using proxy components such as event gateways or message queues is possible in the majority of the cases due to the loosely coupled nature of serverless applications.
However, the crucial point here is that the platform's documentation describes official ways to integrate custom event sources, for example, company's proprietary applications emitting custom events, which hastens the overall development process.

\paragraph{\orchestration}
The next significant aspect characterizing FaaS platforms is related to possible ways of orchestrating multiple functions.
Apart from connecting functions by means of events and message queues, the specification of workflows~\cite{leymann1999production} incorporating multiple functions is another way to orchestrate complex function interactions.
Therefore, support for \orchestration brings additional benefits to platform users.
In this category, we consider only FaaS-oriented orchestrators such as AWS Step Functions or Azure Durable Functions following the definition of a function orchestrator provided in~\Cref{subsec:terminology}.
Essentially, the majority of function orchestrators are separate offerings that are tailored to work with FaaS platforms and require a separate classification framework.
For instance, multiple criteria from the \managerial view can also be used to classify function orchestrators, e.g., installation, licensing, or quotas.
Moreover, other orchestration aspects, e.g., expressiveness and extensibility of the underlying workflow language, require a more detailed analysis.

For the sake of brevity, we present the baseline information relevant for developers willing to start defining function orchestrations.
Firstly, the \textit{Workflow Definition} process might vary significantly~\cite{related:8605772}, e.g., a \textit{custom Domain-Specific-Language~(DSL)} can be used to define a function workflow, or even a \textit{standard language} such as BPEL~\cite{BPEL}.
Another option is to implement an \textit{orchestrating function} in a general-purpose programming language such as Java, which will be responsible for calling other involved functions and aggregating the results.
Here, the execution of orchestrating functions is controlled by the function orchestrator that handles stateful operations or error handling, for instance, making this option not valid for programming languages that are not explicitly supported by the orchestrator.
Another dimension to consider is the presence of documentation for \orchestrationConstructs, e.g., to understand how parallel or sequential task execution can be modeled or whether it is possible to handle errors during the workflow execution.
Additionally, function orchestrators can impose \orchestrationQuota on certain aspects of function workflows execution.
For example, the overall \orchestrationQuotaExecTime might be restricted or the \orchestrationQuotaIO can be limited to a particular value~\cite{Yussupov2019_FaaSPortability}.

\paragraph{\testdeb}
Another crucial set of mechanisms is related to testing and debugging of standalone functions and entire serverless applications.
While FaaS platforms are typically not responsible for the code development, additional ways to test \textit{Functional} and \textit{Non-functional} aspects of deployed functions can be provided, e.g., unit and integration testing, load testing, etc.
For example, a platform might offer mechanisms for facilitating unit testing with platform-specific libraries, or verifying function invocation using dedicated CLI commands.
Moreover, a combination of \textit{Local and Remote} debugging mechanisms might be provided by the platform.
In all these cases, both platform-native and third-party tooling may provide the possible set of solutions.

\paragraph{\observability}
The presence of mechanisms for observing serverless applications is a crucial factor for deciding on the FaaS platform.
For instance, platforms might provide various \observeLog and \observeMon options including both \textit{platform-native tooling} and \textit{third-party tooling}.
Additionally, platforms can provide documented ways to integrate existing tools, e.g., using a push-based or pull-based integration approaches.

\paragraph{\delivery}
The next category groups together dimensions related to facilitating delivering developed applications.
To hasten application deployment, various \deliveryDepAut tools can be supported by the platform.
For example, a platform-native deployment automation tool such as AWS Cloud Formation~\cite{aws:cloud-formation}, or a third-party tools such as the Serverless Framework~\cite{serverless-framework} can be present.
Moreover, presence of documented \deliveryCICD  ways can facilitate the DevOps processes.

\paragraph{\codereuse}
The ability to use existing applications as a basis for implementation as well as having access to exemplary code and configuration snippets can facilitate reducing the time to market.
Essentially, we distinguish two separate dimensions of \codereuse, namely the availability of supported \textit{Function Marketplaces}, and \textit{Code Sample Repositories} which are actively-maintained by the platform's provider or open source community.

\paragraph{\accessmgmt}
Finally, FaaS platforms might provide various \accessmgmt options related to using the platform from both, developer's and user's perspectives.
For example, a platform can provide native or support external \accessAuth mechanisms.
Additionally, there might be supported \accessControl mechanisms for defining the access rules for functions and related resources that interact with them, e.g., forbidding a function to access a particular data store resources.


\section{FaaS Platform Technology Review}
\label{sec:tech-results}
\noindent
In this section, we present the second contribution of this article, \ie the results of the FaaS platforms review using the classification framework introduced in~\Cref{sec:framework}.
As for the presentation of our classification framework, we first discuss the review results relevant for the \managerial view, followed by the \devops view.

\subsection{A Business View on FaaS Platforms}
\label{ssec:tech-review-managerial}
\noindent 
The \managerial view of our classification framework (\cref{ssec:managerial-view}) provides various categories for classifying and comparing existing FaaS platforms at a high-level.
These high-level categories can be of help for researchers and practitioners willing to identify FaaS platforms suitable for hosting existing or new FaaS-based applications by focusing only on those platforms that adhere to high-level projects' requirements.
Such top-down classification approach helps eliminating the need to delve into technical details of platforms that could be ignored beforehand.

In the following, we present and discuss the classification of the ten FaaS platforms selected in \cref{sec:research-design}, based on the categories in the \managerial view of our classification framework presented in \cref{ssec:managerial-view}.

\begin{figure}[h]
\begin{minipage}{.6\columnwidth}
\includegraphics[width=.95\textwidth]{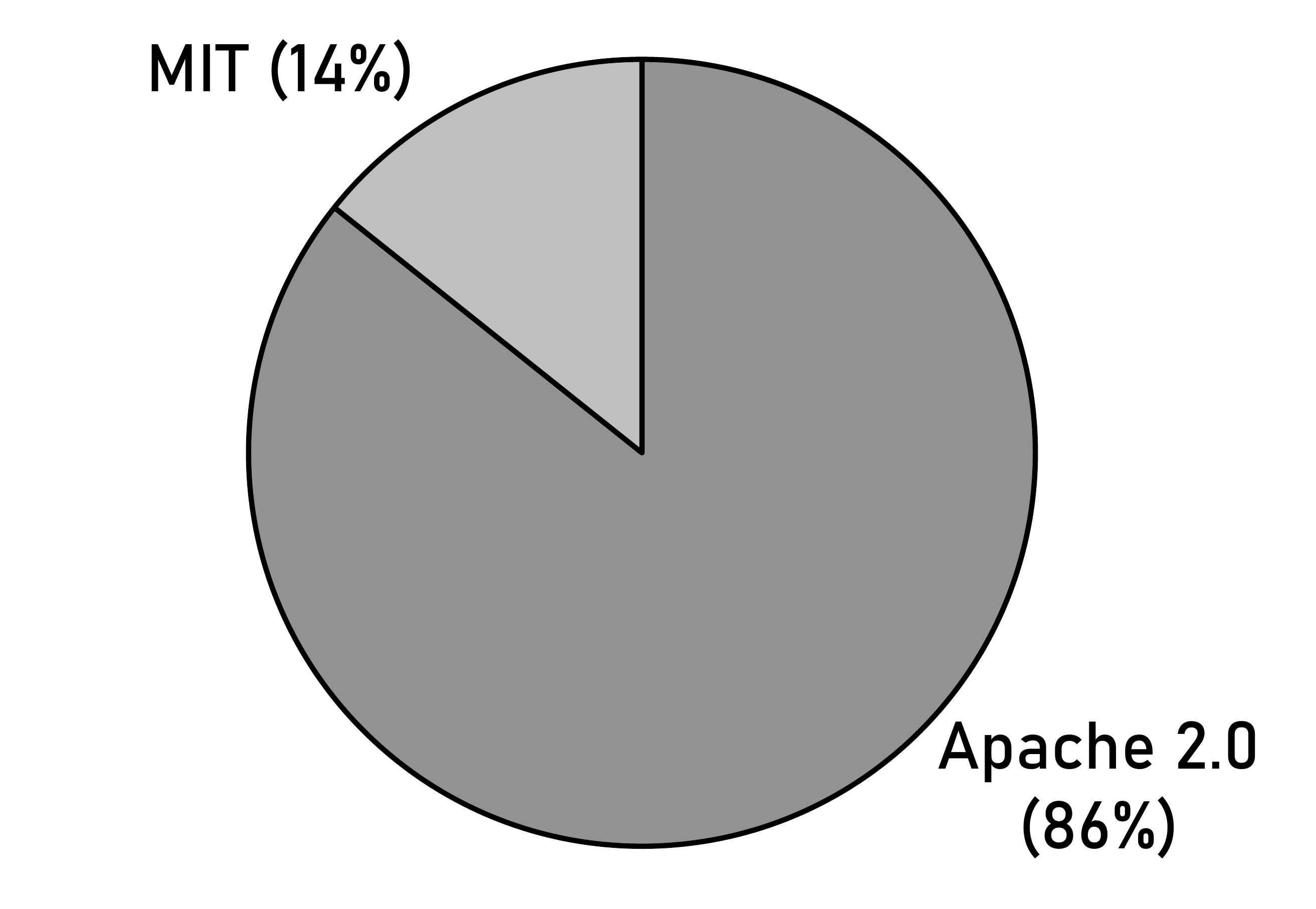}
\end{minipage}
\begin{minipage}{.39\columnwidth}
\footnotesize
\begin{tabular}{|cc|}
	\hline
	\licName & \textit{count} \\
	\hline
	Apache 2.0 & 6 \\
	MIT & 1 \\
	\hline
\end{tabular}
\end{minipage}
\caption{Distribution of \licName{}s among open source FaaS platforms, obtained by associating each \licName with the amount (\textit{count}) of FaaS platforms using it.}
\label{plot:open-license}
\end{figure}

\paragraph{\license}
The different licensing options used by the considered FaaS platforms are listed in \cref{tab:tech-review-managerial-license}.
As expected, all commercial solutions are licensed under provider's own proprietary license.
Non-commercial solutions instead mainly use the permissive \textit{Apache 2.0 License}, as easily observable in \Cref{plot:open-license}. 
The only exception is OpenFaaS, which uses the permissive \textit{MIT License}.
\begin{table}[h]
\centering
\caption{ Classification of FaaS Platforms, based on the \license category in the \managerial view of our classification framework.}
\label{tab:tech-review-managerial-license}
\footnotesize
\begin{tabular}{p{.2\linewidth}p{.42\linewidth}p{.22\linewidth}}
    \hline
                                                & \licName & \licType \\
    \hline
    \rowcolor[HTML]{EFEFEF}
    \textit{Apache Openwhisk}                   & Apache 2.0                    & permissive    \\
    \textit{AWS Lambda}                         & AWS Service Terms             & proprietary   \\
    \rowcolor[HTML]{EFEFEF}
    \textit{Fission}                            & Apache 2.0                    & permissive    \\
    \textit{Fn}                                 & Apache 2.0                    & permissive    \\
    \rowcolor[HTML]{EFEFEF}
    \textit{Google Cloud \newline Functions}    & Google Cloud Platform \newline Terms   & proprietary   \\
    \textit{Knative}                            & Apache 2.0                    & permissive    \\
    \rowcolor[HTML]{EFEFEF}
    \textit{Kubeless}                           & Apache 2.0                    & permissive    \\
    \textit{MS Azure \newline Functions}        & Microsoft Fn SLA              & proprietary   \\
    \rowcolor[HTML]{EFEFEF}
    \textit{Nuclio}                             & Apache 2.0                    & permissive    \\
    \textit{OpenFaaS}                           & MIT                           & permissive    \\
    \hline
\end{tabular}
\end{table}

\begin{center}
\noindent
\setlength{\fboxsep}{2mm}
\fbox{
    \centering
    \begin{minipage}{.9\linewidth}
      \textbf{Main Findings: Licensing}
      \begin{itemize}
          \item[\ding{212}] All open source FaaS platforms use permissive licenses.
          \item[\ding{212}] Most of open source FaaS platforms (6/7) are licensed under Apache 2.0 license.
          \item[\ding{212}] All commercial FaaS platforms use proprietary licenses, with MS Azure Functions also releasing some of its components as open source projects.
      \end{itemize}
\end{minipage}}
\vspace{.2\baselineskip}
\end{center}

\paragraph{\installation}
\cref{tab:tech-review-managerial-installation} classifies the considered FaaS platforms by indicating whether they are available \textit{as-a-service} or whether they can be installed on-premises.
Not surprisingly, commercial FaaS platforms are all offered \textit{as-a-service}. 
MS Azure Functions also supports installing the \textit{Azure Functions Host}, which enables running serverless applications on Linux, Kubernetes, MacOS and Windows.
%
\begin{table}[h]
\centering
\caption{ Classification of FaaS Platforms, based on the \installation category in the \managerial view of our classification framework. The abbreviation \enquote{\NA{}} stays for \enquote{not applicable}.}
\label{tab:tech-review-managerial-installation}
\footnotesize
\begin{threeparttable}
\begin{tabular}{p{.2\linewidth}p{.19\linewidth}p{.45\linewidth}}
    \hline
        & \textit{Type}                 
        & \textit{Target Hosts} \\
    \hline
    \rowcolor[HTML]{EFEFEF}
    \textit{Apache Openwhisk}                   
        & installable
        & Docker, Kubernetes, Linux, \newline MacOS, Mesos \\
    \textit{AWS Lambda}
        & as-a-service
        & \NA \\
    \rowcolor[HTML]{EFEFEF}
    \textit{Fission}
        & installable
        & Kubernetes \\
    \textit{Fn}
        & installable
        & Docker, Linux, MacOS, Unix \\
    \rowcolor[HTML]{EFEFEF}
    \textit{Google Cloud \newline Functions}
        & as-a-service
        & \NA \\
    \textit{Knative}
        & installable
        & Kubernetes \\
    \rowcolor[HTML]{EFEFEF}
    \textit{Kubeless}
        & installable
        & Kubernetes, Linux, MacOS, \newline Windows \\
    \textit{MS Azure \newline Functions}
        & as-a-service, installable 
        & Linux, Kubernetes, MacOS, \newline Windows \\
    \rowcolor[HTML]{EFEFEF}
    \textit{Nuclio}
        & as-a-service, installable
        & Kubernetes \\
    \textit{OpenFaaS}
        & installable
        & Docker\tnote{*}, faasd, Kubernetes, OpenShift\\
    \hline
\end{tabular}
\begin{tablenotes}
\item[*] \textit{In swarm mode \cite{docker:swarm}.}
\end{tablenotes}
\end{threeparttable}
\end{table}
\begin{figure}[h]
\centering
\includegraphics[width=.99\columnwidth]{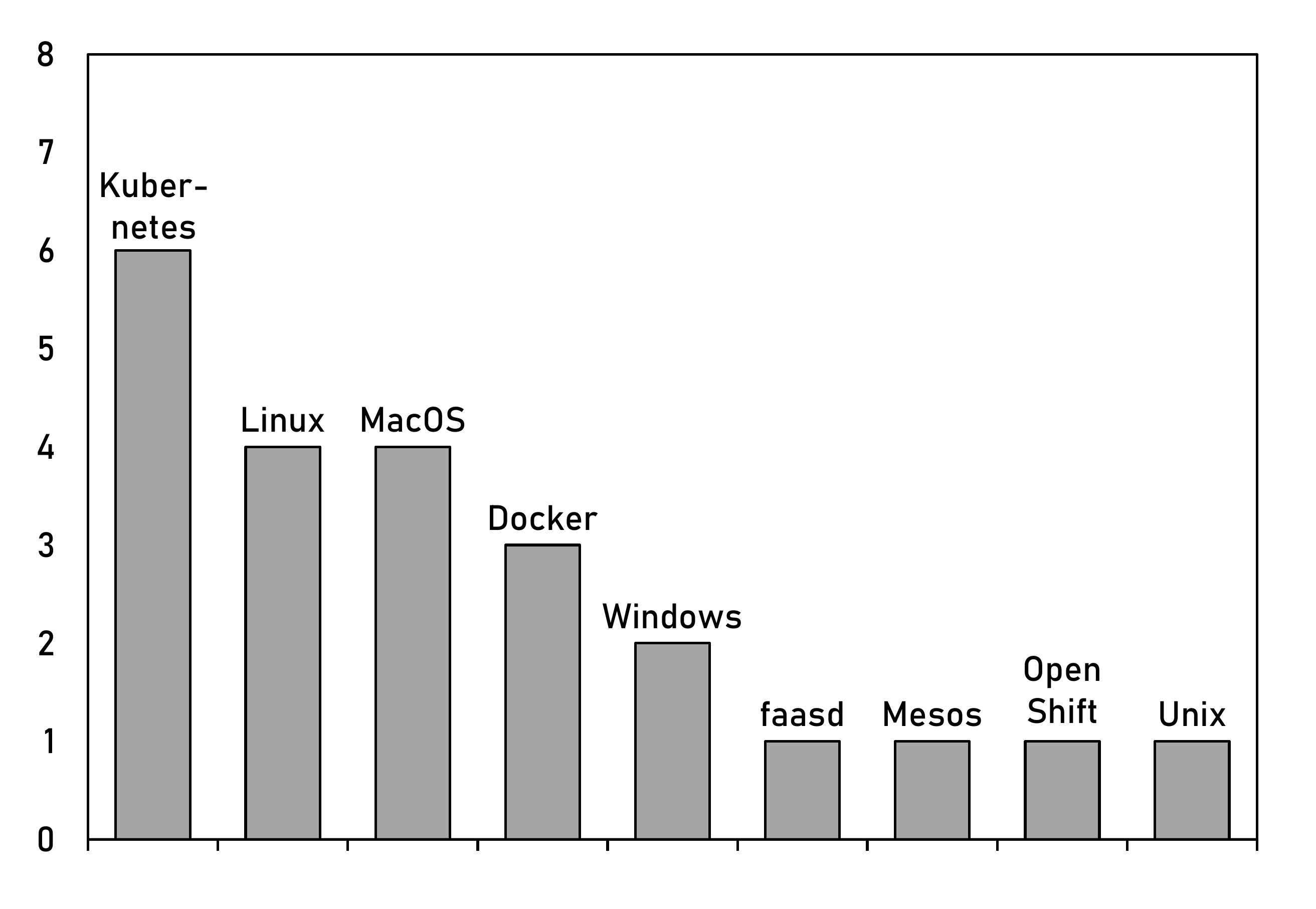}
\caption{Frequencies of values for the dimension \textit{Target Hosts} of the category \textit{Installation}, among the eight installable FaaS platforms.}
\label{plot:target-hosts}
\end{figure}

All open source solutions are \textit{installable} on different hosts, with Nuclio also coming \textit{as-a-service}, \ie allowing to run FaaS application on hosted instances of Nuclio and providing additional features, \eg advanced monitoring and logging solutions, and a 24/7 support \cite{nuclio}.
The target hosts actually supported by each installable platform are listed in \Cref{tab:tech-review-managerial-installation}.
\Cref{plot:target-hosts} instead displays the frequencies of target hosts, \ie it associates each target host with the amount of platforms that can be installed on it.
Kubernetes turns out to be the most supported target host, as all platforms (but Fn) can be installed on Kubernetes hosts.
A reason for this is that installable platforms are either developed by extending Kubernetes itself or to be natively integrated with it.

\begin{center}
\noindent
\setlength{\fboxsep}{2mm}
\fbox{
    \centering
    \begin{minipage}{.9\linewidth}
      \textbf{Main Findings: Installation}
      \begin{itemize}
          \item[\ding{212}] Most of the open source platforms (6/7) are \textit{only} available for on-premises installation.
          \item[\ding{212}] All commercial platforms are offered as-a-service, except some parts of MS Azure Functions that can also be installed on-premises.
          \item[\ding{212}] Most of installable platforms (5/8) support multiple target hosts, with Kubernetes being the most popular target host among them (7/8).
      \end{itemize}
\end{minipage}}
\vspace{.2\baselineskip}
\end{center}

\paragraph{\sourceCode}
The availability of the source code of the considered FaaS platforms (\ie whether they are open or closed sourced) is classified in \cref{tab:tech-review-managerial-sourcecode}, which also lists the open source repository and main programming language in which an open source platform is implemented.
AWS Lambda and Google Cloud Functions are closed source, while MS Azure Function is partly released open source.
Microsoft is indeed maintaining a GitHub repository~\cite{ms:azure-functions-github} where part of the C\# sources of MS Azure Functions are publicly available.
The GitHub repository provides sources of a simplified version MS Azure Functions runtime (\ie \textit{Azure Functions Host}), which can be installed and customised to run functions on-premises. 
The repository also provides open source tools for developing, debugging and testing functions running on MS Azure Functions.
All other platforms in \cref{tab:tech-review-managerial-sourcecode} are instead fully open source, and their source code can be found in corresponding GitHub repositories.

\begin{table}[h]
\centering
\caption{Classification of FaaS Platforms, based on the \sourceCode category in the \managerial view of our classification framework. The abbreviation \enquote{\NA{}} stays for \enquote{not applicable}.}
\label{tab:tech-review-managerial-sourcecode}
\footnotesize
\begin{threeparttable}
\begin{tabular}{p{.2\linewidth}p{.19\linewidth}p{.19\linewidth}p{.21\linewidth}}
    \hline
        & \textit{Availability}                 
        & \textit{Open Source Repository} 
        & \textit{Programming Language} \\
    \hline
    \rowcolor[HTML]{EFEFEF}
    \textit{Apache Openwhisk}                   
        & open source
        & GitHub
        & Scala \\
    \textit{AWS Lambda}
        & closed source
        & \NA
        & \NA \\
    \rowcolor[HTML]{EFEFEF}
    \textit{Fission}
        & open source
        & GitHub
        & Go \\
    \textit{Fn}
        & open source
        & GitHub
        & Go \\
    \rowcolor[HTML]{EFEFEF}
    \textit{Google Cloud \newline Functions}
        & closed source
        & \NA
        & \NA \\
    \textit{Knative}
        & open source
        & GitHub
        & Go \\
    \rowcolor[HTML]{EFEFEF}
    \textit{Kubeless}
        & open source
        & GitHub
        & Go \\
    \textit{MS Azure \newline Functions}
        & open source\tnote{*} 
        & GitHub
        & C\# \\
    \rowcolor[HTML]{EFEFEF}
    \textit{Nuclio}
        & open source
        & GitHub
        & Go \\
    \textit{OpenFaaS}
        & open source
        & GitHub
        & Go \\
    \hline
\end{tabular}
\begin{tablenotes}
    \item[*] \textit{Partially open sourced.} 
\end{tablenotes}
\end{threeparttable}
\end{table}
\begin{figure}[h]
\begin{minipage}{.6\columnwidth}
\includegraphics[width=.95\textwidth]{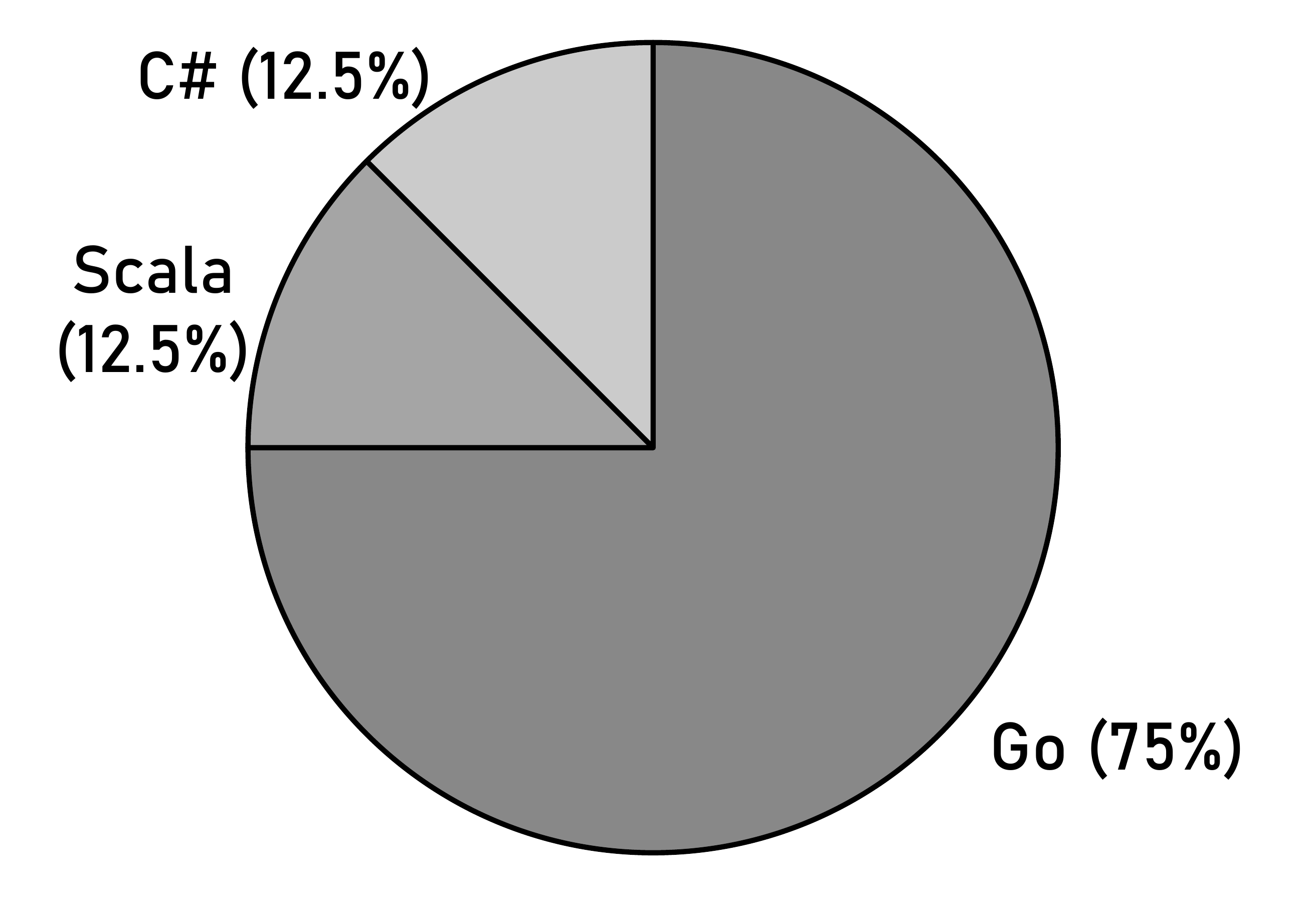}
\end{minipage}
\begin{minipage}{.39\columnwidth}
\footnotesize
\begin{tabular}{|cc|}
	\hline
	\textit{Programming} & \\
	\textit{Language} & \textit{count} \\
	\hline
	Go & 6 \\
	C\# & 1 \\
	Scala & 1 \\
	\hline
\end{tabular}
\end{minipage}
\caption{Distribution of main \textit{Programming Language}s for the classified FaaS platforms that release part of/all their sources on GitHub.}
\label{plot:prog-langs}
\end{figure}

Notably, Go turns out to be the most employed programming languages for developing the FaaS platforms which code is publicly available on GitHub (including MS Azure Functions), as also clearly visible from \Cref{plot:prog-langs}.
All reviewed open source platforms (but Apache Openwhisk) are indeed mainly developed in Go, which emphasizes the relevance of Go for cloud-native development. 
Additionally, this language choice could indicate that platforms are developed by extending Kubernetes, which is also developed in Go, or with the installation and integration with Kubernetes in mind.

\begin{center}
\noindent
\setlength{\fboxsep}{2mm}
\fbox{
    \centering
    \begin{minipage}{.9\linewidth}
      \textbf{Main Findings: Source Code}
      \begin{itemize}
          \item[\ding{212}] The source code of all open source FaaS platforms is hosted on GitHub.
          \item[\ding{212}] Most of the open source FaaS platforms (6/7) are mainly implemented in Go.
          \item[\ding{212}] MS Azure Functions is the only commercial FaaS platform partially open sourcing its components.
      \end{itemize}
\end{minipage}}
\vspace{.2\baselineskip}
\end{center}

\paragraph{\release}
The classification of the considered FaaS platforms based on their \release \relStatus is shown in \cref{tab:tech-review-managerial-release}.
Commercial platforms are obviously all in \textit{production}.
We can also observe that open source platforms show a kind of maturity.
Nuclio indeed already comes as a ready-to-market (\textit{rtm}) solution, while all other open source platforms (except for Knative) come as \textit{stable release}s or \textit{release candidate}s, which are the most mature statuses for in-development software.

\begin{table}[h]
\centering
\caption{ Classification of FaaS Platforms, based on the \release category in the \managerial view of our classification framework.}
\label{tab:tech-review-managerial-release}
\footnotesize
\begin{tabular}{p{.4\linewidth}p{.38\linewidth}}
    \hline
        & \textit{Status} \\
    \hline
    \rowcolor[HTML]{EFEFEF}
    \textit{Apache Openwhisk}                   
        & release candidate \\
    \textit{AWS Lambda}
        & production \\
    \rowcolor[HTML]{EFEFEF}
    \textit{Fission}
        & stable release \\
    \textit{Fn}
        & release candidate \\
    \rowcolor[HTML]{EFEFEF}
    \textit{Google Cloud Functions}
        & production \\
    \textit{Knative}
        & beta  \\ 
    \rowcolor[HTML]{EFEFEF}
    \textit{Kubeless}
        & release  candidate \\
    \textit{MS Azure Functions}
        & production \\
    \rowcolor[HTML]{EFEFEF}
    \textit{Nuclio}
        & rtm \\
    \textit{OpenFaaS}
        & release candidate \\
    \hline
\end{tabular}
\end{table}

\begin{center}
    \noindent
    \setlength{\fboxsep}{2mm}
    \fbox{
        \centering
        \begin{minipage}{.9\linewidth}
            \textbf{Main Findings: Release}
            \begin{itemize}
                \item[\ding{212}] All commercial platforms are production-ready.          
                \item[\ding{212}] Open source platforms are also quite mature, with 4/7 release candidates, 1/7 stable releases and 1/7 ready-to-market (rtm).
            \end{itemize}
    \end{minipage}}
    \vspace{.2\baselineskip}
\end{center}


\begin{table*}[p]
    \centering
    \caption{ Classification of FaaS Platforms, based on the \interface category in the \managerial view of our classification framework.}
    \label{tab:tech-review-managerial-interface}
    \footnotesize
    \def\arraystretch{1.1}
    \definecolor{lightgray}{gray}{0.9}
    \begin{threeparttable}
        \begin{tabularx}{.99\textwidth}{c c m{5em} m{3em} c c m{4.5em} c c m{4.5em} C C}
            \hline
            \multicolumn{2}{l}{}  
            & \centering \textit{Apache Openwhisk} 
            & \centering \textit{AWS Lambda} 
            & \textit{Fission}
            & \textit{Fn} 
            & \centering \textit{Google Cloud Functions} 
            & \textit{Knative} 
            & \textit{Kubeless} 
            & \centering \textit{MS Azure Functions} 
            & \textit{Nuclio} 
            & \textit{OpenFaaS} \\
            
            \hline
            \parbox[t]{1mm}{\multirow{3}{*}{\rotatebox[origin=c]{90}{\textit{Type}}}}
            & \cellcolor{lightgray} cli      
            & \cellcolor{lightgray} \centering \ok
            & \cellcolor{lightgray} \centering \ok
            & \cellcolor{lightgray} \centering \ok
            & \cellcolor{lightgray} \centering \ok
            & \cellcolor{lightgray} \centering \ok
            & \cellcolor{lightgray} \centering \ok
            & \cellcolor{lightgray} \centering \ok
            & \cellcolor{lightgray} \centering \ok
            & \cellcolor{lightgray} \centering \ok
            & \cellcolor{lightgray} \ok\\
            & \cellcolor{white} api 
            & \cellcolor{white} \centering \ok
            & \cellcolor{white} \centering \ok
            & \cellcolor{white} \centering \ko
            & \cellcolor{white} \centering \ok
            & \cellcolor{white} \centering \ok
            & \cellcolor{white} \centering \ok
            & \cellcolor{white} \centering \ko
            & \cellcolor{white} \centering \ok
            & \cellcolor{white} \centering \ko
            & \cellcolor{white} \ok \\ 
            & \cellcolor{lightgray} gui    
            & \cellcolor{lightgray} \centering \ko
            & \cellcolor{lightgray} \centering \ok
            & \cellcolor{lightgray} \centering \ko
            & \cellcolor{lightgray} \centering \ok
            & \cellcolor{lightgray} \centering \ok
            & \cellcolor{lightgray} \centering \ko
            & \cellcolor{lightgray} \centering \ko
            & \cellcolor{lightgray} \centering \ok
            & \cellcolor{lightgray} \centering \ok
            & \cellcolor{lightgray} \ok  \\ 
            \hline
            
            \parbox[t]{1mm}{\multirow{4}{*}{\rotatebox[origin=c]{90}{\textit{App. Man.}}}}
            & \cellcolor{lightgray} create    
            & \cellcolor{lightgray} \centering \ok 
            & \cellcolor{lightgray} \centering \ok 
            & \cellcolor{lightgray} \centering \ok 
            & \cellcolor{lightgray} \centering \ok 
            & \cellcolor{lightgray} \centering \ok 
            & \cellcolor{lightgray} \centering \ok 
            & \cellcolor{lightgray} \centering \ok 
            & \cellcolor{lightgray} \centering \ok 
            & \cellcolor{lightgray} \centering \ok 
            & \cellcolor{lightgray} \ok 
            \\ 
            & \cellcolor{white} retrieve      
            & \cellcolor{white} \centering \ok 
            & \cellcolor{white} \centering \ok 
            & \cellcolor{white} \centering \ok 
            & \cellcolor{white} \centering \ok 
            & \cellcolor{white} \centering \ok 
            & \cellcolor{white} \centering \ok 
            & \cellcolor{white} \centering \ok 
            & \cellcolor{white} \centering \ok 
            & \cellcolor{white} \centering \ok 
            & \cellcolor{white} \ok 
            \\ 
            & \cellcolor{lightgray} update
            & \cellcolor{lightgray} \centering \ok 
            & \cellcolor{lightgray} \centering \ok 
            & \cellcolor{lightgray} \centering \ok 
            & \cellcolor{lightgray} \centering \ok 
            & \cellcolor{lightgray} \centering \ok 
            & \cellcolor{lightgray} \centering \ok 
            & \cellcolor{lightgray} \centering \ok 
            & \cellcolor{lightgray} \centering \ok 
            & \cellcolor{lightgray} \centering \ok 
            & \cellcolor{lightgray} \ok 
            \\ 
            & \cellcolor{white} delete 
            & \cellcolor{white} \centering \ok 
            & \cellcolor{white} \centering \ok 
            & \cellcolor{white} \centering \ok 
            & \cellcolor{white} \centering \ok 
            & \cellcolor{white} \centering \ok 
            & \cellcolor{white} \centering \ok 
            & \cellcolor{white} \centering \ok 
            & \cellcolor{white} \centering \ok 
            & \cellcolor{white} \centering \ok 
            & \cellcolor{white} \ok 
            \\ 
            \hline 
            \parbox[t]{1mm}{\multirow{5}{*}{\rotatebox[origin=c]{90}{\textit{Plat. Adm.}}}}
            & \cellcolor{white} deployment      
            & \cellcolor{white} \centering \ok 
            & \cellcolor{white} \centering \ko 
            & \cellcolor{white} \centering \ok 
            & \cellcolor{white} \centering \ok 
            & \cellcolor{white} \centering \ko 
            & \cellcolor{white} \centering \ok 
            & \cellcolor{white} \centering \ok 
            & \cellcolor{white} \centering \ko 
            & \cellcolor{white} \centering \ok 
            & \cellcolor{white} \ok 
            \\ 
            & \cellcolor{lightgray} configuration    
            & \cellcolor{lightgray} \centering \ko 
            & \cellcolor{lightgray} \centering \ko 
            & \cellcolor{lightgray} \centering \ok 
            & \cellcolor{lightgray} \centering \ok 
            & \cellcolor{lightgray} \centering \ko 
            & \cellcolor{lightgray} \centering \ok 
            & \cellcolor{lightgray} \centering \ok 
            & \cellcolor{lightgray} \centering \ko 
            & \cellcolor{lightgray} \centering \ko 
            & \cellcolor{lightgray} \ko 
            \\ 
            & \cellcolor{white} enactment 
            & \cellcolor{white} \centering \ok 
            & \cellcolor{white} \centering \ko 
            & \cellcolor{white} \centering \ok 
            & \cellcolor{white} \centering \ok 
            & \cellcolor{white} \centering \ko 
            & \cellcolor{white} \centering \ok 
            & \cellcolor{white} \centering \ok 
            & \cellcolor{white} \centering \ko 
            & \cellcolor{white} \centering \ok 
            & \cellcolor{white} \ok 
            \\ 
            & \cellcolor{lightgray} termination    
            & \cellcolor{lightgray} \centering \ko\tnote{*} 
            & \cellcolor{lightgray} \centering \ko 
            & \cellcolor{lightgray} \centering \ko\tnote{*} 
            & \cellcolor{lightgray} \centering \ko\tnote{*} 
            & \cellcolor{lightgray} \centering \ko 
            & \cellcolor{lightgray} \centering \ko\tnote{*} 
            & \cellcolor{lightgray} \centering \ko\tnote{*} 
            & \cellcolor{lightgray} \centering \ko 
            & \cellcolor{lightgray} \centering \ko\tnote{*} 
            & \cellcolor{lightgray} \ko\tnote{*} 
            \\ 
            & \cellcolor{white} undeployment 
            & \cellcolor{white} \centering \ko\tnote{*} 
            & \cellcolor{white} \centering \ko 
            & \cellcolor{white} \centering \ko\tnote{*} 
            & \cellcolor{white} \centering \ko\tnote{*} 
            & \cellcolor{white} \centering \ko 
            & \cellcolor{white} \centering \ko\tnote{*} 
            & \cellcolor{white} \centering \ko\tnote{*} 
            & \cellcolor{white} \centering \ko 
            & \cellcolor{white} \centering \ko\tnote{*} 
            & \cellcolor{white} \ko\tnote{*}  
            \\ 
            \hline
        \end{tabularx}
        \begin{tablenotes}
            \scriptsize
            \item[*] \textit{Termination/undeployment can be achieved by stopping/uninstalling the platform instance with host commands.} 
        \end{tablenotes}
    \end{threeparttable}
\end{table*}
\begin{table*}[p]
    \centering
    \caption{ Classification of FaaS Platforms, based on the \community category in the \managerial view of our classification framework. \enquote{\NA{}} stays for \enquote{not applicable}.}
    \label{tab:tech-review-managerial-community}
    \footnotesize
    \def\arraystretch{1.1}
    \definecolor{lightgray}{gray}{0.9}
    \begin{threeparttable}
        \begin{tabularx}{.99\textwidth}{c c m{5em} m{3em} c c m{4.5em} c c m{4.5em} C C}
            \hline
            \multicolumn{2}{l}{}  
            & \centering \textit{Apache Openwhisk} 
            & \centering \textit{AWS Lambda} 
            & \textit{Fission}
            & \textit{Fn} 
            & \centering \textit{Google Cloud Functions} 
            & \textit{Knative} 
            & \textit{Kubeless} 
            & \centering \textit{MS Azure Functions}
            & \textit{Nuclio} 
            & \textit{OpenFaaS} \\
            
            \hline
            \parbox[t]{1mm}{\multirow{4}{*}{\rotatebox[origin=c]{90}{\textit{GitHub}}}}
            & \cellcolor{lightgray} Stars    
            & \cellcolor{lightgray} \centering 4.6k 
            & \cellcolor{lightgray} \centering \NA 
            & \cellcolor{lightgray} \centering 5k 
            & \cellcolor{lightgray} \centering 4.5k 
            & \cellcolor{lightgray} \centering \NA 
            & \cellcolor{lightgray} \centering 2.7k\tnote{*} 
            & \cellcolor{lightgray} \centering 5.5k 
            & \cellcolor{lightgray} \centering \NA 
            & \cellcolor{lightgray} \centering 3.2k 
            & \cellcolor{lightgray} 17.1k 
            \\ 
            & \cellcolor{white} Forks      
            & \cellcolor{white} \centering 890 
            & \cellcolor{white} \centering \NA 
            & \cellcolor{white} \centering 451 
            & \cellcolor{white} \centering 328 
            & \cellcolor{white} \centering \NA 
            & \cellcolor{white} \centering 563\tnote{*} 
            & \cellcolor{white} \centering 558 
            & \cellcolor{white} \centering \NA 
            & \cellcolor{white} \centering 318 
            & \cellcolor{white} 1.4k 
            \\ 
            & \cellcolor{lightgray} Issues    
            & \cellcolor{lightgray} \centering 260 
            & \cellcolor{lightgray} \centering \NA 
            & \cellcolor{lightgray} \centering 191 
            & \cellcolor{lightgray} \centering 121 
            & \cellcolor{lightgray} \centering \NA 
            & \cellcolor{lightgray} \centering 274\tnote{*}
            & \cellcolor{lightgray} \centering 158 
            & \cellcolor{lightgray} \centering \NA 
            & \cellcolor{lightgray} \centering 56 
            & \cellcolor{lightgray} 63 
            \\ 
            & \cellcolor{white} Commits      
            & \cellcolor{white} \centering 2.7k 
            & \cellcolor{white} \centering \NA 
            & \cellcolor{white} \centering 1.1k 
            & \cellcolor{white} \centering 3.4k 
            & \cellcolor{white} \centering \NA 
            & \cellcolor{white} \centering 3.8k\tnote{*} 
            & \cellcolor{white} \centering 1k 
            & \cellcolor{white} \centering \NA 
            & \cellcolor{white} \centering 1.2k 
            & \cellcolor{white} 1.8k 
            \\ 
            & \cellcolor{lightgray} Contributors   
            & \cellcolor{lightgray} \centering 173 
            & \cellcolor{lightgray} \centering \NA 
            & \cellcolor{lightgray} \centering 91 
            & \cellcolor{lightgray} \centering 85 
            & \cellcolor{lightgray} \centering \NA 
            & \cellcolor{lightgray} \centering 165\tnote{*} 
            & \cellcolor{lightgray} \centering 82 
            & \cellcolor{lightgray} \centering \NA 
            & \cellcolor{lightgray} \centering 51 
            & \cellcolor{lightgray} 138 
            \\ 
            \hline 
            \parbox[t]{1mm}{\rotatebox[origin=c]{90}{\textit{SO}}}
            & \cellcolor{white} Questions 
            & \cellcolor{white} \centering 188 
            & \cellcolor{white} \centering 15.1k  
            & \cellcolor{white} \centering 6 
            & \cellcolor{white} \centering 20 
            & \cellcolor{white} \centering 8.5k  
            & \cellcolor{white} \centering 57 
            & \cellcolor{white} \centering 7 
            & \cellcolor{white} 6.3k \centering  
            & \cellcolor{white} \centering 3 
            & \cellcolor{white} 28 
            \\ 
            \hline
        \end{tabularx}
        \begin{tablenotes}
            \scriptsize
            \item[*] \textit{Values for the function hosting component of Knative, \ie Knative Serving.}
        \end{tablenotes}
    \end{threeparttable}
\end{table*}
\begin{table*}[p]
    \centering
    \caption{ Classification of FaaS Platforms, based on the \documentation category in the \managerial view of our classification framework.}
    \label{tab:tech-review-managerial-documentation}
    \footnotesize
    \def\arraystretch{1.1}
    \definecolor{lightgray}{gray}{0.9}
    \begin{threeparttable}
        \begin{tabularx}{.99\textwidth}{c c m{5em} m{3em} c c m{4.5em} c c m{4.5em} C C}
            \hline
            \multicolumn{2}{l}{}  
            & \centering \textit{Apache Openwhisk} 
            & \centering \textit{AWS Lambda} 
            & \textit{Fission}
            & \textit{Fn} 
            & \centering \textit{Google Cloud Functions} 
            & \textit{Knative} 
            & \textit{Kubeless} 
            & \centering \textit{MS Azure Functions} 
            & \textit{Nuclio} 
            & \textit{OpenFaaS} \\
            
            \hline
            \parbox[t]{1mm}{\multirow{2}{*}{\rotatebox[origin=c]{90}{\textit{App.}}}}
            & \cellcolor{lightgray} development    
            & \cellcolor{lightgray} \centering \ok 
            & \cellcolor{lightgray} \centering \ok 
            & \cellcolor{lightgray} \centering \ok 
            & \cellcolor{lightgray} \centering \ok 
            & \cellcolor{lightgray} \centering \ok 
            & \cellcolor{lightgray} \centering \ko 
            & \cellcolor{lightgray} \centering \ok 
            & \cellcolor{lightgray} \centering \ok 
            & \cellcolor{lightgray} \centering \ko 
            & \cellcolor{lightgray} \ko 
            \\ 
            & \cellcolor{white} deployment      
            & \cellcolor{white} \centering \ok 
            & \cellcolor{white} \centering \ok 
            & \cellcolor{white} \centering \ok 
            & \cellcolor{white} \centering \ok 
            & \cellcolor{white} \centering \ok 
            & \cellcolor{white} \centering \ok 
            & \cellcolor{white} \centering \ok 
            & \cellcolor{white} \centering \ok 
            & \cellcolor{white} \centering \ok 
            & \cellcolor{white} \ok 
            \\ 
            \hline
            \parbox[t]{1mm}{\multirow{4}{*}{\rotatebox[origin=c]{90}{\textit{Platform}}}}
            & \cellcolor{lightgray} usage    
            & \cellcolor{lightgray} \centering \ok 
            & \cellcolor{lightgray} \centering \ok 
            & \cellcolor{lightgray} \centering \ok 
            & \cellcolor{lightgray} \centering \ok 
            & \cellcolor{lightgray} \centering \ok 
            & \cellcolor{lightgray} \centering \ok 
            & \cellcolor{lightgray} \centering \ok 
            & \cellcolor{lightgray} \centering \ok 
            & \cellcolor{lightgray} \centering \ok 
            & \cellcolor{lightgray} \ok 
            \\ 
            & \cellcolor{white} development      
            & \cellcolor{white} \centering \ok 
            & \cellcolor{white} \centering \ko 
            & \cellcolor{white} \centering \ko\tnote{*} 
            & \cellcolor{white} \centering \ko\tnote{*} 
            & \cellcolor{white} \centering \ko 
            & \cellcolor{white} \centering \ko\tnote{*} 
            & \cellcolor{white} \centering \ok 
            & \cellcolor{white} \centering \ko 
            & \cellcolor{white} \centering \ko\tnote{*} 
            & \cellcolor{white} \ok 
            \\ 
            & \cellcolor{lightgray} deployment   
            & \cellcolor{lightgray} \centering \ok 
            & \cellcolor{lightgray} \centering \ko 
            & \cellcolor{lightgray} \centering \ok 
            & \cellcolor{lightgray} \centering \ok 
            & \cellcolor{lightgray} \centering \ko 
            & \cellcolor{lightgray} \centering \ok 
            & \cellcolor{lightgray} \centering \ok 
            & \cellcolor{lightgray} \centering \ok 
            & \cellcolor{lightgray} \centering \ok 
            & \cellcolor{lightgray} \ok 
            \\ 
            & \cellcolor{white} architecture      
            & \cellcolor{white} \centering \ok 
            & \cellcolor{white} \centering \ko 
            & \cellcolor{white} \centering \ko 
            & \cellcolor{white} \centering \ko 
            & \cellcolor{white} \centering \ko 
            & \cellcolor{white} \centering \ko 
            & \cellcolor{white} \centering \ok 
            & \cellcolor{white} \centering \ko 
            & \cellcolor{white} \centering \ok 
            & \cellcolor{white} \ok 
            \\ 
            \hline 
        \end{tabularx}
        \begin{tablenotes}
            \item[*] \textit{Only providing guidelines/code of conduct for contributing to the project.}
        \end{tablenotes}
    \end{threeparttable}
\end{table*}
\begin{table*}[p]
    \centering
    \caption{ Classification of FaaS Platforms, based on the \restrictions category in the \managerial view of our classification framework. Letters \textsf{L} and \textsf{U} are used to denote the possible values \textit{limited} and \textit{unbounded}, respectively.}
    \label{tab:tech-review-managerial-quotas}
    \footnotesize
    \def\arraystretch{1.1}
    \definecolor{lightgray}{gray}{0.9}
    \begin{threeparttable}
        \begin{tabularx}{.99\textwidth}{c c m{5em} m{3em} c c m{4.5em} c c m{4.5em} C C}
            \hline
            \multicolumn{2}{l}{}  
            & \centering \textit{Apache Openwhisk} 
            & \centering \textit{AWS Lambda} 
            & \textit{Fission}
            & \textit{Fn} 
            & \centering \textit{Google Cloud Functions} 
            & \textit{Knative} 
            & \textit{Kubeless} 
            & \centering \textit{MS Azure Functions} 
            & \textit{Nuclio} 
            & \textit{OpenFaaS} \\
            
            \hline
            \parbox[t]{1mm}{\multirow{2}{*}{\rotatebox[origin=c]{90}{\textit{Depl.}}}}
            & \cellcolor{lightgray} \textit{Pack. Size}    
            & \cellcolor{lightgray} \centering \textsf{L} 
            & \cellcolor{lightgray} \centering \textsf{L} 
            & \cellcolor{lightgray} \centering \textsf{U} 
            & \cellcolor{lightgray} \centering \textsf{U} 
            & \cellcolor{lightgray} \centering \textsf{L} 
            & \cellcolor{lightgray} \centering \textsf{U} 
            & \cellcolor{lightgray} \centering \textsf{U} 
            & \cellcolor{lightgray} \centering \textsf{L} 
            & \cellcolor{lightgray} \centering \textsf{U} 
            & \cellcolor{lightgray} \textsf{U} 
            \\ 
            & \cellcolor{white} \textit{Code Size}      
            & \cellcolor{white} \centering  \textsf{L} 
            & \cellcolor{white} \centering  \textsf{U} 
            & \cellcolor{white} \centering  \textsf{U} 
            & \cellcolor{white} \centering  \textsf{U} 
            & \cellcolor{white} \centering  \textsf{U} 
            & \cellcolor{white} \centering  \textsf{U} 
            & \cellcolor{white} \centering  \textsf{U} 
            & \cellcolor{white} \centering  \textsf{U} 
            & \cellcolor{white} \centering  \textsf{U} 
            & \cellcolor{white} \textsf{U} 
            \\ 
            \hline
            \parbox[t]{1mm}{\multirow{4}{*}{\rotatebox[origin=c]{90}{\textit{Runtime}}}}
            & \cellcolor{lightgray} \textit{CPU}    
            & \cellcolor{lightgray} \centering \textsf{U} 
            & \cellcolor{lightgray} \centering \textsf{U} 
            & \cellcolor{lightgray} \centering \textsf{U}\tnote{*} 
            & \cellcolor{lightgray} \centering \textsf{U} 
            & \cellcolor{lightgray} \centering \textsf{L} 
            & \cellcolor{lightgray} \centering \textsf{U} 
            & \cellcolor{lightgray} \centering \textsf{U}\tnote{*} 
            & \cellcolor{lightgray} \centering \textsf{U} 
            & \cellcolor{lightgray} \centering \textsf{U} 
            & \cellcolor{lightgray} \textsf{U}\tnote{*} 
            \\ 
            & \cellcolor{white} \textit{Memory}      
            & \cellcolor{white} \centering  \textsf{L} 
            & \cellcolor{white} \centering  \textsf{L} 
            & \cellcolor{white} \centering  \textsf{U}\tnote{*} 
            & \cellcolor{white} \centering  \textsf{U} 
            & \cellcolor{white} \centering  \textsf{L} 
            & \cellcolor{white} \centering  \textsf{U} 
            & \cellcolor{white} \centering  \textsf{U}\tnote{*} 
            & \cellcolor{white} \centering  \textsf{U} 
            & \cellcolor{white} \centering  \textsf{U} 
            & \cellcolor{white} \textsf{U}\tnote{*} 
            \\ 
            & \cellcolor{lightgray} \textit{Storage}    
            & \cellcolor{lightgray} \centering \textsf{U} 
            & \cellcolor{lightgray} \centering \textsf{L} 
            & \cellcolor{lightgray} \centering \textsf{U} 
            & \cellcolor{lightgray} \centering \textsf{U} 
            & \cellcolor{lightgray} \centering \textsf{U} 
            & \cellcolor{lightgray} \centering \textsf{U} 
            & \cellcolor{lightgray} \centering \textsf{U} 
            & \cellcolor{lightgray} \centering \textsf{L} 
            & \cellcolor{lightgray} \centering \textsf{U} 
            & \cellcolor{lightgray} \textsf{U} 
            \\ 
            & \cellcolor{white} \textit{Exec. Time}      
            & \cellcolor{white} \centering  \textsf{L} 
            & \cellcolor{white} \centering  \textsf{L} 
            & \cellcolor{white} \centering  \textsf{U}\tnote{*} 
            & \cellcolor{white} \centering  \textsf{U} 
            & \cellcolor{white} \centering  \textsf{L} 
            & \cellcolor{white} \centering  \textsf{U}\tnote{*} 
            & \cellcolor{white} \centering  \textsf{U}\tnote{*} 
            & \cellcolor{white} \centering  \textsf{L} 
            & \cellcolor{white} \centering  \textsf{U}\tnote{*} 
            & \cellcolor{white} \textsf{U}\tnote{*} 
            \\ 
            \hline
        \end{tabularx}
        \begin{tablenotes}
            \scriptsize
            \item[*] \textit{Unbounded by default, but application deployments can be configured to set quotas.}
        \end{tablenotes}
    \end{threeparttable}
\end{table*}

\paragraph{\interface}
\cref{tab:tech-review-managerial-interface} classifies considered FaaS platforms under the \interface category of our classification framework, \ie by showing whether they offer a CLI, API or GUI, and which capabilities are featured through such interfaces. 
While we can observe that all platforms support all CRUD operations for serverless applications, they considerably vary in terms of interface type and of operations offered to manage the platform.
A first distinction occurs between commercial platforms and open source platforms, with the former being the only providing all types of interfaces.
In contrast, all open source solutions provide a command-line interface (\textit{CLI}) to access and administer the platform, with only 4/7 also providing an HTTP-based \textit{API} and only 2/7 featuring a \textit{GUI}.

\begin{figure}[h]
\centering
\includegraphics[width=.99\columnwidth]{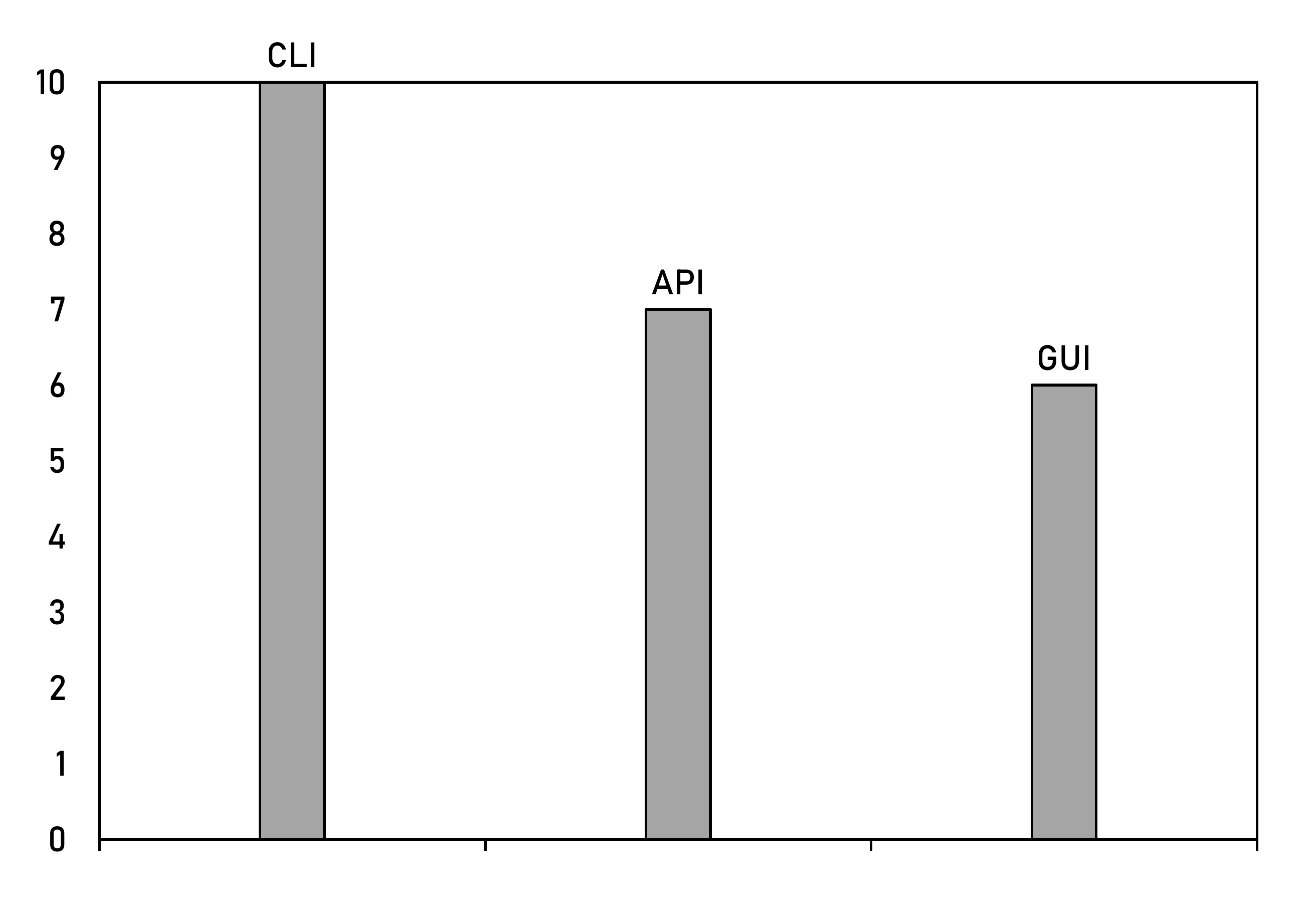}
\caption{Frequencies of values for the dimension \textit{Type} of the category \textit{Interface}, among the ten classified FaaS platforms.}
\label{plot:intf-types}
\end{figure}

\Cref{plot:intf-types} slightly elaborates on \interface \intType{}s, by showing their overall frequencies, \ie by associating each \intType of \interface with the amount of FaaS platforms supporting it. 
It clearly emerges that CLI and API are the most important types to be  supported, and this is clearly motivated by need for programming how to automate the management of serverless applications (and of installable platforms, as well). 
GUI is however also supported by more than a half of reviewed platforms (6/10), as it eases manually interacting with the platform wherever needed.

Finally, it is worth noting that a neat distinction between commercial platforms and open source platforms occurs as per what regards the administration of the platform itself.
Commercial platforms do not provide any operation to administer the platform itself, which is a logical result of being offered as-a-service~(\cref{tab:tech-review-managerial-installation}).
Conversely, given that all open source platforms come as installable solutions (\cref{tab:tech-review-managerial-installation}), they also provide operations for deploying and enacting the platform, with some also providing operations to configure the platform (\eg to set quotas, as shown in \cref{tab:tech-review-managerial-quotas}).
Finally, they all neither support nor provide means for terminating or undeploying the platform (\eg in the form of runnable scripts or binaries).
Installable FaaS platforms indeed rely on what available on the target host for being terminated or undeployed, \eg if deployed with Docker or Kubernetes, they rely on the latters' functionalities to stop and delete the running Docker containers or Kubernetes deployments.

\begin{center}
    \noindent
    \setlength{\fboxsep}{2mm}
    \fbox{
        \centering
        \begin{minipage}{.9\linewidth}
            \textbf{Main Findings: Interface}
            \begin{itemize}
                \item[\ding{212}] All FaaS platforms provide a CLI, with most of them (7/10) being also accessible via an API. A GUI is also offered by more than a half of reviewed platforms~(6/10).
                \item[\ding{212}] All platforms support CRUD operations for managing serverless applications.
                \item[\ding{212}] Open source platforms vary considerably in the way they can be administered.
            \end{itemize}
    \end{minipage}}
    \vspace{.2\baselineskip}
\end{center}

\begin{figure}[h]
\centering
\includegraphics[width=.99\columnwidth]{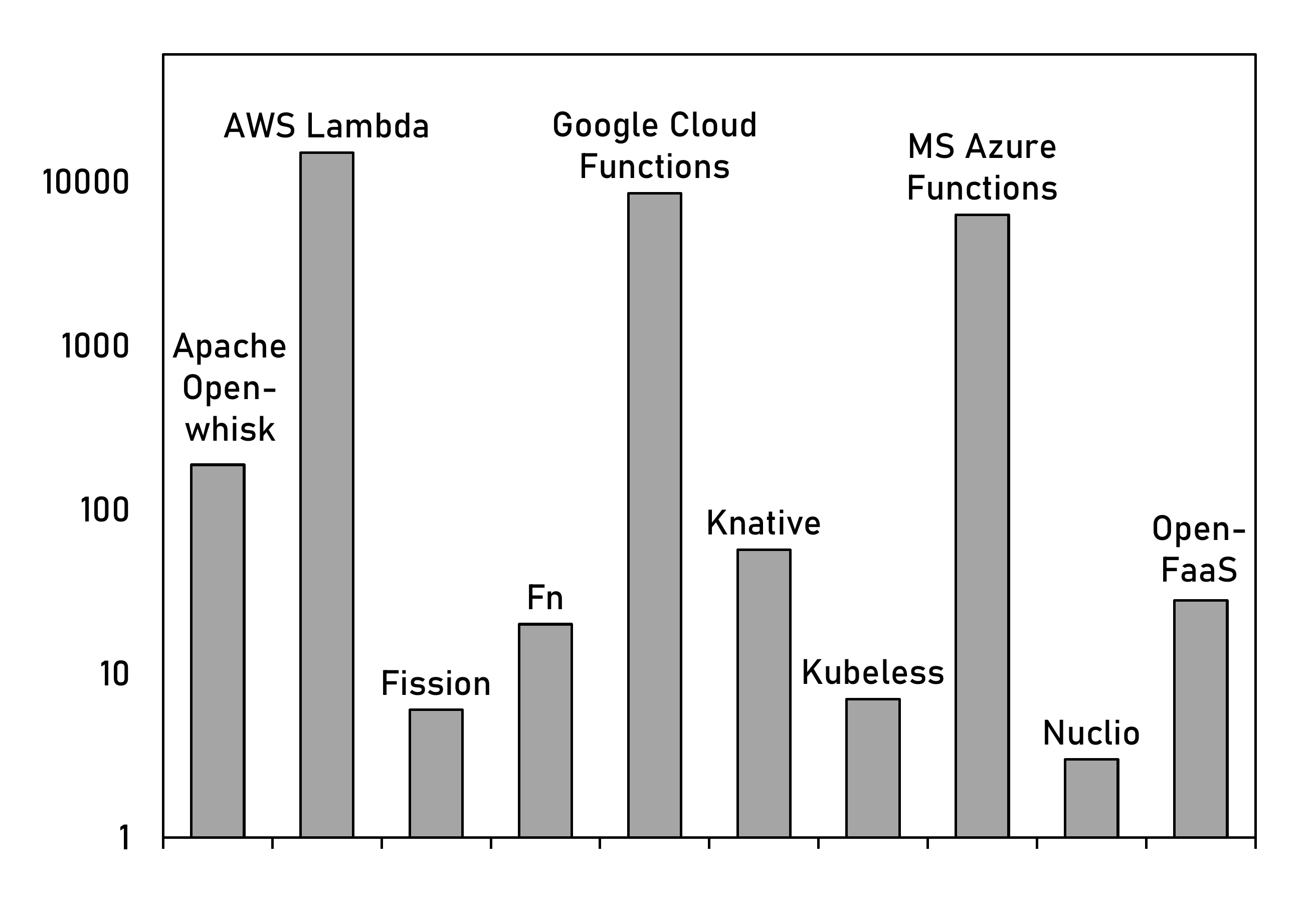}
\caption{Values for the dimension \commStackoverflow of the category \community (displayed on a logarithmic $y$-axis).}
\label{plot:stackoverflow}
\end{figure}

\paragraph{\community}
As an additional indicator of the platform's popularity, in~\Cref{tab:tech-review-managerial-community} we present the details on community involved in the development of considered FaaS platforms\footnote{In the table, we omit GitHub metrics for MS Azure Functions, as only some of its sub-components are released open source.}. 
Notably, the amount of questions on StackOverflow demonstrates that commercial FaaS platforms are considerably more popular and used in comparison with the reviewed open source platforms (\Cref{plot:stackoverflow}).
This is influenced by the difference in the maturity of the platforms themselves, by the underlying infrastructure, and by natively supported services. 

Among open source FaaS platforms, OpenFaaS is by far the most popular in terms of stars, and it shows also a quite active community (in terms of forks, commits and contributors).
Kubeless, Fission, Apache OpenWhisk and Fn respectively follow in term of popularity, if measured in amount of stars.
The Knative platform is instead that with the most active community of contributors, with its Serving component showing a total of 3.8k commits, by far higher if compared with all other open source platforms.
Finally, Apache Openwhisk is actively maintained by 173 contributors, which constitute the biggest community of contributors among investigated open source platforms.
Despite all such information just provides an indicator of popularity, it can still be important while choosing a platform for prioritising those with higher popularity or bigger community of contributors, as both indicate an active maintenance of the platform itself.

\begin{center}
    \noindent
    \setlength{\fboxsep}{2mm}
    \fbox{
        \centering
        \begin{minipage}{.9\linewidth}
            \textbf{Main Findings: Community}
            \begin{itemize}
                \item[\ding{212}] OpenFaaS, Apache Openwhisk, and Knative have the highest ratings on GitHub in terms of stars, contributors, and commits, respectively.
                \item[\ding{212}] Stackoverflow questions show a drastic difference in interest between commercial and open source platforms.
            \end{itemize}
    \end{minipage}}
    \vspace{.2\baselineskip}
\end{center}

\paragraph{\documentation}
\cref{tab:tech-review-managerial-documentation} illustrates the classification of considered FaaS platforms based on the documentation they provide on development and deployment of supported applications, and on the platform itself.
All considered platforms widely document how to use the platform itself and how to deploy serverless applications.
Most of considered platforms (except for Knative, Nuclio and OpenFaaS) also concretely document how to develop serverless application that can run on the platform, hereby included the available runtime environments and how to exploit the support provided by the platform (see \cref{ssec:devops-view} for further details).
In addition, each installable platform documents its deployment, by providing installation guidelines.

\begin{figure}[h]
\centering
\includegraphics[width=.99\columnwidth]{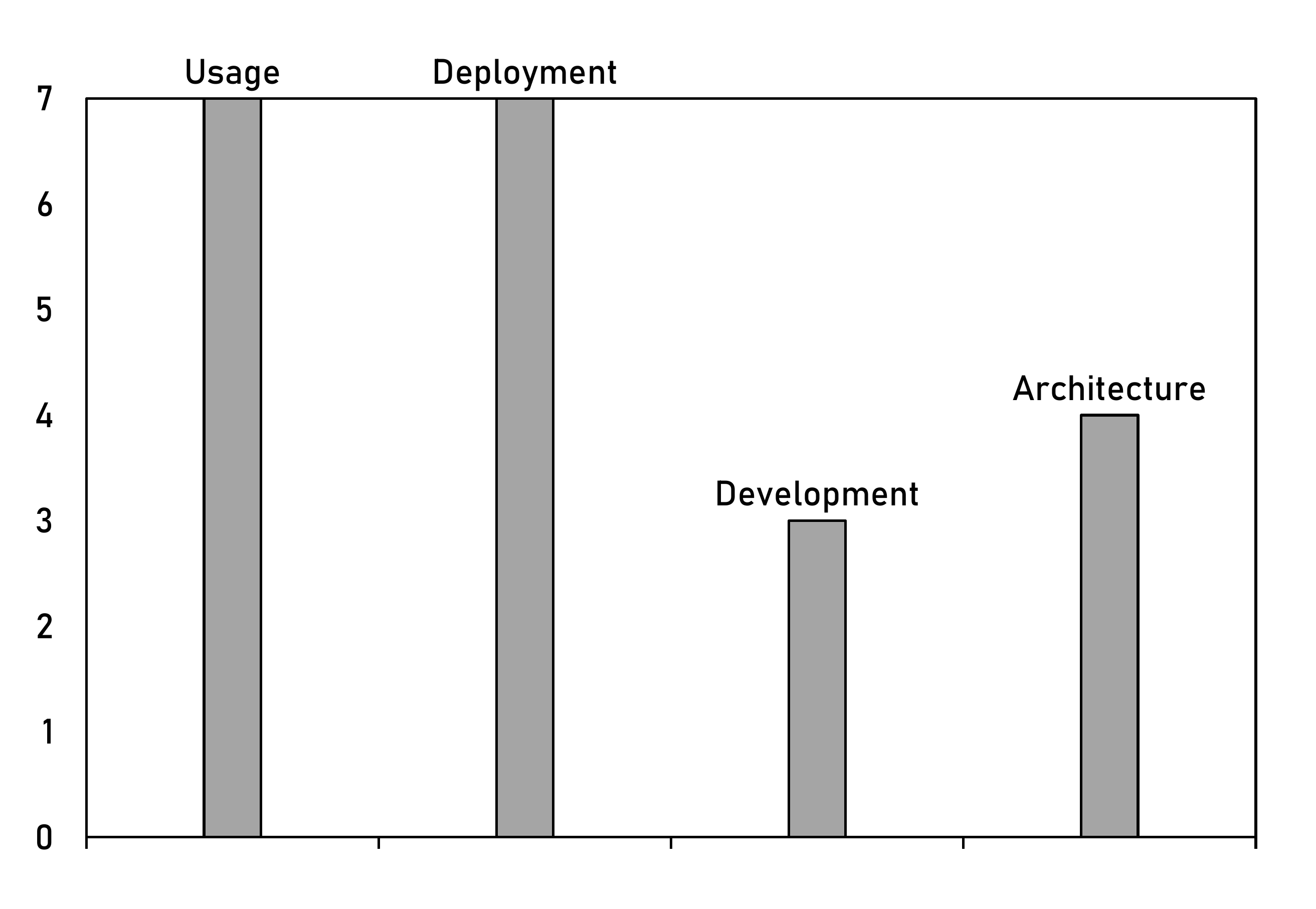}
\caption{Frequencies of values for the dimension \docPlatform of the category \documentation, among the seven open source platforms.}
\label{plot:platf-docs}
\end{figure}

Notable differences can instead be observed when looking at how considered FaaS platforms document their development and architecture.
Apart from the three commercial solutions, which obviously do not  publicly document their development or architecture, one would expect all open source solutions to document the architecture of the platform and its development.
However, the classification of open source FaaS platforms in \Cref{tab:tech-review-managerial-documentation} (for which a visual representation is displayed in \Cref{plot:platf-docs}) clearly shows that this is not the case.
Only Apache OpenWhisk, Kubeless and OpenFaaS fully document \textit{both} their architecture and development.
Also Nuclio documents its architecture, but it only provides guidelines to contribute to its development.
Finally, Fission, Fn and Knative only provide guidelines or code of conduct for contributing to the platform, without giving information on their architecture. 

\begin{center}
\noindent
\setlength{\fboxsep}{2mm}
\fbox{
    \centering
    \begin{minipage}{.9\linewidth}
      \textbf{Main Findings: Documentation}
      \begin{itemize}
          \item[\ding{212}] All platforms provide application deployment and platform usage documentation.
          \item[\ding{212}] More than a half of open source platforms (4/7) do not document their development, with 3/7 also not documenting their architecture.
      \end{itemize}
\end{minipage}}
\vspace{.2\baselineskip}
\end{center}

\paragraph{\restrictions}
The classification of considered FaaS platforms in \cref{tab:tech-review-managerial-quotas} indicates whether such platforms establish quotas related to various aspect of application's lifecycle, e.g., runtime or deployment.
Commercial solutions provide upperbounds to deployable packages' size, and they set timeouts for function executions.
In addition, AWS Lambda sets quotas for the memory and storage space available for each function, Google Cloud Functions instead sets quotas for CPU usage and memory, while MS Azure Functions sets an upperbound to the available storage space.
All such quotas vary based on the chosen service level, with each as-a-service FaaS platform setting different quotas for different service levels.

Open source, installable FaaS platforms instead mainly come without predefined quotas, with all of them (except for Apache OpenWhisk) being unbounded in all considered dimensions.
Fission, Kubeless, Nuclio and OpenFaaS however allow to set quotas on computing resources, so as to configure and manage resource consumption on target hosts.
The only open source FaaS Platform coming with pre-defined quotas is Apache OpenWhisk, which natively sets an upperbound to the deployable package and code sizes, and the memory consumption and execution time of running functions.
Notably, for all installable FaaS platforms, additional quotas can be set by exploiting the support provided by the target host (\eg if installing a FaaS platform on Kubernetes, the latter can be configured to set quotas to running deployments). 

\begin{center}
\noindent
\setlength{\fboxsep}{2mm}
\fbox{
    \centering
    \begin{minipage}{.9\linewidth}
      \textbf{Main Findings: Quotas}
      \begin{itemize}
          \item[\ding{212}] Most of the open source platforms (6/7) do not impose quotas, with 4/7 anyhow allowing to set customised quotas.
          \item[\ding{212}] Quotas in commercial platforms can be changed by switching to different subscription plans.
      \end{itemize}
\end{minipage}}
\vspace{.2\baselineskip}
\end{center}

\subsection{A Technical View on FaaS Platforms}
\label{ssec:tech-review-dev}
\noindent
As shown in~\Cref{sec:framework}, FaaS platforms can be classified using numerous categories that comprise heterogeneous sets of criteria related to their technicalities.
In the following, we elaborate on the results of reviewing the ten selected FaaS platforms (\cref{tab:platforms-documentation}) using the \devops view in our classification framework.

\begin{table*}[ht]
    \begin{threeparttable}[b]
        \centering
        \caption{Classification of considered FaaS Platforms, based on the \devruntime and \devruntimeextension dimensions of the \dev category in the \devops view of our classification framework.}
        \label{tab:tech-review-dev}
        \footnotesize
        \def\arraystretch{1.1}
        \definecolor{lightgray}{gray}{0.9}
        \begin{tabularx}{\textwidth}{@{} c c m{4.8em} m{3em} c c m{4.5em} c c m{4.5em} m{3em} C}
            \hline
            \multicolumn{2}{l}{}  
            & \centering \textit{Apache Openwhisk} 
            & \centering \textit{AWS Lambda} 
            & \textit{Fission}
            & \textit{Fn}
            & \centering \textit{Google Cloud Functions}
            & \textit{Knative}
            & \textit{Kubeless}
            & \centering \textit{MS Azure Functions}
            & \textit{Nuclio} 
            & \textit{OpenFaaS} \\
            
            \hline
            \parbox[t]{1.5mm}{\multirow{12}{*}{\rotatebox[origin=c]{90}{\devruntime}}}
            
            & \cellcolor{white} \scriptsize Ballerina 
            & \cellcolor{white} \centering \ok 
            & \cellcolor{white} \centering \ko 
            & \cellcolor{white} \centering \ko 
            & \cellcolor{white} \centering \ko 
            & \cellcolor{white} \centering \ko 
            & \cellcolor{white} \centering \ko 
            & \cellcolor{white} \centering \ok 
            & \cellcolor{white} \centering \ko 
            & \cellcolor{white} \centering \ko 
            & \cellcolor{white} \ko \\         
            
            & \cellcolor{lightgray} \scriptsize Custom Binary 
            & \cellcolor{lightgray} \centering \ok 
            & \cellcolor{lightgray} \centering \ko 
            & \cellcolor{lightgray} \centering \ok 
            & \cellcolor{lightgray} \centering \ko 
            & \cellcolor{lightgray} \centering \ko 
            & \cellcolor{lightgray} \centering \ko 
            & \cellcolor{lightgray} \centering \ko 
            & \cellcolor{lightgray} \centering \ko 
            & \cellcolor{lightgray} \centering \ok 
            & \cellcolor{lightgray} \ko \\         
            
            & \cellcolor{white} \scriptsize Docker Image
            & \cellcolor{white} \centering \ok 
            & \cellcolor{white} \centering \ko 
            & \cellcolor{white} \centering \ko 
            & \cellcolor{white} \centering \ok 
            & \cellcolor{white} \centering \ko 
            & \cellcolor{white} \centering \ok 
            & \cellcolor{white} \centering \ok 
            & \cellcolor{white} \centering \ok 
            & \cellcolor{white} \centering \ok 
            & \cellcolor{white} \ok  \\        
            
            & \cellcolor{lightgray} \scriptsize Go
            & \cellcolor{lightgray} \centering \ok 
            & \cellcolor{lightgray} \centering \ok 
            & \cellcolor{lightgray} \centering \ok 
            & \cellcolor{lightgray} \centering \ok 
            & \cellcolor{lightgray} \centering \ok 
            & \cellcolor{lightgray} \centering \ko 
            & \cellcolor{lightgray} \centering \ok 
            & \cellcolor{lightgray} \centering \ko 
            & \cellcolor{lightgray} \centering \ok 
            & \cellcolor{lightgray} \ok \\         
            
            & \cellcolor{white} \scriptsize Java
            & \cellcolor{white} \centering \ok 
            & \cellcolor{white} \centering \ok 
            & \cellcolor{white} \centering \ok 
            & \cellcolor{white} \centering \ok 
            & \cellcolor{white} \centering \ko 
            & \cellcolor{white} \centering \ko 
            & \cellcolor{white} \centering \ok 
            & \cellcolor{white} \centering \ok 
            & \cellcolor{white} \centering \ok 
            & \cellcolor{white} \ok  \\        
            
            & \cellcolor{lightgray}\scriptsize MS .NET
            & \cellcolor{lightgray} \centering \ok 
            & \cellcolor{lightgray} \centering \ok 
            & \cellcolor{lightgray} \centering \ok 
            & \cellcolor{lightgray} \centering \ok 
            & \cellcolor{lightgray} \centering \ko 
            & \cellcolor{lightgray} \centering \ko 
            & \cellcolor{lightgray} \centering \ok 
            & \cellcolor{lightgray} \centering \ok 
            & \cellcolor{lightgray} \centering \ok 
            & \cellcolor{lightgray} \ok \\         
            
            & \cellcolor{white} \scriptsize NodeJS
            & \cellcolor{white} \centering \ok 
            & \cellcolor{white} \centering \ok 
            & \cellcolor{white} \centering \ok 
            & \cellcolor{white} \centering \ok 
            & \cellcolor{white} \centering \ok 
            & \cellcolor{white} \centering \ko 
            & \cellcolor{white} \centering \ok 
            & \cellcolor{white} \centering \ok 
            & \cellcolor{white} \centering \ok 
            & \cellcolor{white} \ok  \\        
            
            & \cellcolor{lightgray} \scriptsize Perl
            & \cellcolor{lightgray} \centering \ko 
            & \cellcolor{lightgray} \centering \ko 
            & \cellcolor{lightgray} \centering \ok 
            & \cellcolor{lightgray} \centering \ko 
            & \cellcolor{lightgray} \centering \ko 
            & \cellcolor{lightgray} \centering \ko 
            & \cellcolor{lightgray} \centering \ko 
            & \cellcolor{lightgray} \centering \ko 
            & \cellcolor{lightgray} \centering \ko 
            & \cellcolor{lightgray} \ko \\         

            & \cellcolor{white} \scriptsize PHP
            & \cellcolor{white} \centering \ok 
            & \cellcolor{white} \centering \ko 
            & \cellcolor{white} \centering \ok 
            & \cellcolor{white} \centering \ko 
            & \cellcolor{white} \centering \ko 
            & \cellcolor{white} \centering \ko 
            & \cellcolor{white} \centering \ok 
            & \cellcolor{white} \centering \ok 
            & \cellcolor{white} \centering \ko 
            & \cellcolor{white} \ok  \\        
            
            & \cellcolor{lightgray} \scriptsize Python
            & \cellcolor{lightgray} \centering \ok 
            & \cellcolor{lightgray} \centering \ok 
            & \cellcolor{lightgray} \centering \ok 
            & \cellcolor{lightgray} \centering \ok 
            & \cellcolor{lightgray} \centering \ok 
            & \cellcolor{lightgray} \centering \ko 
            & \cellcolor{lightgray} \centering \ok 
            & \cellcolor{lightgray} \centering \ok 
            & \cellcolor{lightgray} \centering \ok 
            & \cellcolor{lightgray} \ok \\         
            
            & \cellcolor{white} \scriptsize Ruby
            & \cellcolor{white} \centering \ok 
            & \cellcolor{white} \centering \ok 
            & \cellcolor{white} \centering \ok 
            & \cellcolor{white} \centering \ok 
            & \cellcolor{white} \centering \ko 
            & \cellcolor{white} \centering \ko 
            & \cellcolor{white} \centering \ok 
            & \cellcolor{white} \centering \ko 
            & \cellcolor{white} \centering \ko 
            & \cellcolor{white} \ok  \\        
            
            & \cellcolor{lightgray} \scriptsize Rust
            & \cellcolor{lightgray} \centering \ok 
            & \cellcolor{lightgray} \centering \ko 
            & \cellcolor{lightgray} \centering \ko 
            & \cellcolor{lightgray} \centering \ko 
            & \cellcolor{lightgray} \centering \ko 
            & \cellcolor{lightgray} \centering \ko 
            & \cellcolor{lightgray} \centering \ko 
            & \cellcolor{lightgray} \centering \ko 
            & \cellcolor{lightgray} \centering \ko 
            & \cellcolor{lightgray} \ko \\         
            
            & \cellcolor{white} \scriptsize Shell
            & \cellcolor{white} \centering \ok 
            & \cellcolor{white} \centering \ok 
            & \cellcolor{white} \centering \ok 
            & \cellcolor{white} \centering \ko 
            & \cellcolor{white} \centering \ko 
            & \cellcolor{white} \centering \ko 
            & \cellcolor{white} \centering \ko 
            & \cellcolor{white} \centering \ok 
            & \cellcolor{white} \centering \ok 
            & \cellcolor{white} \ko \\        
            
            & \cellcolor{lightgray} \scriptsize Swift
            & \cellcolor{lightgray} \centering \ok 
            & \cellcolor{lightgray} \centering \ko 
            & \cellcolor{lightgray} \centering \ko 
            & \cellcolor{lightgray} \centering \ko 
            & \cellcolor{lightgray} \centering \ko 
            & \cellcolor{lightgray} \centering \ko 
            & \cellcolor{lightgray} \centering \ko 
            & \cellcolor{lightgray} \centering \ko 
            & \cellcolor{lightgray} \centering \ko 
            & \cellcolor{lightgray} \ko \\         
            
            \hline
            
            \multicolumn{2}{m{7em}}{\centering \textit{Runtime Customization}}
            & \centering \ok
            & \centering \ok
            & \centering \ok
            & \centering \ok 
            & \centering \ko 
            & \centering \ok  
            & \centering \ok 
            & \centering \ok 
            & \centering \ok
            & \ok \\
            \hline
            
        \end{tabularx}
    \end{threeparttable}
\end{table*}

\begin{table*}[t]
		 \caption{Classification of considered FaaS Platforms, based on the \dev category in the \devops view of our classification framework. The abbreviation \enquote{\NS{}} stays for \enquote{not specified}, meaning that no related information is in the documentation.}    
    \label{tab:tech-review-dev-idesdk}
    \begin{threeparttable}[b]
        \centering
        \footnotesize
        \begin{tabularx}{.99\textwidth}{m{14em} C m{17em}}
            \hline
            & \textit{\devIDE} 
            & \centering\arraybackslash \textit{\devSDK} \\
            \hline
            \rowcolor[HTML]{EFEFEF}
            \textit{Apache Openwhisk}                   
            & Visual Studio Code\tnote{*}, Xcode\tnote{*} 
            & \centering\arraybackslash Go, NodeJS, Python, Swift \\
            \textit{AWS Lambda}
            & AWS Cloud9, Eclipse, IntelliJ, PyCharm, Visual~Studio,~Visual~Studio~Code, Visual~Studio~Team~Services
            & \centering\arraybackslash Go, Java, MS .NET, NodeJS,~Python,~Ruby \\
            \rowcolor[HTML]{EFEFEF}
            \textit{Fission}
            & \NS
            & \centering\arraybackslash \NS \\
            \textit{Fn}
            & \NS
            & \centering\arraybackslash Go \\
            \rowcolor[HTML]{EFEFEF}
            \textit{Google Cloud Functions}
            & \NS
            & \centering\arraybackslash Dart, Go, Java, MS .NET, NodeJS,~Python,~Ruby \\
            \textit{Knative}
            & \NS
            & \centering\arraybackslash \NS \\
            \rowcolor[HTML]{EFEFEF}
            \textit{Kubeless}
            & Visual Studio Code
            & \centering\arraybackslash \NS \\
            \textit{MS Azure Functions}
            & Visual Studio, Visual Studio Code
            & \centering\arraybackslash Java, MS .NET \\
            \rowcolor[HTML]{EFEFEF}
            \textit{Nuclio}
            & Jupyter Notebooks
            & \centering\arraybackslash Go, Java, MS .NET, Python \\
            \textit{OpenFaaS}
            & \NS
            & \centering\arraybackslash \NS \\
            \hline
        \end{tabularx}
        \begin{tablenotes}
            \scriptsize
            \item[*] \textit{Deprecated / No longer maintained}
        \end{tablenotes}
    \end{threeparttable}
\end{table*}

\paragraph{\dev}
One of the main technical aspects related to development is whether a platform supports a required function runtime, which eventually allows developing functions in a chosen programming language such as NodeJS or Java.
\Cref{tab:tech-review-dev} shows the details of function runtime support for the list of reviewed platforms sorted alphabetically, and \Cref{plot:function-runtimes} displays the frequencies of supported runtimes, \ie it associates each programming language with the amount of FaaS platforms supporting the corresponding runtime.

\begin{figure}[h]
\centering
\includegraphics[width=.99\columnwidth]{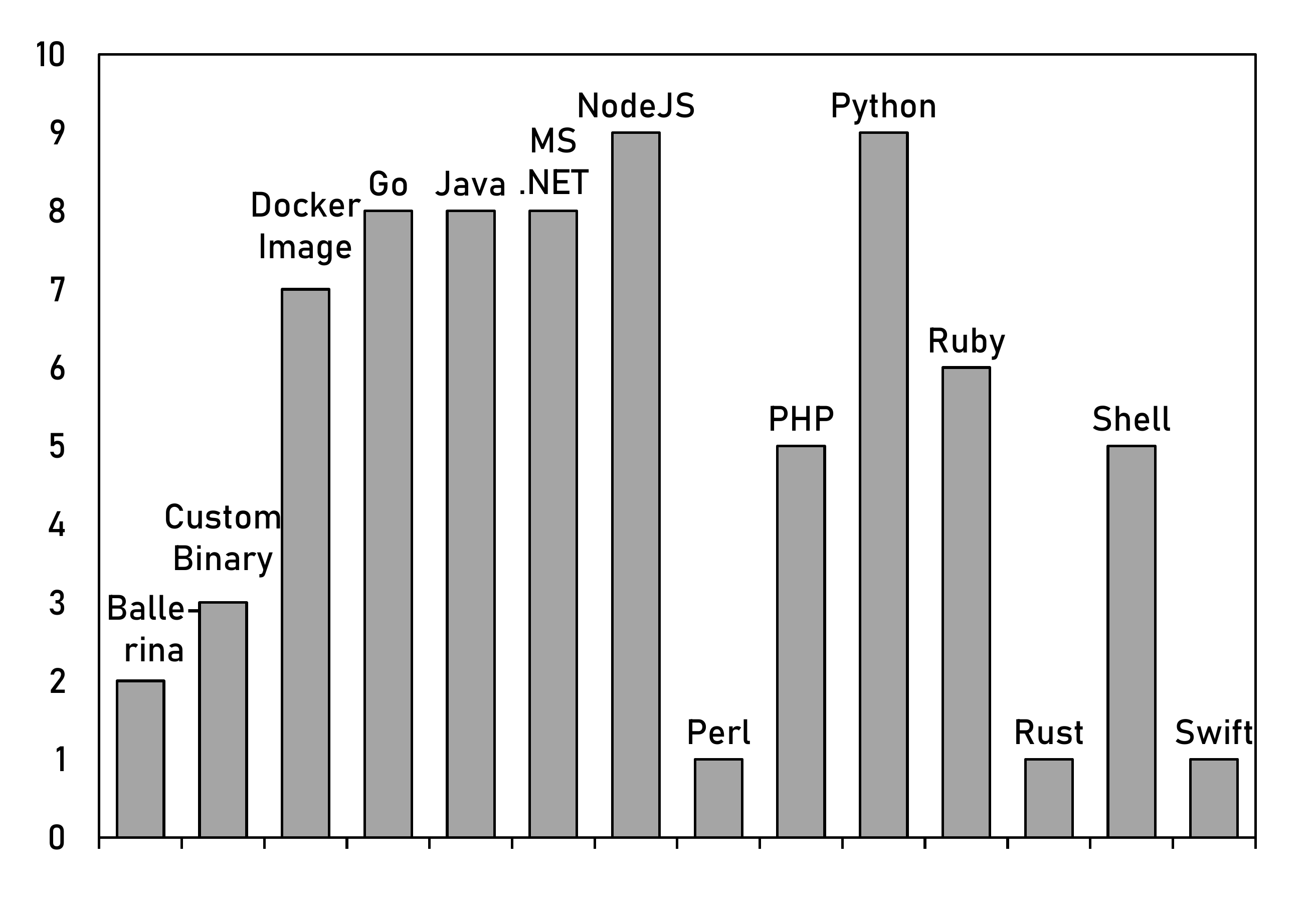}
\caption{Frequencies of values for the dimension \devruntime of the category \dev, among the ten classified FaaS~platforms.}
\label{plot:function-runtimes}
\end{figure}

The most popular languages are \textit{NodeJS} and \textit{Python} support for which is described by 9/10 platforms.
Both NodeJS and Python are interpretable languages with a huge community and ecosystem of libraries, making these languages a perfect choice for fast-paced FaaS-based development.
The next place is shared by \textit{Go}, \textit{Java}, and \textit{.NET} support for which is stated by eight out of ten reviewed platforms.
As shown previously, Go has become an inherent part of the cloud-native development world.
Go is a compiled and fast language, which is also used to implement multiple FaaS platforms, which is an additional reason why Go is supported by most of them.
Moreover, both Java and .NET are mature, well-established languages with a large ecosystem of general-purpose libraries making them a good additions to the main list of supported languages.
The third and fourth place among supported function runtimes are then occupied by \textit{Docker Images} and \textit{Ruby}, respectively.
While Ruby is another good example of a popular programming language for web development, a large support for Docker Images worth a special highlight.

Essentially, supporting Docker Images as a function runtime allows developing and running functions in any programming language assuming that all required dependencies can be included together with the function's logic and the resulting container is compatible with the FaaS platform, i.e., a platform is able to invoke the function and pass the event data.
However, implementing function as container images requires more effort, since the underlying container has typically to implement a specific interface allowing the platform to interact with the function within the container.
Docker image can also encapsulate custom binaries, support for which is stated separately by several reviewed platforms, e.g., Nuclio and Kubeless.
The main difference is in which kind of artifacts are used as an input for deploying the function, i.e., a Dockerfile or the binary with a specification of its invocation details.

Additionally, supporting Docker images is one of the possible way to enable the \textit{Runtime Customization} since a function implemented in a non-supported programming language can be invoked by the platform.
However, there are other ways to customize function runtimes.
For example, in some platforms the runtime support is implemented via dedicated container images, e.g., for running Java8 applications.
In such cases, extending the runtime might also mean building a modified runtime container image on top of an existing image, e.g., to add a particular library dependency.
Most of the reviewed platforms provide a description of runtime customization options. 

\begin{figure}[t]
\centering
\includegraphics[width=.99\columnwidth]{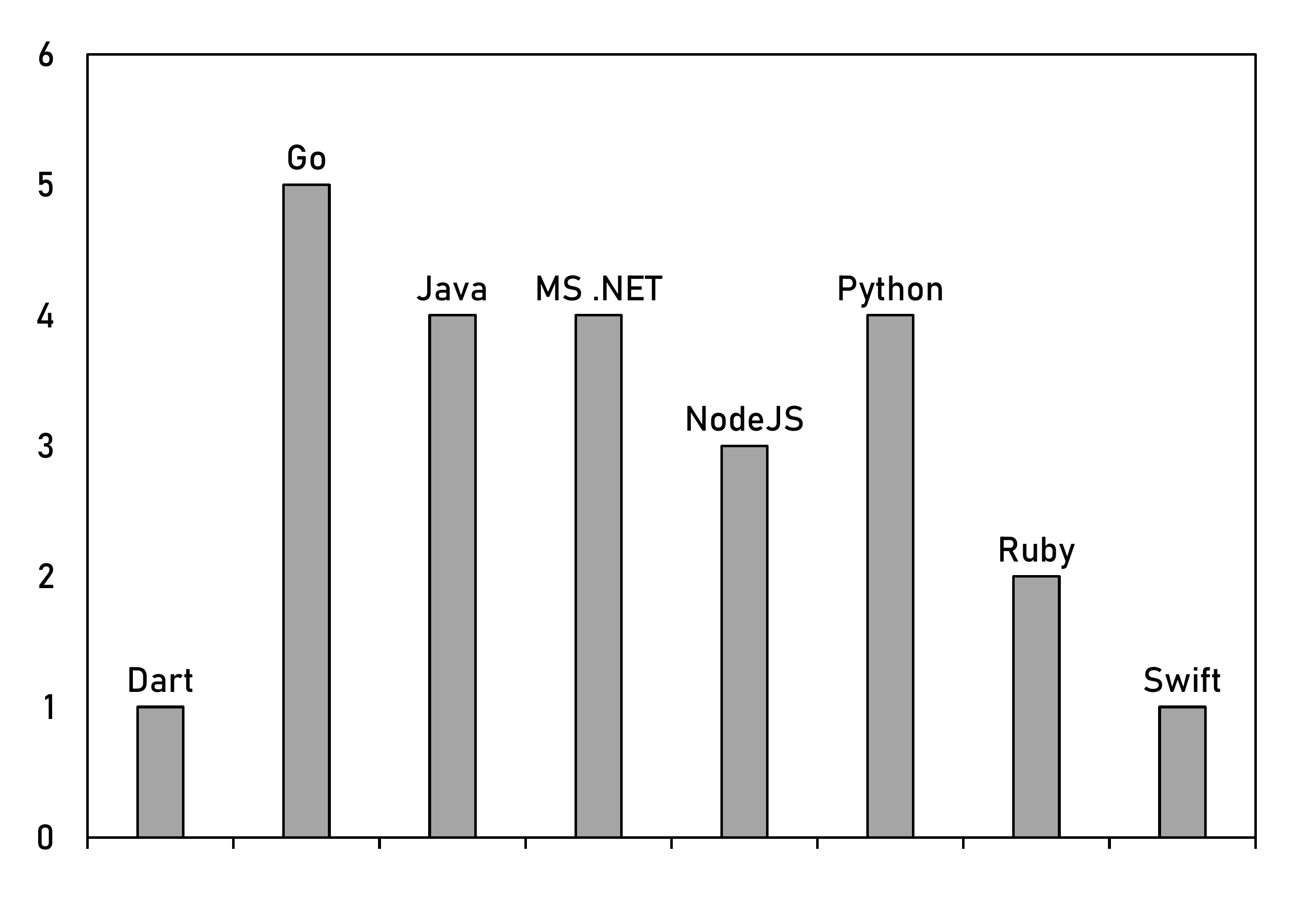}
\caption{Frequencies of values for the dimension \devSDK of the category \dev, among the six FaaS platforms providing some language-specific client library.}
\label{plot:function-runtimes}
\end{figure}

The classification of considered FaaS platforms based on the \dev category is completed by checking whether each FaaS platform provides plugins for IDEs and rich text editors, as well as language-specific client libraries for programmatic access to the API of the platform.
\Cref{tab:tech-review-dev-idesdk} shows the details about supported \devIDE (\eg in the form of plugins) and \devSDK.
One can readily observe that the presence of platform-specific IDEs or of plugins for existing IDEs is rarely the case, especially for open source FaaS platforms.

Most of the considered platforms (6/10) instead provide client libraries for existing programming languages, which wrap the platforms' APIs to facilitate the development of serverless applications (\eg by easing the use of platform-specific libraries to work with typed data or events). 
\Cref{plot:function-runtimes} displays the frequencies of programming languages for which a client library is provided, \ie it associates a programming language with the amount of FaaS platforms providing a client library for it (among the six platforms that are known to provide a client library for a programming language - \Cref{tab:tech-review-dev-idesdk}). 
The figure again highlights the importance of Go for serverless and cloud-native application development, and it confirm that Go, Java, MS .NET, NodeJS and Python are the most programming languages most supported by reviewed FaaS platforms.

\begin{center}
\noindent
\setlength{\fboxsep}{2mm}
\fbox{
    \centering
    \begin{minipage}{.9\linewidth}
      \textbf{Main Findings: Development}
      \begin{itemize}
          \item[\ding{212}] No programming language is supported by \textit{all} reviewed FaaS platforms.
          \item[\ding{212}] Go, Java, MS .NET, NodeJS and Python are the most supported programming languages. Docker images are also popular and help using custom programming languages.
          \item[\ding{212}] IDEs and text editor plugins are mainly offered by commercial platforms.
          \item[\ding{212}] Most platforms offer client libraries, with open source platforms supporting less languages.
      \end{itemize}
\end{minipage}}
\vspace{.2\baselineskip}
\end{center}

\paragraph{\vers}
The results for versioning support with respect to single functions and application that represent a combination of multiple functions and resources are shown in~\Cref{tab:tech-review-dev-versioning}.

\begin{table}[h]
    \begin{threeparttable}[b]
        \centering
        \caption{Classification of considered FaaS Platforms, based on the \vers category in the \devops view of our classification framework. \textsf{D} and \textsf{I} are used to denote the possible values \textit{dedicated mechanisms} and \textit{implicit versioning}, respectively. The abbreviation \enquote{\NS{}} stays for \enquote{not specified}, meaning that a platform does not explicitly mention the versioning of serverless applications.}
        \label{tab:tech-review-dev-versioning}
        \footnotesize
        \def\arraystretch{1.1}
        \begin{tabularx}{.99\columnwidth}{m{10em} C m{5em}}
            \hline
            & \textit{\versapps} & \centering\arraybackslash \textit{\versfuncs} \\
            \hline
            \rowcolor[HTML]{EFEFEF}
            \textit{Apache Openwhisk}                   
            & \textsf{I}
            & \centering\arraybackslash \textsf{I} \\
            \textit{AWS Lambda}
            & \textsf{D}
            & \centering\arraybackslash \textsf{D} \\
            \rowcolor[HTML]{EFEFEF}
            \textit{Fission}
            & \textsf{I}
            & \centering\arraybackslash \textsf{I} \\
            \textit{Fn}
            & \textsf{I}
            & \centering\arraybackslash \textsf{I} \\
            \rowcolor[HTML]{EFEFEF}
            \textit{Google Cloud \newline Functions}
            & \textsf{I}
            & \centering\arraybackslash \textsf{I} \\
            \textit{Knative}
            & \NS
            & \centering\arraybackslash \textsf{D} \\
            \rowcolor[HTML]{EFEFEF}
            \textit{Kubeless}
            & \centering \NS
            & \centering\arraybackslash \textsf{I} \\
            \textit{MS Azure \newline Functions}
            & \textsf{I}
            & \centering\arraybackslash \textsf{I} \\
            \rowcolor[HTML]{EFEFEF}
            \textit{Nuclio}
            & \NS
            & \centering\arraybackslash \textsf{D} \\
            \textit{OpenFaaS}
            & \NS
            & \centering\arraybackslash \textsf{I} \\
            \hline
        \end{tabularx}
    \end{threeparttable}
\end{table}

One interesting fact is that some reviewed platforms tend to omit the notion of a serverless application and rather focus on deploying single functions.
For example, Knative does not differentiate between functions and serverless applications, since a Service in Knative represents a container. 
While in theory a service could also contain multiple functions, the documentation does not provide any details on such deployment strategy and how internal functions can be distinguished.
As a result, application versioning in such platforms is not really possible even implicitly, e.g., by establishing certain naming conventions for applications.
Secondly, most platforms do not offer dedicated versioning mechanisms, which makes only implicit versioning possible.
In some cases, there are rudimentary versioning capabilities, which, however, cannot be considered dedicated versioning mechanisms.
For example, in Apache Openwhisk there is an internal version property in metadata, which is assigned automatically after the deployment but cannot be managed.
The only documentation describing both function- and application-level versioning is AWS Lambda, where applications can be defined and managed using AWS SAM templates.

\begin{table*}[!p]
					\centering
    			\rotatebox{90}{
        \begin{threeparttable}[b]
        \centering
        \caption{Classification of considered FaaS Platforms, based on the \eventsources category in the \devops view of our classification framework. \NS stays for \enquote{not specified}, meaning that no related information is provided in the documentation.}
        \label{tab:tech-review-eventsources}
        \scriptsize
        \def\arraystretch{1.2}
        \definecolor{lightgray}{gray}{0.9}
        \begin{tabular}
            {@{} 
                c 
                m{15mm} @{}
                m{20mm} 
                m{15mm} 
                m{15mm} 
                m{08mm} 
                m{15mm} 
                m{20mm} 
                m{15mm} 
                m{15mm} 
                m{15mm} 
                m{15mm} 
            }
            \hline
            \multicolumn{2}{l}{}
            & \centering \textit{Apache Openwhisk} 
            & \centering \textit{AWS Lambda} 
            & \centering \textit{Fission}
            & \centering \textit{Fn} 
            & \centering \textit{Google Cloud Functions} 
            & \centering \textit{Knative} 
            & \centering \textit{Kubeless} 
            & \centering \textit{MS Azure Functions} 
            & \centering \textit{Nuclio} 
            & \centering\arraybackslash \textit{OpenFaaS} \\
            \hline
            
            \parbox[t]{1.5mm}{\multirow{4}{*}[-1.5em]{\rotatebox[origin=c]{90}{\eventendpoint}}}
            
            & \centering Synchronous Call
            & \centering HTTP 
            & \centering HTTP 
            & \centering HTTP 
            & \centering HTTP 
            & \centering HTTP, RPC\tnote{2} 
            & \centering HTTP 
            & \centering HTTP 
            & \centering HTTP 
            & \centering HTTP 
            & \centering\arraybackslash HTTP \\         
            
            & \cellcolor{lightgray} \centering Asynchronous \newline Call
            & \cellcolor{lightgray} \centering HTTP 
            & \cellcolor{lightgray} \centering HTTP 
            & \cellcolor{lightgray} \centering \NS 
            & \cellcolor{lightgray} \centering \NS 
            & \cellcolor{lightgray} \centering \NS 
            & \cellcolor{lightgray} \centering \NS 
            & \cellcolor{lightgray} \centering \NS 
            & \cellcolor{lightgray} \centering \NS 
            & \cellcolor{lightgray} \centering \NS 
            & \cellcolor{lightgray} \centering\arraybackslash HTTP \\         
            
            & \cellcolor{white} \scriptsize \centering Endpoint Customization
            & \cellcolor{white} \centering \ok 
            & \cellcolor{white} \centering \ok 
            & \cellcolor{white} \centering \ok 
            & \cellcolor{white} \centering \ok 
            & \cellcolor{white} \centering \ok 
            & \cellcolor{white} \centering \ok 
            & \cellcolor{white} \centering \ok 
            & \cellcolor{white} \centering \ok 
            & \cellcolor{white} \centering \ok 
            & \cellcolor{white} \centering\arraybackslash \NS  \\        
            
            & \cellcolor{lightgray} \scriptsize \centering TLS Support        
            & \cellcolor{lightgray} \centering \ok 
            & \cellcolor{lightgray} \centering \ok 
            & \cellcolor{lightgray} \centering \NS 
            & \cellcolor{lightgray} \centering \NS 
            & \cellcolor{lightgray} \centering \ok 
            & \cellcolor{lightgray} \centering \ok 
            & \cellcolor{lightgray} \centering \ok 
            & \cellcolor{lightgray} \centering \ok 
            & \cellcolor{lightgray} \centering \NS 
            & \cellcolor{lightgray} \centering\arraybackslash \ok \\         
            
            \hline
            
            \parbox[t]{2mm}{\multirow{2}{*}{\rotatebox[origin=c]{90}{\begin{minipage}{6em}\eventdatastore\end{minipage}}}}
            
            & \centering File Level
            & \centering \NS 
            & \centering AWS S3 
            & \centering \NS 
            & \centering \NS 
            & \centering Google Cloud Storage 
            & \centering Google Cloud Storage 
            & \centering \NS 
            & \centering Azure Blob Storage   
            & \centering \NS 
            & \centering\arraybackslash Min.io \\         
            
            & \cellcolor{lightgray} \centering Database \newline Mode
            & \cellcolor{lightgray} \centering IBM Cloudant    
            & \cellcolor{lightgray} \centering Amazon DynamoDB 
            & \cellcolor{lightgray} \centering \NS             
            & \cellcolor{lightgray} \centering \NS             
            & \cellcolor{lightgray} \centering Cloud Firestore\tnote{3}, Firebase Realtime Database\tnote{3} 
            & \cellcolor{lightgray} \centering Apache CouchDB, AWS DynamoDB 
            & \cellcolor{lightgray} \centering \NS 
            & \cellcolor{lightgray} \centering Azure Cosmos DB 
            & \cellcolor{lightgray} \centering \NS 
            & \cellcolor{lightgray} \centering\arraybackslash \NS \\         
            \hline
            
            \multicolumn{2}{c}{\centering \textit{\eventscheduler}}
            & \centering \ok 
            & \centering \ok 
            & \centering \ok 
            & \centering \NS 
            & \centering \ok 
            & \centering \ok 
            & \centering \ok 
            & \centering \ok 
            & \centering \ok 
            & \centering\arraybackslash \ok \\         
            \hline
            
            \multicolumn{2}{c}{\cellcolor{lightgray}\centering \textit{\eventmq}}
            & \cellcolor{lightgray} \centering \NS                                   
            & \cellcolor{lightgray} \centering AWS SQS, AWS SNS                      
            & \cellcolor{lightgray} \centering NATS, Azure Storage Queue             
            & \cellcolor{lightgray} \centering \NS                                   
            & \cellcolor{lightgray} \centering Google Cloud Pub/Sub                  
            & \cellcolor{lightgray} \centering AWS SQS, AWS SNS, Google Cloud PubSub 
            & \cellcolor{lightgray} \centering \NS                 
            & \cellcolor{lightgray} \centering Azure Queue Storage, Azure Service Bus 
            & \cellcolor{lightgray} \centering RabbitMQ, MQTT, NATS        
            & \cellcolor{lightgray} \centering\arraybackslash AWS SQS, AWS SNS \\                         
            \hline
            
            \multicolumn{2}{m{20mm}}{\centering \textit{\eventstreaming}}
            & \centering Apache Kafka, IBM Message Hub 
            & \centering Amazon Kinesis                
            & \centering Apache Kafka                  
            & \centering \NS                           
            & \centering Google Cloud Pub/Sub                              
            & \centering Apache Kafka, AWS Kinesis, Google Cloud Pub/Sub   
            & \centering AWS Kinesis, NATS, Apache Kafka                   
            & \centering Azure Event Hubs                                 
            & \centering Apache Kafka, AWS Kinesis, Iguazio Data Science Platform, Azure Event Hubs                             
            & \centering\arraybackslash Apache Kafka \\                         
            \hline
            
            \multicolumn{2}{m{20mm}}{\cellcolor{lightgray} \centering \textit{\eventspecservice}}
            & \cellcolor{lightgray} \centering GitHub, Slack\tnote{1}, Weather Company Data\tnote{1}, IBM Push Notifications\tnote{1}, Websocket API\tnote{1}  
            & \cellcolor{lightgray} \centering Amazon Cognito, Lex, Alexa, Kinesis Data Firehose 
            & \cellcolor{lightgray} \centering Kubernetes Watch          
            & \cellcolor{lightgray} \centering \NS                       
            & \cellcolor{lightgray} \centering Google Cloud Operations   
            & \cellcolor{lightgray} \centering Apache Camel, Kubernetes event, GitHub, BitBucket, GitLab, Google Cloud Scheduler, AWS CodeCommit, AWS Cognito, FTP/SFTP, Heartbeat events, Websocket                 
            & \cellcolor{lightgray} \centering \NS                               
            & \cellcolor{lightgray} \centering Microsoft Graph, Event Grid       
            & \cellcolor{lightgray} \centering \NS                 
            & \cellcolor{lightgray} \centering\arraybackslash IFTTT, VMware vCenter \\                         
            \hline
            
            \multicolumn{2}{m{20mm}}{\centering \textit{\eventintegration}}
            & \centering hooks, polling, connection patterns 
            & \centering \NS 
            & \centering plugins 
            & \centering \NS  
            & \centering \NS  
            & \centering plugins  
            & \centering plugins  
            & \centering Custom I/O bindings  
            & \centering \NS  
            & \centering\arraybackslash connector plugins \\          
            \hline            
        \end{tabular}
        \begin{tablenotes}
            \scriptsize
            \item[1] \textit{Only one-directional integration}
            \item[2] \textit{To be used for testing purposes}
            \item[3] \textit{Feature is in a pre-release state}
        \end{tablenotes}
    \end{threeparttable}
    }
\end{table*}

\vspace{1\baselineskip}
\begin{center}
\noindent
\setlength{\fboxsep}{2mm}
\fbox{
    \centering
    \begin{minipage}{.9\linewidth}
      \textbf{Main Findings: Versioning}
      \begin{itemize}
          \item[\ding{212}] Most open source platforms (6/7) employ implicit versioning of functions, while most commercial platforms (2/3) provide dedicated mechanisms for versioning functions.
          \item[\ding{212}] The versioning of serverless applications is not explicitly mentioned by 40\% of the reviewed FaaS platforms.
      \end{itemize}
\end{minipage}}
\vspace{.2\baselineskip}
\end{center}



\paragraph{\eventsources}
The detailed results of the review for this category are shown in~\Cref{tab:tech-review-eventsources}.
There are several observations related to the \textit{Endpoint} event sources category that can be highlighted.
Firstly, all platforms provide capabilities to synchronously invoke functions exposed as endpoints using HTTP protocol, whereas the majority~(7/10) of reviewed platforms do not document support for asynchronous invocation of endpoints.
In addition, almost all platforms~(9/10) provide mechanisms to customize endpoint names and the majority of platforms allow securing the endpoint communication using HTTPS.

Scheduled function invocation is also supported by the majority of platforms~(9/10), as described by the \textit{Scheduler} category.
The scheduling is typically time-based, and it is implemented by exploiting cron jobs. 
The actual definition of time-based scheduling however slightly varies among considered FaaS platforms, mainly in terms of the required formatting.

Different considerations instead apply to triggers related to \textit{Data Store}s and \textit{Message Queue}s.
Most open source FaaS platforms (4/7) indeed do not document any support for event sources in the \textit{Data Store} category, neither for file level data stores nor for relational and NoSQL databases.
There are some exceptions: Knative's documentation lists commercial and open source data stores as supported, Apache Openwhisk states support for IBM's NoSQL database called Cloudant, and OpenFaaS described how Min.io\footnote{Min.io (\url{https://min.io}) is an object storage solution, which is positioned as an open source alternative to AWS S3.}. can be integrated.
The main reason for this is that open source FaaS platforms come as distinct and standalone products, which typically do not come with a rich set of natively integrated services at a first place, unlike commercial FaaS platforms.
Not surprisingly, all commercial FaaS platforms provide built-in support for multiple data stores originating from the same provider, e.g., AWS Lambda can be seamlessly integrated with the object storage service offering called AWS S3, and Amazon's NoSQL database called DynamoDB.

In the \textit{Message Queue} category, similar as with data stores support, commercial FaaS platforms focus on provider-specific message queues such as AWS SQS and AWS SNS for AWS Lambda or Azure Queue Storage for Azure Functions.
As can be seen in~\Cref{tab:tech-review-eventsources}, the support for message queues is stated by the larger part~(4/7) of the reviewed FaaS platforms.
This list includes both open source messaging solutions such as RabbitMQ, and commercial solutions such as Google Cloud Pub/Sub~\cite{google:pubsub}.
Likewise, for the \textit{Stream Processing Platform} category, commercial FaaS platforms provide support for provider-specific services such as Google Cloud Pub/Sub for Google Cloud Functions or Azure Event Hubs for Azure Functions.
In case of open source platforms, Apache Kafka is the most supported~(6/7) stream processing platform, whereas in some cases support for commercial stream processing offerings is provided as well.

For the \textit{Special-purpose Service} category, there is no clear pattern on the choice of supported services.
On the one hand, commercial providers focus on integrating their FaaS offerings with provider-specific services such as AWS Cognito or Microsoft Graph.
On the other hand, open source platforms support a variety of external services, \eg GitHub~\cite{github}, Slack~\cite{slack}, or IFTTT~\cite{ifttt}.
To a large extent, the availability of certain services is driven by the needs of open source contributors, making platforms very heterogeneous under this dimension. 

Finally, as shown in the \textit{Event Source Integration} category the reviewed platforms mention multiple possible ways to integrate custom event sources, e.g., plugin development, webhooks, or polling.
Essentially, one of the most common ways to integrate custom event sources that is not always listed explicitly is by means of messaging.
However, since out-of-the-box integration typically requires less effort, it is more beneficial to have an explicitly-supported integration for a desired event source.

\begin{center}
\noindent
\setlength{\fboxsep}{2mm}
\fbox{
    \centering
    \begin{minipage}{.9\linewidth}
      \textbf{Main Findings: Event Sources}
      \begin{itemize}
          \item[\ding{212}] All platforms support synchronous, HTTP-based function invocation, whereas asynchronous calls are supported by 3/10 platforms.
          \item[\ding{212}] More than a half of open source platforms~(4/7) do not support data store event sources.
          \item[\ding{212}] Schedulers and stream processing platforms are supported by most platforms~(9/10). Messaging is also widely supported~(7/10).
          \item[\ding{212}] Support for special-purpose services varies significantly from platform to platform.
          \item[\ding{212}] More than a half of reviewed platforms~(6/10) support the integration of custom event sources.
      \end{itemize}
\end{minipage}}
\vspace{.2\baselineskip}
\end{center}
\begin{table*}[t]
    \caption{Classification of considered FaaS Platforms, based on the \orchestration category in the \devops view of our classification framework. \textsf{C} denotes \textit{custom DSL}, and \textsf{O} denotes \textit{orchestrating function}, with the list of supported programming languages for developing orchestrating functions given in square braces. The abbreviations \enquote{\NS{}} and \enquote{\NA{}} stay for \enquote{not specified} and \enquote{not applicable}, respectively.}
    \label{tab:tech-review-dev-orchestrators}
    \begin{threeparttable}
        \centering
        \footnotesize
        \begin{tabularx}{.99\textwidth}{m{13em} C m{12em} m{7em} m{10em}}
            \hline
            & \centering Orchestrator 
            & \centering Workflow Definition 
            & \centering Control Flow Described 
            & \centering\arraybackslash Quotas\\
            \hline
            \rowcolor[HTML]{EFEFEF}
            \textit{Apache Openwhisk}
            & \centering Openwhisk Composer
            & \centering \textsf{O [JavaScript]} 
            & \centering \ok 
            & \centering\arraybackslash execution time \\ 
            \textit{AWS Lambda}
            & \centering AWS Step Functions
            & \centering \textsf{C} 
            & \centering \ok 
            & \centering\arraybackslash execution time, I/O~size\\
            \rowcolor[HTML]{EFEFEF}
            \textit{Fission}
            & \centering Fission Workflows
            & \centering \textsf{C}
            & \centering \ok 
            & \centering\arraybackslash \NS \\
            \textit{Fn}
            & \centering Fn Flow
            & \centering \textsf{O [Java]}
            & \centering \ok 
            & \centering\arraybackslash \NS \\
            \rowcolor[HTML]{EFEFEF}            
            \textit{Google Cloud Functions}
            & \centering \centering \NS
            & \centering \NA
            & \centering \NA 
            & \centering\arraybackslash \NA \\
            \textit{Knative}
            & \centering Knative Eventing
            & \centering \textsf{C}
            & \centering \ok\tnote{*}
            & \centering\arraybackslash \NS \\
            \rowcolor[HTML]{EFEFEF}
            \textit{Kubeless}
            & \centering \NS
            & \centering \NA 
            & \centering \NA 
            & \centering\arraybackslash \NA \\
            \textit{MS Azure Functions}
            & \centering Azure Durable Functions
            & \centering \textsf{O [C\#, JavaScript]}
            & \centering \ok
            & \centering\arraybackslash \NS \\
            \rowcolor[HTML]{EFEFEF}
            \textit{Nuclio}
            & \centering \NS
            & \centering \NA 
            & \centering \NA 
            & \centering\arraybackslash \NA \\
            \textit{OpenFaaS}
            & \centering \NS
            & \centering \NA 
            & \centering \NA 
            & \centering\arraybackslash \NA \\
            \hline
        \end{tabularx}
        \begin{tablenotes}
            \scriptsize
            \item[*] \textit{Only sequence and parallel execution are supported.}
        \end{tablenotes}
    \end{threeparttable}
\end{table*}
\paragraph{\orchestration}
\Cref{tab:tech-review-dev-orchestrators} presents the results of the platforms review with respect to the \textit{Function Orchestration} category.
Orchestrating multiple functions is an important task, with a the majority of reviewed FaaS platforms~(6/10) providing a dedicated function orchestrator aimed to facilitate this task.
Furthermore, in most cases orchestrators are provided as standalone components or service offerings, \eg Openwhisk Composer or Azure Durable Functions.
While it is also possible to use general-purpose workflow engines, e.g., Google Composer is a general-purpose workflow engine based on Apache Airflow which can be used to compose functions via generic HTTP operators, in this review we focus on dedicated function orchestrators.

Half of the reviewed orchestrators allows defining workflows using so-called \textit{orchestrating functions}, which define the required control flow involving multiple functions.
With this approach, orchestrators typically provide a smaller set of supported programming languages for defining workflows, in comparison with supported function runtimes.
Moreover, a set of supported control flow constructs such as sequences, exclusive and parallel gateways, is not always defined by the language itself.
For example, Azure Durable Functions relies entirely on the constructs of the underlying language, i.e., JavaScript or C\#, whereas Openwhisk Composer while also allows defining orchestrations using JavaScript-based functions, also introduces a set of so-called combinators, i.e., module-specific commands representing workflow constructs.
For instance, a loop in Openwhisk Composer-based workflow is defined via a module-specific command \texttt{composer.repeat} instead of the regular \texttt{for} loop.

AWS Lambda, Fission and Knative instead allow orchestrating functions by relying on DSL-based workflow definitions, \eg, AWS Step Functions provides a custom DSL for composing functions on AWS Lambda.
The list of supported control flow constructs is defined by the DSL itself, and it can vary significantly.
For example, the Eventing component of Knative allows defining sequences and parallel executions, while the orchestrators provided by AWS Lambda and Fission feature more expressive power enabling the description of more complex workflows.

Finally, most platforms do not specify any quotas for limiting the execution of the workflows orchestrating functions.
The only exceptions are Apache Openwhisk and AWS Step Functions, which allow to set timeouts for establishing maximum task execution time, with AWS Step Functions also allowing to limit I/O size.

\begin{center}
\noindent
\setlength{\fboxsep}{2mm}
\fbox{
    \centering
    \begin{minipage}{.9\linewidth}
      \textbf{Main Findings: Function Orchestration}
      \begin{itemize}
          \item[\ding{212}] More than a half of reviewed FaaS platforms~(6/10) support function orchestration by providing a dedicated function orchestrator.
          \item[\ding{212}] 50\% of offered orchestrators rely on custom DSLs for workflow definitions, and the other 50\% exploit orchestrating functions.
          \item[\ding{212}] All FaaS platforms supporting function orchestration document control flow constructs.
      \end{itemize}
\end{minipage}}
\end{center}

\paragraph{\testdeb}
\Cref{tab:tech-review-dev-testing} illustrates the review results regarding platform-specific testing mechanisms.%
\begin{table}[!b]
    \begin{threeparttable}
        \centering
        \caption{Classification of testing mechanisms for considered FaaS Platforms.
            \textsf{F} and \textsf{N} denote \textit{functional} and \textit{non-functional} testing mechanisms, respectively.
            The abbreviation \enquote{\NS{}} stays for \enquote{not specified}, meaning that no related information is documented.
        }
        \label{tab:tech-review-dev-testing}
        
        \footnotesize
        \begin{tabularx}{.99\columnwidth}{m{15mm} m{18mm} C}
            \hline
            & \centering Testing mechanisms
            & \centering\arraybackslash Documented features \\
            \hline
            \rowcolor[HTML]{EFEFEF}
            \textit{Apache Openwhisk}
            & \centering \textsf{F}
            & \centering\arraybackslash function invocation testing \\
            \textit{AWS Lambda}
            & \centering \textsf{F}, \textsf{N}\tnote{*}
            & \centering\arraybackslash function invocation testing, testing using external tools \\
            \rowcolor[HTML]{EFEFEF}
            \textit{Fission}
            & \centering \textsf{F}
            & \centering\arraybackslash function invocation testing \\
            \textit{Fn}
            & \centering \textsf{F}
            & \centering\arraybackslash unit testing of Java functions \\
            \rowcolor[HTML]{EFEFEF}            
            \textit{Google Cloud Functions}
            & \centering \textsf{F}\tnote{*}
            & \centering\arraybackslash testing using external tools\\
            \textit{Knative}
            & \centering \NS
            & \centering\arraybackslash \NS \\
            \rowcolor[HTML]{EFEFEF}
            \textit{Kubeless}
            & \centering \NS
            & \centering\arraybackslash \NS \\
            \textit{MS Azure \newline Functions}
            & \centering \textsf{F}\tnote{*}
            & \centering\arraybackslash testing using external tools \\
            \rowcolor[HTML]{EFEFEF}
            \textit{Nuclio}
            & \centering \NS
            & \centering\arraybackslash \NS \\
            \textit{OpenFaaS}
            & \centering \NS
            & \centering\arraybackslash \NS \\
            \hline
        \end{tabularx}
        \begin{tablenotes}
            \scriptsize
            \item[*] \textit{Only guidelines provided.}
        \end{tablenotes}
    \end{threeparttable}
\end{table}
\begin{table}[!h]
        \begin{threeparttable}
        \centering
        \caption{Classification of debugging mechanisms for considered FaaS Platforms.
            \textsf{L} and \textsf{R} denote \textit{local} and \textit{remote} debugging, respectively.
            The abbreviation \enquote{\NS{}} stays for \enquote{not specified}, meaning that no related information is documented.
        }
        \label{tab:tech-review-dev-debugging}
        \def\arraystretch{.9}
        \footnotesize
        \begin{tabularx}{.99\columnwidth}{m{16mm} m{18mm} C}
            \hline
            & \centering Debugging mechanisms
            & \centering\arraybackslash Documented features \\
            \hline
            \rowcolor[HTML]{EFEFEF}
            \textit{Apache Openwhisk}
            & \centering \NS\tnote{*}
            & \centering\arraybackslash \NS \\
            \textit{AWS Lambda}
            & \centering \textsf{L}
            & \centering\arraybackslash Step-through debugging using IDE plugins and AWS~SAM~CLI \\
            \rowcolor[HTML]{EFEFEF}
            \textit{Fission}
            & \centering \textsf{L}
            & \centering\arraybackslash Log-based debugging using platform's CLI tool\\
            \textit{Fn}
            & \centering \textsf{L}
            & \centering\arraybackslash Log-based debugging using platform's CLI tool \\
            \rowcolor[HTML]{EFEFEF}            
            \textit{Google Cloud Functions}
            & \centering \textsf{L}
            & \centering\arraybackslash Step-through debugging using dedicated functions 
            development framework \\
            \textit{Knative}
            & \centering \textsf{L}\tnote{**}
            & \centering\arraybackslash Log-based debugging using Kubernetes \\
            \rowcolor[HTML]{EFEFEF}
            \textit{Kubeless}            
            & \centering \textsf{L}\tnote{**}
            & \centering\arraybackslash Log-based debugging using Kubernetes \\
            \textit{MS Azure \newline Functions}
            & \centering \textsf{L}
            & \centering\arraybackslash Step-through debugging using dedicated functions 
            development framework \\
            \rowcolor[HTML]{EFEFEF}
            \textit{Nuclio}
            & \centering \NS
            & \centering\arraybackslash \NS \\
            \textit{OpenFaaS}
            & \centering \textsf{L}
            & \centering\arraybackslash Log-based debugging using platform's CLI tool \\
            \hline
        \end{tabularx}
        \begin{tablenotes}
            \scriptsize
            \item[*] \textit{A debugging tool for NodeJS applications is available, but deprecated.}
            \item[**] \textit{Only through features of the underlying hosting environment.}
        \end{tablenotes}
    \end{threeparttable}
\end{table}
The majority of reviewed FaaS platforms~(6/10) documents some features related to functional testing. 
As testing of the function code is not a direct responsibility of a FaaS platform, most platforms just provide mechanisms for invoking deployed functions for testing purposes, \eg using a dedicated CLI command or GUI.
In particular, Apache Openwhisk provides a tool for testing the invocation of NodeJS functions, AWS Lambda provides a GUI to invoke deployed functions for testing purposes, and Fn offers a function development kit~(FDK) for Java, which facilitates implementing unit tests.
AWS Lambda, Google Cloud Functions and MS Azure also provide guidelines on how to exploit external tooling for for functional testing of serverless applications, typically referring to the platform-specific IDE plugins they provide (formerly reported in \Cref{tab:tech-review-dev-idesdk}).
AWS Lambda also illustrate some aspects of non-functional testing, and in particular related to load testing of serverless applications.

The review results related to provided debugging mechanisms are shown in~\Cref{tab:tech-review-dev-debugging}.
One observation is that all reviewed FaaS platforms document mechanisms mainly related to local debugging.
Not surprisingly, the commercial platforms provide advanced debugging solutions, as all of them provide dedicated tooling for step-through debugging of functions.
Open source platforms instead mainly document log-based debugging, which can be done using either platform-specific tooling or tools supported by the underlying hosting platform, e.g., the \texttt{kubctl} tooling from Kubernetes.
Only in few cases~(2/10), platforms do not explicitly provide any information related to debugging, even if they may still support log-based debugging or debugging through available external tools.
Overall, both testing and debugging mechanisms provided by open source FaaS platforms are rather rudimentary, whereas commercial platforms often offer more features.
\begin{center}    
    \noindent
    \setlength{\fboxsep}{2mm}
    \fbox{
        \centering
        \begin{minipage}{.9\linewidth}
            \textbf{Main Findings: Testing and Debugging}            
            \begin{itemize}
                \setlength\itemsep{.15mm}
                \item[\ding{212}] The majority of reviewed FaaS platforms~(6/10) documents possible options for functional testing. Commercial platforms also provide more advanced options.
                \item[\ding{212}] Most of reviewed FaaS platforms~(8/10) support local debugging of functions.
                \item[\ding{212}] Open source FaaS platforms mainly support testing throughout function invocations, and log-based debugging.          
            \end{itemize}
    \end{minipage}}
\end{center}


\paragraph{\observability}
This category covers logging and monitoring dimensions of FaaS platforms, and the corresponding classification of considered platforms is shown in~\Cref{tab:tech-review-dev-observability}.
As with other tooling-related categories, commercial platforms provide standalone solutions for logging and monitoring.
For example, Amazon offers AWS CloudWatch~\cite{aws:cloud-watch} and AWS CloudTrail~\cite{aws:cloud-trail} for monitoring and logging of functions hosted on AWS Lambda.
Likewise, Microsoft provides Azure Application Insights~\cite{azure:app-insights} and Google offers its Operations suite~(formerly Stackdriver)~\cite{google:operations}.
In contrast, open source platforms rely on external tooling, e.g., using a combination of Prometheus~\cite{prometheus} and Grafana~\cite{grafana} for monitoring, or combining ElasticSearch~\cite{elasticsearch} and Kibana~\cite{elasticsearch:kibana} for logging.
Nuclio and Knative also support the integration with commercial monitoring and logging services, \ie Azure Application Insights and Google Cloud Operations, respectively.

\begin{table}[h]
    \begin{threeparttable}[b]
        \centering
        \caption{Classification of considered FaaS Platforms, based on the \observability category in the \devops view.}
        \label{tab:tech-review-dev-observability}
        \footnotesize
        \def\arraystretch{.9}
        \begin{tabularx}{.99\columnwidth}{
                m{17mm} @{}
                m{22mm} 
                m{24mm} 
                m{12mm}
            }
            \hline
            & \centering Monitoring 
            & \centering Logging 
            & \centering\arraybackslash Tooling Integr.\\
            \hline
            \rowcolor[HTML]{EFEFEF}
            \textit{Apache Openwhisk}                   
            & \centering Kamon, Prometheus, Datadog
            & \centering Logback (slf4j)
            & \centering\arraybackslash \NS \\
            \textit{AWS Lambda}
            & \centering AWS CloudWatch
            & \centering AWS CloudTrail, CloudWatch, CloudFormation
            & \centering\arraybackslash \NS \\
            \rowcolor[HTML]{EFEFEF}
            \textit{Fission}
            & \centering Istio + Prometheus
            & \centering Istio + Jaeger 
            & \centering\arraybackslash using a service mesh \\
            \textit{Fn}
            & \centering Prometheus, Zipkin, Jaeger
            & \centering Docker container logs 
            & \centering\arraybackslash push-based \\
            \rowcolor[HTML]{EFEFEF}            
            \textit{Google Cloud \newline Functions}
            & \centering Google Cloud Operations
            & \centering Google Cloud Operations
            & \centering\arraybackslash \NS \\
            \textit{Knative}
            & \centering Prometheus + Grafana, Zipkin, Jaeger
            & \centering ElasticSearch + Kibana, Google Cloud Operations 
            & \centering\arraybackslash push-based \\
            \rowcolor[HTML]{EFEFEF}
            \textit{Kubeless}
            & \centering Prometheus + Grafana
            & \centering \NS
            & \centering\arraybackslash \NS \\
            \textit{MS Azure \newline Functions}
            & \centering Azure Application Insights 
            & \centering Azure Application Insights 
            & \centering\arraybackslash \NS \\
            \rowcolor[HTML]{EFEFEF}
            \textit{Nuclio}
            & \centering Prometheus, Azure Application Insights
            & \centering stdout, Azure Application Insights
            & \centering\arraybackslash push-based, pull-based \\
            \textit{OpenFaaS}
            & \centering OpenFaaS Gateway + Prometheus
            & \centering Kubernetes cluster API, Swarm cluster API, Loki
            & \centering\arraybackslash pull-based \\
            \hline
        \end{tabularx}
    \end{threeparttable}
\end{table}

Most reviewed open source platforms~(5/7) document possible ways to integrate external logging and monitoring tools.
Essentially, the integration can be achieved either by configuring the FaaS platform to send events and logs to an external endpoint~(push-based) or vice versa, i.e., when an external component pulls events and logs from the platform~(pull-based).
Additionally, Fission allows enabling Istio, a service mesh, and installing add-ons, e.g., for monitoring and distributed tracing.

\begin{center}
\noindent
\setlength{\fboxsep}{2mm}
\fbox{
    \centering
    \begin{minipage}{.9\linewidth}
      \textbf{Main Findings: Observability}
      \begin{itemize}
          \item[\ding{212}] All commercial platforms offer dedicated tooling for the monitoring and logging of functions.
          \item[\ding{212}] All open source platforms rely on integration of third-party tooling for supporting the monitoring and logging of functions.
      \end{itemize}
\end{minipage}}
\vspace{.2\baselineskip}
\end{center}

\paragraph{\delivery}
The results of the review related to the aspects of application delivery are shown in~\Cref{tab:tech-review-dev-delivery}, with several observations that can be made on deployment automation.
Firstly, most platforms rely on some form of declarative deployment modeling~\cite{Endres2017_DeclImp-Patterns}, either by using proprietary tools and formats (\eg Azure Resource Manager or AWS SAM) or by relying on the deployment automation featured by their underlying hosting environments (\eg declarative deployments using custom resource definitions for Kubernetes-based platforms).
In open source platforms, the deployment is automated typically by using a platform-native CLI tool that takes a declarative application specification as an input.
Another thing worth mentioning is that the Serverless Framework, a popular solution for automating the deployment of FaaS-based applications, is explicitly mentioned only by Kubeless.

\begin{table}[t]
    \begin{threeparttable}[b]
        \centering
        \caption{Classification of considered FaaS Platforms, based on the \delivery category in the \devops view of our classification framework. \textsf{P} and \textsf{T} denote \textit{Platform-native tooling} and \textit{third party tooling}, respectively. The abbreviation \enquote{\NS{}} stays for \enquote{not specified}, meaning that no related information is documented.}
        \label{tab:tech-review-dev-delivery}
        \footnotesize
        \begin{tabularx}{.99\columnwidth}{m{5em} m{11em} C}
            \hline
            & \centering Deployment Automation & CI/CD \\
            \hline
            \rowcolor[HTML]{EFEFEF}
            \textit{Apache Openwhisk}                   
            & \centering wskdeploy~(\textsf{P})
            & \NS \\
            \textit{AWS Lambda}
            & \centering AWS Cloud Formation~(\textsf{P}), AWS SAM~(\textsf{P})
            & AWS CodePipeline(\textsf{P}) \\
            \rowcolor[HTML]{EFEFEF}
            \textit{Fission}
            & \centering Kubernetes~(\textsf{P})\tnote{*}
            & \NS \\
            \textit{Fn}
            & \centering Fn CLI~(\textsf{P})
            & \NS \\
            \rowcolor[HTML]{EFEFEF}            
            \textit{Google Cloud Functions}
            & \centering \NS~\tnote{**}
            & Cloud Build~(\textsf{P}) \\
            \textit{Knative}
            & \centering Kubernetes~(\textsf{P})\tnote{*}
            & \NS \\
            \rowcolor[HTML]{EFEFEF}
            \textit{Kubeless}            
            & \centering Kubernetes~(\textsf{P})\tnote{*}, Serverless Framework~(\textsf{T})
            & \NS \\
            \textit{MS Azure \newline Functions}
            & \centering Azure Resource Manager~(\textsf{P}) 
            & Azure Pipelines~(\textsf{P}), Azure App Service~(\textsf{P}), Jenkins~(\textsf{T}) \\
            \rowcolor[HTML]{EFEFEF}
            \textit{Nuclio}
            & \centering nuctl~(\textsf{P})\tnote{*}
            & \NS \\
            \textit{OpenFaaS}
            & \centering faas-cli~(\textsf{P})
            & OpenFaaS Cloud~(\textsf{P}), Jenkins~(\textsf{T}) \\
            \hline
        \end{tabularx}
        \begin{tablenotes}
            \scriptsize
            \item[*] \textit{Using Kubernetes specification with Custom Resource Definitions.}
            \item[**] \textit{Cloud Deployment Manager does not support function resources, and gcloud CLI only allows deploying functions manually.}
        \end{tablenotes}
    \end{threeparttable}
\end{table}

Most platforms~(6/10) do not provide any information on CI/CD pipelines integration.
Moreover, only OpenFaaS describes possible integration with Jenkins, by also providing a CI/CD-optimized version of the platform that already comes with out-of-the-box integration, \ie OpenFaaS Cloud.
Commercial platforms instead typically have a dedicated CI/CD service, \ie AWS CodePipeline, Cloud Build, or Azure Pipelines.

\begin{center}
\noindent
\setlength{\fboxsep}{2mm}
\fbox{
    \centering
    \begin{minipage}{.9\linewidth}
      \textbf{Main Findings: Application Delivery}
      \begin{itemize}
          \item[\ding{212}] 9/10 reviewed platforms follow a declarative approach to automate application deployment.
          \item[\ding{212}] Commercial platforms natively support CI/CD throughout dedicated tooling.
          \item[\ding{212}] From open source platforms only OpenFaaS documents integration with CI/CD pipelines.
      \end{itemize}
\end{minipage}}
\vspace{.2\baselineskip}
\end{center}


\paragraph{\codereuse}
The classification of considered FaaS platforms concerning aspects of code reuse are shown in~\Cref{tab:tech-review-dev-delivery}.
Notably, only AWS Lambda and MS Azure Functions are equipped with a function marketplace where to pick already existing, runnable functions. 
Amazon indeed provides the Serverless Application Repository, which includes various serverless use case applications, whereas Microsoft provides a more general-purpose marketplace called Azure Marketplace, which also offers applications based on Azure Functions.
At the time of writing, Google's marketplace instead does not offer applications based on Google Cloud Functions.
Going beyond marketplaces all platforms instead provide one or more code sample repositories, which typically reside in the same or in a separate project under the same GitHub organization.

\begin{table}[t]
    \begin{threeparttable}[b]
        \centering
        \caption{Classification of considered FaaS Platforms, based on the \codereuse category in the \devops view of our classification framework. The abbreviation \enquote{\NS{}} stays for \enquote{not specified}, meaning that no related information is documented.}
        \label{tab:tech-review-dev-codereuse}
        \footnotesize
        \begin{tabularx}{.99\columnwidth}{m{11em}@{} m{6em} C}
            \hline
            & \centering Function Marketplaces & Official Code Sample Repositories \\
            \hline
            \rowcolor[HTML]{EFEFEF}
            \textit{Apache Openwhisk}                   
            & \centering \NS
            & \ok \\ 
            \textit{AWS Lambda}
            & \centering AWS Serverless Application Repository
            & \ok \\ 
            \rowcolor[HTML]{EFEFEF}
            \textit{Fission}
            & \centering \NS
            & \ok \\ 
            \textit{Fn}
            & \centering \NS
            & \ok \\ 
            \rowcolor[HTML]{EFEFEF}            
            \textit{Google Cloud Functions}
            & \centering \NS
            & \ok \\ 
            \textit{Knative}
            & \centering \NS
            & \ok \\ 
            \rowcolor[HTML]{EFEFEF}
            \textit{Kubeless}
            & \centering \NS
            & \ok \\ 
            \textit{MS Azure \newline Functions}
            & \centering Azure Marketplace
            & \ok \\ 
            \rowcolor[HTML]{EFEFEF}
            \textit{Nuclio}
            & \centering \NS
            & \ok \\ 
            \textit{OpenFaaS}
            & \centering \NS
            &  \ok \\ 
            \hline
        \end{tabularx}
    \end{threeparttable}
\end{table}

\begin{center}
\noindent
\setlength{\fboxsep}{2mm}
\fbox{
    \centering
    \begin{minipage}{.9\linewidth}
      \textbf{Main Findings: Code Reuse}
      \begin{itemize}
          \item[\ding{212}] AWS Lambda and MS Azure Functions offer a marketplace where to pick ready-to-use functions and serverless applications.
          \item[\ding{212}] All platforms provide examples of functions in code sample repositories.
      \end{itemize}
\end{minipage}}
\vspace{.2\baselineskip}
\end{center}

\paragraph{\accessmgmt}
Finally, the results of the review related to the aspects of access management covering support for authentication mechanisms and access control are shown in~\Cref{tab:tech-review-dev-accessmgmt}.
With respect to available authentication mechanisms, commercial platforms offer dedicated services such AWS IAM or Google Cloud IAM that provide a powerful and highly-flexible way to control authentication.
Azure Functions instead supports usage of a Microsoft account, or authentication via a trusted third party credentials such as Facebook, Google, or Twitter.
On the contrary, most open source platforms do not cover such aspects, either by outsourcing this task to the underlying hosting environment (\eg Kubernetes) or implementing basic authentication using shared secrets.
In terms of access control capabilities, all commercial providers support defining access rules for who can invoke functions and how functions can access specific resources.
The documentation of most open source platforms does not provide details on such advanced topics.
The only exception is OpenFaaS, which allows configuring functions as read-only, hence preventing them to modify the underlying file system, and Kubeless which describes Kubernetes-based mechanisms.
Notably, even if some Kubernetes-based platforms do not explicitly mention the support for access control mechanisms, it is still possible to reuse Kubernetes-native mechanisms to control the accesses to functions and/or resources.

\begin{table}[t]
    \begin{threeparttable}[b]
        \centering
        \caption{Classification of considered FaaS Platforms, based on the \accessmgmt category in the \devops view of our classification framework. The abbreviation \enquote{\NS{}} stays for \enquote{not specified}, meaning that no related information is documented.}
        \label{tab:tech-review-dev-accessmgmt}
        \footnotesize
        \def\arraystretch{.9}
        \begin{tabularx}{.99\columnwidth}{m{18mm} C m{15mm}}
            \hline
                & \centering Authentication Mechanisms 
                & \centering\arraybackslash Access Control \\
            \hline
            \rowcolor[HTML]{EFEFEF}
            \textit{Apache Openwhisk}                   
                & \centering function invocation using shared secrets
                & \centering\arraybackslash \NS \\ 
            \textit{AWS Lambda}
                & \centering AWS IAM
                & \centering\arraybackslash resources, functions \\ 
            \rowcolor[HTML]{EFEFEF}
            \textit{Fission}
                & \centering \NS
                & \centering\arraybackslash \NS \\ 
            \textit{Fn}
                & \centering \NS
                & \centering\arraybackslash \NS \\ 
            \rowcolor[HTML]{EFEFEF}            
            \textit{Google Cloud Functions}
                & \centering Cloud IAM
                & \centering\arraybackslash resources, functions \\ 
            \textit{Knative}
                & \centering \NS
                & \centering\arraybackslash \NS \\
            \rowcolor[HTML]{EFEFEF}
            \textit{Kubeless}
                & \centering \ok\tnote{**}
                & \centering\arraybackslash \ok\tnote{**} \\ 
            \textit{MS Azure \newline Functions}
                & \centering Azure Active Directory, Facebook, Google, Microsoft account, Twitter
                & \centering\arraybackslash resources, functions \\ 
            \rowcolor[HTML]{EFEFEF}
            \textit{Nuclio}
                & \centering \NS
                & \centering\arraybackslash \NS \\
            \textit{OpenFaaS}
                & \centering function invocation using shared secrets, built-in basic authentication or proprietary OAuth2 API access plugins 
                & \centering\arraybackslash resources\tnote{*} \\ 
            \hline
        \end{tabularx}
        \begin{tablenotes}
            \scriptsize
            \item[*] \textit{Functions can be made read-only to forbid modifying the file system.}
            \item[**] \textit{Kubernetes-based mechanisms are described in the documentation.}
        \end{tablenotes}
    \end{threeparttable}
\end{table}

\begin{center}
\noindent
\setlength{\fboxsep}{2mm}
\fbox{
    \centering
    \begin{minipage}{.9\linewidth}
      \textbf{Main Findings: Access Management}
      \begin{itemize}
          \item[\ding{212}] Commercial platforms natively support authentication and resource access control.
          \item[\ding{212}] Open source platforms typically rely on features offered by the hosting environment to enforce authentication and resource access control.
      \end{itemize}
\end{minipage}}
\vspace{.2\baselineskip}
\end{center}

\section{FaaStener: Platform Selection Support System}
\label{sec:portal}
\begin{figure}[h]
    \centering
    \includegraphics[width=.99\columnwidth]{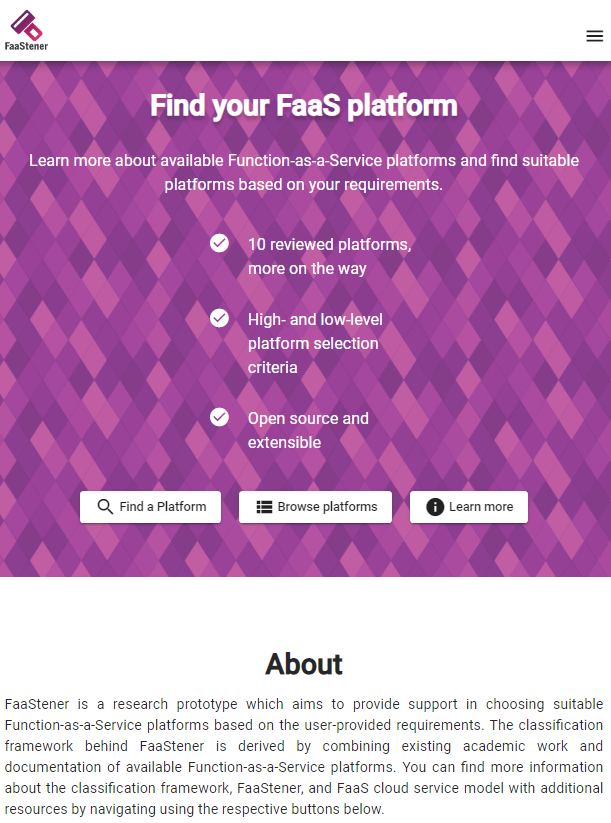}
    \caption{\faastener: A FaaS platforms selection support system}
    \label{fig:portal}
\end{figure}

\noindent
To facilitate the storage and usage of collected data for the FaaS selection process, we implemented an open source \textit{FaaS Platform Selection Support System}, called \faastener\footnote{The sources of \faastener are publicly available on GitHub at \url{https://github.com/faastener/faastener}, while running instance of \faastener can be accessed at \url{https://faastener.github.io}.}.
\faastener is a web-based application developed in Angular and Angular Material components library\footnote{\textit{Angular}: \url{https://angular.io}. \textit{Angular Material}: \url{https://material.angular.io}.} providing a graphical user interface for interacting with the platform's data (\cref{fig:portal}).

\faastener enables users to browse the information for a chosen reviewed platform, to perform a multi-attribute search for suitable FaaS platform based on the specified requirements (e.g., taken from the business, technical, or a combination of both views) and to visualize the corresponding results, which significantly improves the usability of our classification framework and technology review.
Another advantage is the possibility to perform feature-wise comparison of multiple platforms by simultaneously looking at their classification in separate tabs of the browser.

It is finally worth noting that \faastener has been intentionally designed to ease the maintenance and extension of the systematic knowledge base it enables browsing, which is currently constituted by the results of our technology review.
The technology review results are indeed encoded as JSON files, which structure represents the overall structure of our classification framework.
Since \faastener is open sourced, updating information on an already reviewed FaaS platform or adding a new platform just require to update or upload the corresponding JSON files, by exploiting the mechanism of pull requests in GitHub, which allows to accept contributions from all interested parties.
To facilitate the contribution process, \faastener provides the relevant documentation and resource links in a dedicated site section.


\section{Threats to validity}
\label{sec:threats-to-validity}
\noindent
Wohlin et al.~\cite{Wohlin2012_ExpSE} provide a standardised classification of threats of validity potentially affecting secondary studies.
Four of such potential threats may apply to our study, namely threats to \textit{external} validity, to \textit{construct} and \textit{internal} validities, and to \textit{conclusions} validity.
We hereafter discuss them and illustrates the countermeasure we adopted to mitigate them.

External validity concerns the applicability of a set of results in a more general context~\cite{Wohlin2012_ExpSE}.
Since we focused on self-declared information available on the official websites and GitHub repositories of FaaS platforms, our results and observations may only be partly applicable to the broader practices and information available on FaaS platforms, hence threatening external validity.
To reinforce the validity of our findings, we organised 3 feedback sessions during our analysis of the documentation available on the websites and GitHub repositories of considered FaaS platforms.
We analyzed the discussion following-up from each feedback session, and we exploited this qualitative data to fine-tune our classification framework, the actual classification of considered platforms, and teh applicability of our findings.
We also made our data easily accessible throughout the \textsc{FaaStener}, which is open source, and whose GitHub repository includes all artifacts we produced during our analysis.
We believe that this can help in making our classification framework, technology review and observations more explicit and applicable in practice.

Construct and internal validity instead concern the generalizability of the constructs under studies and the validity of the methods actually exploited to study and analyze data, respectively~\cite{Wohlin2012_ExpSE}.
These inherently includes the possible biases affecting our study.
To mitigate both above threats, we adopted various inter-rater reliability assessment rounds, aimed at avoiding biases by triangulation (\cref{sec:research-design}).
We indeed performed various iterations among the authors both for (i) refining the initially obtained classification framework and obtaining that show in \cref{sec:framework}, and for (ii) cross-checking the actual classification of considered FaaS platforms until a full agreement among all authors was achieved.

Finally, threats to conclusions validity may apply depending on the degree to which the conclusions of a study are reasonably based on the available data~\cite{Wohlin2012_ExpSE}. 
To mitigate this threat, we exploited theme coding and inter-rater reliability assessment to limit observer and interpretation biases, both while developing the classification framework and while actually classifying the considered FaaS platforms.
The ultimate goal in both cases was indeed to perform a sound analysis of the data we retrieved from official websites and GitHub repositories of considered FaaS platforms.
In addition, the conclusions drawn in this paper were independently drawn by the authors, and they were then double-checked against the available information in joint discussion sessions.

\section{Related Work}
\label{sec:related}
\noindent 
There already exist some research efforts qualitatively classifying and reviewing FaaS platforms. 
Such efforts were already listed in \cref{sec:research-design}, as they were reviewed during Step 1 of our study and exploited in Step 2 to derive the initial version of our classification framework.
Throughout the subsequent steps, we extended the set of criteria considered while classifying FaaS platforms, as well as the set of considered platforms themselves.

A concrete example in this direction is the review by Kritikos and Skrzypek~\cite{related:8605774}, who conducted an assessment of serverless frameworks using a set of criteria corresponding to various phases of a serverless application's lifecycle, e.g., design, development, deployment, or execution.
In their work, Kritikos and Skrzypek distinguish between provisioning and abstraction frameworks.
The former are responsible for provisioning serverless applications by enabling a \enquote{mini-serverless platform}, whereas the latter abstract away the technical details of two or more serverless platforms.
As a result, such Kubernetes-based FaaS platforms as Fission and Kubeless are classified as provisioning frameworks, and the framework Serverless is described as an abstraction framework.
The Authors then evaluate and compare both provisioning and abstraction frameworks by considering 14 qualitative criteria.
Our effort goes a step further in two directions by (i) considering a wider set of criteria to analyse and classify FaaS platforms, and by (ii) considering a wider set of FaaS platforms.

Other examples of criteria-based reviews of FaaS platforms are those by Lee et al.~\cite{related:8457830}, Lynn et al.~\cite{related:8241104}, and Mohanty et al.~\cite{related:8591002},  
Lee et al.~\cite{related:8457830} evaluate the performance characteristics of production FaaS platforms, and include a feature-wise comparison of the chosen products, i.e., FaaS platforms from Amazon, Microsoft, Google, and IBM.
Lynn et al.~\cite{related:8241104} evaluate seven enterprise serverless platforms with the term serverless being used interchangeably with FaaS.
The chosen products are reviewed based on the set of defined criteria and include, e.g., AWS Lambda, Microsoft Azure Functions, and Google Cloud Functions.
Mohanty et al.~\cite{related:8591002} conduct a feature-wise comparison consisting of 15 features together with performance evaluation of four open source FaaS platforms, namely Kubeless, OpenWhisk, Fission, and OpenFaaS.
In this work, the Authors put more focus on performance evaluation of open source platforms.

Some works focus on evaluating FaaS-related, complementary concepts such as function orchestrators and serverless application repositories.
L{\'o}pez et al.~\cite{related:8605772} evaluate and compare three commercial FaaS orchestrations systems, namely AWS Step Functions, IBM Composer, and Azure Durable Functions.
The chosen function orchestrators are reviewed based on the set of defined criteria, and their runtime overhead is evaluated experimentally.
Spillner~\cite{related:spillner2019quantitative} investigates how FaaS-based applications are specified, stored, and offered using the AWS Serverless Applications Repository.
Various statistics are presented related to different aspects of functions' lifecycle.

All the aforementioned research efforts, together with other existing criteria-based reviews of FaaS platforms~\cite{related:snowballing:rajan2018serverless,related:snowballing:palade2019evaluation,related:snowballing:kumar2019serverless,related:snowballing:kalnauz2019productivity,related:snowballing:gand2020serverless,related:snowballing:bortolini2019investigating}, provide valuable contributions by taking various angles for reviewing FaaS platforms and related relevant concepts.
At the same time, the features considered in all such efforts for classifying FaaS platforms, as well as the classified FaaS platforms themselves, do not overlap, hence scattering knowledge across different pieces of literature.
We instead aim to provide a more comprehensive classification framework and technical review, by (i)~extending the set features already considered in literature for classifying FaaS platforms, (ii)~enabling to classify FaaS platforms at two different levels of detail, based on the involved stakeholders, and (iii)~providing a more comprehensive technical review for the most popular FaaS platforms based on our classification framework. 
In addition, we also provide a first selection support system enabling researchers and practitioners to look for the FaaS platforms most suited to their needs.

Finally, it is worth relating our work to several research efforts tackling the problem of FaaS platforms benchmarking.
For instance, Wang et al.~\cite{related:Wang:2018:PBC:3277355.3277369} analyze the performance, resource management, and architectures of the FaaS platforms from Amazon, Microsoft, and Google.
Authors implement measurement functions to discover hidden architecture details of these platforms and collect performance-related data.
Our work differs from that by Wang et al.~\cite{related:Wang:2018:PBC:3277355.3277369} since we focus on feature-wise classification and reviewing of FaaS platforms, rather than on their benchmarking.
Similar considerations apply to other research efforts benchmarking FaaS platforms~\cite{related:bm:10.1007/978-3-319-99819-0_11,related:bm:10.1007/978-3-319-75178-8_34,kuhlenkamp:evaluation-faas-platforms,related:bm:doi:10.1002/cpe.4792}.


\section{Conclusions \& Future Work}
\label{sec:conclusion}
\noindent 
With the ultimate goal of supporting researchers and practitioners in classifying existing FaaS platforms and choosing those most suited to their needs, we presented a classification framework, technology review and selection support system for FaaS platforms.
Our \textit{FaaS Platform Classification Framework} enables the characterisation of FaaS platforms under two different perspectives.
Such perspectives are clearly shown in the \textit{FaaS Platform Technology Review} we presented based on such framework, which provides both a business and a technical view on ten existing FaaS Platforms.
The \textit{FaaS Platform Selection Support System} then materialises the results presented in our technology review in the form of a web-based application enabling researchers and practitioners to submit multi-attribute queries for searching for FaaS platforms satisfying their needs, and to compare different platforms based on their managerial or technical views. 

Apart from enabling a thorough comparison of existing FaaS platforms, our contributions result in the stemming out of various novel research and innovation directions. 
For instance, we observed a lack of solutions for testing, debugging and versioning FaaS-based project, especially for separately testing, debugging and versioning of FaaS-based applications from the functions used to implement them. 
Portability is another example of research direction stemming out from our study, as FaaS platforms resulted to be quite heterogeneous in featured triggers, in supported orchestration, monitoring and logging solutions, and in deployment automation technologies.
This obviously hampers the migration of a FaaS-based application from a FaaS platform to another, hence requiring solutions for enhancing their portability.

In addition of the above listed directions for future work, we actually plan to extend the results and support provided by the proposed classification framework, technology review and selection support system.
We intend to extend the classification framework defined in this work with other dimensions (\eg security- or performance-related aspects), and to extend the technology review both by reviewing already considered platforms with the novel introduced dimensions and by considering other existing FaaS platforms and/or related tooling (\eg function orchestrators).
In addition, we plan to combine the categories and dimensions forming our classification framework into advanced metrics for evaluating and comparing FaaS platforms, and to exploit such advanced metrics to extend our selection support system into a full-fledged decision support system, which will be capable of providing smart informed recommendations to researchers and practitioners needing a FaaS platform for run their applications.

\medskip \noindent
\textbf{Acknowledgements.}
This work is partially funded by the European Union's Horizon 2020 research and innovation project \emph{RADON}~(825040), and by the projects \emph{AMaCA} (POR-FSE) and \emph{DECLware} (University of Pisa, PRA\_2018\_66).
We also wish to thank Roman Bunz, Simon Hauser, and Leon Kiefer~(listed alphabetically), who contributed to setting up this work during an initial exploratory study of existing FaaS platforms.

\section*{References}
\bibliography{bibliography}

\smallskip \noindent
{\footnotesize All links were last followed on the 27th March 2020.}

\end{document}